\documentclass[trackchanges,twocolumn]{aastex701}

\newcommand{\eg}{e.g., }
\newcommand{\ie}{i.e., }
\newcommand{\Msun}{M_{\odot}}

\newcommand{\Rcsm}{r_{\rm CSM}}
\newcommand{\Menv}{M_{\rm env}}

\newcommand{\XH}{X({\rm H})_{\rm env}}

\newcommand{\kms}{km~s$^{-1}$}

\newcommand{\Nifs}{$^{56}$Ni}
\newcommand{\Mni}{M{\rm (^{56}Ni)}}

\newcommand{\Mdot}{\dot{M}}
\newcommand{\vw}{v_{\rm wind}}
\newcommand{\vi}{v_{\rm 0}}
\newcommand{\vf}{v_{\rm \infty}}

\newcommand{\Eexp}{E_{\rm exp}}
\newcommand{\Myr}{\Msun~{\rm yr}^{-1}}

\newcommand{\vph}{v_{\rm ph}}

\newcommand{\Rout}{r_{\rm out}}

\newcommand{\tPT}{t_{\rm PT}}
\newcommand{\texp}{t_{\rm exp}}
\def\gsim{\mathrel{\rlap{\lower 4pt \hbox{\hskip 1pt $\sim$}}\raise 1pt
\hbox {$>$}}}
\def\lsim{\mathrel{\rlap{\lower 4pt \hbox{\hskip 1pt $\sim$}}\raise 1pt
\hbox {$<$}}}

\newcommand{\hc}{H_{\rm 0}}
\newcommand{\omegac}{\Omega_{\rm 0}}
\newcommand{\grmid}{(g-r)_{\tPT / 2}}
\newcommand{\BVmid}{(B-V)_{\tPT / 2}}
\newcommand{\stella}{{\tt STELLA}}
\newcommand{\ha}{\rm H \alpha}
\newcommand{\vha}{v_{\rm \ha}}
\newcommand{\feii}{\rm Fe~{\sc ii}}
\newcommand{\vfe}{v_{\rm Fe~{\sc II}}}
\newcommand{\gpcyr}{\rm Gpc^{-3}~\rm yr^{-1}}
\newcommand{\recii}{r({\rm \frac{ECSN}{SN~II}})}
\newcommand{\reccc}{r({\rm \frac{ECSN}{CCSN}})}
\newcommand{\recsn}{R_{\rm ECSN}}
\newcommand{\dmec}{\Delta M_{\rm ECSN}}
\newcommand{\ebv}{E(B-V)}

\begin{document}
%TC:ignore
\title{Electron-Capture Supernova Candidates from Light Curves: Implications for Their Progenitors and Explosion Properties}

\author[orcid=0009-0008-0189-863X]{Masato Sato}
\affiliation{Department of Earth Science and Astronomy, The University of Tokyo, 3-8-1 Komaba, Meguro-ku, Tokyo 153-8902, Japan}
\email[show]{masato.sato.astro@gmail.com}

\author[orcid=0000-0001-8537-3153]{Nozomu Tominaga}
\affiliation{National Astronomical Observatory of Japan, National Institutes of Natural Sciences, 2-21-1 Osawa, Mitaka, Tokyo 181-8588, Japan}
\affiliation{Astronomical Science Program, Graduate Institute for Advanced Studies, SOKENDAI, 2-21-1 Osawa, Mitaka, Tokyo 181-8588, Japan}
\affiliation{Department of Physics, Faculty of Science and Engineering, Konan University, 8-9-1 Okamoto, Kobe, Hyogo 658-8501, Japan}
\email[]{nozomu.tominaga@nao.ac.jp}

\author[0000-0002-5726-538X]{Sergei I. Blinnikov}
\affiliation{NRC Kurchatov Institute, 123182 Moscow, Russia}
% \affiliation{Dukhov Automatics Research Institute (VNIIA), 127055 Moscow, Russia}
\affiliation{Sternberg Astronomical Institute, MSU, 119234 Moscow, Russia}
\email[]{sblinnikov@bk.ru}

\author[0000-0002-0564-1101]{Marat Sh. Potashov}
\affiliation{NRC Kurchatov Institute, 123182 Moscow, Russia}
\affiliation{Sternberg Astronomical Institute, MSU, 119234 Moscow, Russia}
\affiliation{Keldysh Institute of Applied Mathematics RAS, 4 Miusskaya Square, 125047 Moscow, Russia}
\email[]{marat.potashov@gmail.com}

\author[0000-0003-1169-1954]{Takashi J. Moriya}
\affiliation{National Astronomical Observatory of Japan, National Institutes of Natural Sciences, 2-21-1 Osawa, Mitaka, Tokyo 181-8588, Japan}
\affiliation{Astronomical Science Program, Graduate Institute for Advanced Studies, SOKENDAI, 2-21-1 Osawa, Mitaka, Tokyo 181-8588, Japan}
\affiliation{School of Physics and Astronomy, Monash University, Clayton, Victoria 3800, Australia}
\email[]{takashi.moriya@nao.ac.jp}

\author[0000-0002-1125-9187]{Daichi Hiramatsu}
% \affiliation{Center for Astrophysics \textbar~Harvard \& Smithsonian}
% \affiliation{The NSF AI Institute for Artificial Intelligence and Fundamental Interactions}
\affiliation{Department of Astronomy, University of Florida, 211 Bryant Space Science Center, Gainesville, FL 32611-2055, USA}
% \email[]{daichi.hiramatsu@cfa.harvard.edu}
\email[]{dhiramatsu@ufl.edu}

\author[0000-0003-3459-2270]{Francisco Förster}
\affiliation{Data and Artificial Intelligence Initiative (IDIA), Faculty of Physical and Mathematical Sciences, Universidad de Chile, Santiago, Chile}
\affiliation{Millennium Institute of Astrophysics, Nuncio Monseñor Sotero Sanz 100, Of. 104, Providencia, Santiago, Chile}
\affiliation{Center for Mathematical Modeling, Universidad de Chile, Beauchef 851, Santiago 8370456, Chile}
\email[]{francisco.forster@gmail.com}

\author[0000-0003-0227-3451]{Joseph P Anderson}
\affiliation{European Southern Observatory, Alonso de Córdova 3107, Vitacura, Casilla 19001, Santiago, Chile}
\email[]{janderso@eso.org}

\begin{abstract}

Core-collapse supernovae are explosions of massive stars. 
% Core-collapse supernovae are terminal explosions of massive stars. 
While most massive stars end as iron-core-collapse supernovae, less massive stars are expected to explode as electron-capture supernovae (ECSNe), defining the low-mass boundary of core-collapse supernovae. 
% ECSNe were proposed about 40 years ago, and first-principles simulations consistently predict their successful explosions. 
ECSNe were proposed $\sim 40$ years ago, and first-principles simulations predict their successful explosions with low energies of $\sim 10^{50}$~erg. 
% ECSNe were proposed about 40 years ago, and first-principles simulations predict their successful explosions with low energies of $\sim 10^{50}$~erg. 
% Observational identification and characterization of ECSNe are crucial for the completion of stellar-evolution theory.
% To date, only one convincing candidate, SN~2018zd, has been proposed in addition to the historical SN~1054, the progenitor of the Crab Nebula. 
% Nevertheless, only one convincing candidate, SN~2018zd, has been proposed other than the historical SN~1054, the progenitor of the Crab Nebula. 
Nevertheless, only one convincing candidate, SN~2018zd, has been proposed other than SN~1054, the progenitor of the Crab Nebula. 
We search for ECSN candidates among Type~II SNe from the literature and a public Zwicky Transient Facility sample, using a color-based diagnostic, selecting ten candidates with blue colors at the middle of the plateau. 
% We search for ECSN candidates among Type~II SNe from the literature and a public Zwicky Transient Facility sample, using a color-based diagnostic, selecting ten ECSN candidates with blue colors at the middle of the plateau. 
% We search for ECSN candidates among Type~II SNe reported in the literature and within the public Zwicky Transient Facility (ZTF) sample, using a state-of-the-art color-based diagnostic, and select ten blue objects as ECSN candidates, composed of three \textit{gold candidates}, for which mid-plateau spectra disfavor the strong CSM interaction, and seven \textit{silver candidates}, for which mid-plateau spectra are not available. 
% We classify three of them as \textit{gold candidates}, for which mid-plateau spectra disfavor strong circumstellar-medium interaction that would otherwise make the SN appear bluer, and seven as \textit{silver candidates}, for which such spectra are not available. 
% We classify three as \textit{gold}, for which a spectrum around middle of the plateau disfavors strong circumstellar-medium interaction that would otherwise make the SN appear bluer, and seven as \textit{silver}, for which such spectra are not available. 
We classify three as \textit{gold}, for which a spectrum around the middle of the plateau disfavors strong circumstellar-medium interaction that would make the SN bluer, and seven as \textit{silver} without such spectra. 
Comparing the observed multicolor light curves with radiation-hydrodynamical models, 
we infer the explosion energies, $(0.4-1.7)\times10^{50}$~erg for the \textit{gold candidates} and $(0.4-2.7)\times10^{50}$~erg including the \textit{silver candidates}, consistent with first-principles predictions and the mass-loss rates, $3\times10^{-3} - 3 \times 10^{-2}~\Myr$ for the \textit{gold candidates}, which remain similar when the \textit{silver candidates} are included, higher than those expected for the early super-asymptotic-giant-branch phase.
The ECSN occurrence ratios among SNe~II are inferred as $3.0^{+10.6}_{-2.9}$ and $15.7^{+17.3}_{-12.7}~\%$ from the \textit{gold} and \textit{silver candidates}, respectively, which we interpret as lower and upper limits.
% The ECSN occurrence ratios among SNe~II, $\recii$, are inferred as $3.0^{+10.6}_{-2.9}$ and $15.7^{+17.3}_{-12.7}~\%$ from \textit{gold} and \textit{silver candidates}, respectively, which we interpret as lower and upper limit, respectively.
% The candidates suggest an occurrence ratio of ECSNe among SNe~II of $0< r {\rm (\frac{ECSN}{SN~II})} \lesssim 15.7~\%$. 
% The candidates suggest that the occurrence ratio of ECSNe among CCSNe is $0< r {\rm (\frac{ECSN}{CCSN})} \lesssim 9.2~\%$. 
% To more robustly identify ECSNe and refine the estimates of the occurrence ratio, spectroscopic follow-ups of future ECSN candidates, particularly around the mid-plateau phase are essential.
To robustly identify ECSNe and refine this ratio, spectroscopic follow-ups of ECSN candidates around the middle of the plateau are essential.

\end{abstract}

%TC:ignore

% \keywords{\uat{Galaxies}{573} --- \uat{Cosmology}{343} --- \uat{High Energy astrophysics}{739} --- \uat{Interstellar medium}{847} --- \uat{Stellar astronomy}{1583} --- \uat{Solar physics}{1476}}


\section{Introduction \label{sec:intro}}

 Core-collapse supernovae (CCSNe) are the terminal explosions of massive stars, triggered by the collapse of their cores. Collapsing cores are divided into two mass-dependent types: Iron (Fe) cores, for which progenitors are massive enough to ignite static Si burning, 
collapse because of photo-disintegration and the stars explode as Fe-core-collapse supernovae (FeCCSNe, \citealt{Colgate1966-cq, Bethe1985-kh, Woosley2002-sx}).
% Oxygen-neon-magnesium (ONeMg) cores, for which progenitors are not massive enough to ignite static O burning, are supported by electron degeneracy pressure. 
Oxygen-neon-magnesium (ONeMg) cores, for which progenitors are not massive enough to ignite Ne burning, are supported by electron degeneracy pressure. 
Degenerate ONeMg cores collapse owing to electron capture by Mg and Ne, and the stars explode as electron-capture supernovae (ECSNe, \citealt{Miyaji1980-mf, Nomoto1982-zj, Nomoto1984-au, Nomoto1987-fr}).
 The progenitors of ECSNe are likely to be less massive than those of FeCCSNe, defining the low-mass end of the CCSN-progenitor distribution.
 % However, the stellar mass range that forms ECSNe is not clear and may be narrow or even non-existent for solar metallicity \citep{Poelarends2008-hq, Langer2011-zv, Doherty2015-wv, Limongi2023-db}.

% If the progenitors of FeCCSNe retain a massive hydrogen-rich (H-rich) envelope, they are red supergiant (RSG) stars \citep{Smartt2009-lj}.
% Stellar-evolution models predict quasi-steady RSG winds with mass-loss rates of order $10^{-6}-10^{-5}~\Myr$ using empirical relation \citep{Jager1988-xd, Beasor2020-ag, Beasor2023-tc}, broadly consistent with estimates for nearby RSG stars in the Large Magellanic Cloud \citep{Goldman2016-hb} and Small Magellanic Cloud \citep{Yang2023-dn}.
% However, several recent studies suggest that most RSG stars experience substantially enhanced mass loss in the final years before explosion, with inferred rates reaching $\sim10^{-4}-10^{-2}\Myr$ and confined within compact circumstellar shells with the radii $\lesssim10^{15}-10^{16}$~cm \citep{Yaron2017-lq,Morozova2017-oj, Forster2018-ox,Bruch2021-mg,Burrows2021-ts}. 
% Such enhanced mass loss is often attributed to envelope instabilities or energy injection associated with advanced nuclear burning, though its physical origin remains uncertain \citep{Yoon2010-ad,Quataert2016-ql,Moriya2014-oh,Sengupta2025-vv,Suzuki2025-ss}.

The progenitors of ECSNe are super-asymptotic-giant-branch (super-AGB) stars if they retain a hydrogen-rich (H-rich) envelope.
The terminal fate of a super-AGB star is governed by the competition between mass loss from the envelope and the growth of its degenerate core \citep{Siess2007-ea,Doherty2017-vu,Limongi2023-db}. 
If the mass loss dominates, the envelope is completely stripped before the core reaches the Chandrasekhar mass, leaving an ONe white dwarf. 
If the core growth proceeds faster, the core mass approaches to the Chandrasekhar limit and collapses via electron captures, leading to an ECSN.
Recent stellar-evolution models suggest that the mass-loss rate during the early super-AGB phase is around $10^{-5}–10^{-4}~\Myr$ \citep{Limongi2023-db}, 
but how this rate evolves toward the end of the super-AGB phase and whether episodic enhancements of mass loss occur remain uncertain.
% If both the mass-loss and core-growth rates are sustained, the upper-mass limit for a star to explode as an ECSN is $9.22~\Msun$.
If both the mass-loss and core-growth rates are sustained, the estimate of lower-mass limit for a star to explode as an ECSN is $8.50~\Msun$ \citep{Limongi2023-db}.
% Combined with the estimate of the lower-mass limit of $8.50~\Msun$ -- the minimum mass for a star to experience off-center C ignition -- this implies 
Combined with the estimate of the upper-mass limit of $9.22~\Msun$ -- the minimum mass for a star to ignite Ne burning -- this implies 
that ECSNe could account for $\sim10\%$ of all CCSNe, assuming a standard initial mass function \citep{Salpeter1955-fe}.
% an ECSN occurence ratio of $\sim 10 \%$ among CCSNe, assuming standard initial mass function \citep{Salpeter1955-fe}.
However, some studies suggest that the stellar mass window for ECSNe may be non-existent for solar metallicity \citep{Langer2011-zv} because of a strong metal-driven mass loss.
Consequently, the initial mass range for ECSNe remains highly uncertain and likely depends sensitively on metallicity.

ECSN explosions have been successfully simulated under the assumption of spherical symmetry owing to the low-density envelopes of their progenitors \citep{Kitaura2006-ia} in contrast to FeCCSNe for which multidimensional effects such as convection and turbulence are crucial for successful explosions \citep{Takiwaki2012-jo}. 
First-principles simulations revealed a low explosion energy, $\sim 10^{50}$~erg, and small amount of \Nifs, $0.002-0.004~\Msun$, for ECSNe \citep{Kitaura2006-ia, Janka2008-ai, Wanajo2009-yo}. 
The elemental abundance ratios are consistent with the Crab Nebula, a remnant of SN~1054 \citep{Davidson1982-xe, Nomoto1982-zj, Wanajo2009-yo, Temim2024-cc}.
% The elemental abundance ratios of ECSNe, such as high Ni/Fe ratio of $\sim 1-3$ are consistent with the Crab Nebula, a remnant of SN~1054 \citep{Davidson1982-xe, Nomoto1982-zj, Wanajo2009-yo}.
% Moreover, \citet{Tominaga2013, Moriya2014-oh} calculated ECSN light curves and found that they exhibit a plateau with a bolometric brightness of $-15$ to $-16$ mag (as bright as that of SNe~II of RSG progenitors), a faint tail of $\sim -11$ mag at the onset, and a large drop of $\sim 4$ mag from plateau to tail. These calculations were performed using SAGB progenitor models with radii ranging $6.5 \times 10^{13}-7.3 \times 10^{13}$ cm.

% Both types of explosions are observed as H-rich Type~II supernovae (SNe~II) if the progenitors retain a H-rich envelope.
ECSNe are observed as H-rich Type~II supernovae (SNe~II)\footnote{
In this paper, ``SNe~II'' does not include Type IIn or IIb SNe.
% In this paper, ``SNe~II'' denotes Type IIP and IIL SNe only, and does not include other subtypes (\eg Type IIn or IIb).
} if the progenitors retain a H-rich envelope, similarly to FeCCSNe from H-rich red-supergiant (RSG) progenitors \citep{Tominaga2013, Moriya2014-oh}.
When an SN II occurs in the absence of a dense circumstellar medium (CSM), the SN exhibits an initial shock-breakout flash.
% When the progenitor explodes without a dense circumstellar medium (CSM), the SN exhibits an initial shock-breakout flash.
After the shock breakout, a plateau phase follows, powered by the release of energy from the shock-heated H-rich envelope.
The duration and luminosity of the plateau reflect the explosion energy, pre-supernova radius, and envelope mass of the progenitor \citep{Litvinova1985-ty, Popov1993-cu, Eastman1994-hs, Kasen2009-qk}.
% The subsequent tail phase is generally powered by the radioactive decay of $^{56}$Co, the daughter nucleus of $^{56}$Ni synthesized during explosive nucleosynthesis, and its luminosity is approximately proportional to the ejected $^{56}$Ni mass, but it can also be powered or augmented by ejecta-CSM interaction when the explosion occurs within a dense and extended CSM \citep{Moriya2014-oh}.
The subsequent tail phase is generally powered by the radioactive decay of $^{56}$Co, the daughter nucleus of $^{56}$Ni synthesized during explosive nucleosynthesis, and its luminosity is approximately proportional to the ejected $^{56}$Ni mass.
% In contrast, if the explosion occurs within a dense CSM, an early blue brightening -- so-called wind breakout -- is observed, 
If the explosion occurs within a dense CSM, an early blue brightening -- so-called wind breakout -- is observed instead of the canonical shock breakout, 
% as the shock breakout from the SN diffuses out into the CSM \citep{Chevalier2011-ds,Moriya2011-fr,Morozova2017-oj,Moriya2018-ve}
as the breakout radiation diffuses through the CSM \citep{Chevalier2011-ds,Moriya2011-fr,Morozova2017-oj,Moriya2018-ve}.
% Moreover, in the presence of dense CSM, ejecta-CSM interaction can further contribute to the luminosity by efficiently converting kinetic energy into radiation from early breakout-plateau phases to later tail phase, depending on the extent of the dense CSM.
Moreover, depending on the extent of the CSM, ejecta-CSM interaction can further contribute to the luminosity by efficiently converting kinetic energy into radiation, from early times to the tail phase.

The bolometric light curves of ECSNe exhibit plateaus of $-15$ to $-16$ mag \citep{Tominaga2013}, similarly to those of H-rich FeCCSNe with $\sim-15$ to $-18$ mag \citep{Sukhbold2016}, despite their low explosion energies, owing to the extended progenitor envelope structure.
By contrast, the tails of ECSNe are faint, $\sim -11$ mag at the onset, due to the low \Nifs~yields, resulting in a large drop of $\sim 4$ mag from plateau to tail whereas FeCCSNe exhibit tails of $\sim-12.5$ to $-15.5$ mag at the onset, and a drop of $\sim2.5$ mag from plateau to tail.
% ECSNe also display distinctly bluer plateau colors than low-mass FeCCSNe without significant CSM interaction, reflecting the low-density and extended envelope structure of super-AGB progenitors due to their degenerate cores, which place the photosphere deeper than the H-recombination front during the plateau \citep{Kozyreva2021-ci, Sato2024-kt}.
ECSNe also display distinctly bluer plateau colors than low-mass FeCCSNe without significant CSM interaction, reflecting the low-density and extended envelope structure of super-AGB progenitors due to their degenerate cores \citep{Kozyreva2021-ci, Sato2024-kt}.
Leveraging this blue-plateau signature, \citet{Sato2024-kt} proposed a new color-based diagnostic to discriminate ECSNe from low-mass FeCCSNe.

% SNe~II have been extensively discovered and investigated by modern transient surveys and follow-up programs, such as the Carnegie Supernova Project (CSP, \citealt{Hamuy2006-wi}), Panoramic Survey Telescope and Rapid Response System (Pan-STARRS, \citealt{Chambers2016-tp}), Palomar Transient Factory (PTF, \citealt{Law2009-te}), Lick Observatory Supernova Search (LOSS, \citealt{Leaman2011-jd, Li2011-bx}), ASAS-SN (ASAS-SN, \citealt{Shappee2014-hh}), Asteroid Terrestrial-impact Last Alert System (ATLAS, \citealt{Tonry2018-xe}), and Zwicky Transien Facility (ZTF, \citealt{Bellm2018-cn}).
SNe~II have been extensively discovered and investigated by modern transient surveys and follow-up programs, such as the Zwicky Transient Facility (ZTF, \citealt{Bellm2018-cn}).
The upcoming Vera C. Rubin Observatory’s Legacy Survey of Space and Time (LSST; \citealt{LSST-Science-Collaboration2009-rt}) is expected to further accelerate such discoveries.
From the Lick Observatory Supernova Search (LOSS, \citealt{Leaman2011-jd, Li2011-rr}) survey, the volumetric CCSN rate was estimated as 
% $6.2 \times 10^4~\gpcyr$, 
$(7.1 \pm 1.1) \times 10^4~\gpcyr$, 
with SNe~II comprising about $58\%$ of all CCSNe \citep{Li2011-bx}, broadly consistent with other surveys \citep{Das2025-fu, Pessi2025-eg}. 
% Observations of SN II light curves and spectra reveal substantial diversity in their explosion energies, $\sim 10^{50}–10^{51}$~erg, and synthesized \Nifs~masses, $\sim10^{-3}–10^{-1}~\Msun$ \citep{Martinez2022-lv, Subrayan2023-md, Das2025-wj, Anderson2019-zb}. 
% In addition, recent early-time observations indicate that many progenitors are surrounded by dense CSM, with inferred mass-loss rates of $\dot{M}=10^{-4}$–$10^{-2}~\Myr$ \citep{Yaron2017-lq, Morozova2017-oj, Forster2018-ox, Bruch2021-mg, Bruch2023-nl, Subrayan2023-md}.
% % The LOSS survey have estimated the occurence rate of CCSNe as $6.2 \times 10^4~\gpcyr$ and the SN II rate among the CCSNe as $58~\%$ \citep{Li2011-bx}, broadly consistent with other surveys \citep{Das2025-fu, Pessi2025-eg}.
% % % The explosion energy of SNe~II are inferred to range $10^{50}-10^{51}$~erg from comparisons of observed light curves and radiation hydrodynamical light-curve models \citep{Martinez2022-lv, Subrayan2023-md, Das2025-wj}.
% % Observations of the light curves and spectra of SNe~II reveal diversity in properties such as the explosion energy, $\sim 10^{50}-10^{51}$ erg, and $\Mni$, $\sim 10^{-3}-10^{-1}~\Msun$ \citep{Martinez2022-lv, Subrayan2023-md, Das2025-wj, Anderson2019-zb}.
% % Furthermore, recent observations of early light curves and spectra indicate the presence of dense CSM around progenitors, with estimated mass-loss rates ranging from $\Mdot = 10^{-4}$ to $10^{-2}$ $\Myr$ \citep{Yaron2017-lq, Morozova2017-oj, Forster2018-ox, Bruch2021-mg, Bruch2023-nl, Subrayan2023-md}.

Despite the robust explosion prediction in first-principles simulations and the extensive survey of SNe~II, observations of ECSNe have remained elusive. 
% By contrast, observations of ECSNe have remained elusive despite their robust explosion in the first-principles simulations. 
Recently, \citet{Hiramatsu2021-er} proposed SN~2018zd as a promising ECSN candidate on the basis of several observational features and indications which are consistent with theoretical expectations, \ie light-curve morphology, low explosion energy, the presence of dense CSM, nucleosynthetic yields, and progenitor identification (see also \citealt{Van_Dyk2022-ti}).
% A complementary search within the ZTF Census of the Local Universe (ZTF-CLU; \citealt{De2020-kq}) program, a galaxy-targeted, volume-limited effort that aims for spectroscopic completeness of ZTF transients to $m \lesssim 20$~mag within $100''$ of CLU galaxies \citep{Cook2019-np} out to $\sim200$~Mpc, found no secure new ECSN candidates using the color-based diagnostic of \citet{Sato2024-kt} \citep{Das2025-wj}. 
A complementary search within the ZTF Census of the Local Universe (ZTF-CLU; \citealt{De2020-kq}) program, a galaxy-targeted, volume-limited effort that aims for spectroscopic completeness of ZTF transients to a magnitude less than $20$~mag within $100''$ of CLU galaxies \citep{Cook2019-np} out to $\sim200$~Mpc, found no secure new ECSN candidates using the color-based diagnostic \citep{Das2025-wj}. 
These studies motivate a broader, systematic exploration to observationally constrain the occurrence rate, progenitor properties, and explosion physics of ECSNe.

Here, we undertake a broad search for ECSN candidates among previously reported SNe~II in the literature and among ZTF public results, using our color-based diagnostic. 
% For the subsample obtained from ZTF, we apply a volume correction to infer the volumetric occurrence rate of ECSNe and their progenitor initial-mass range. 
Using our candidates, we infer a volumetric occurrence rate of ECSNe. 
We also compare the observed multicolor light curves with radiation-hydrodynamical models to infer explosion and progenitor properties.

This paper is structured as follows. 
Section~\ref{sec:obsmethod} introduces the SN II sample, their observational quantities, and the selection method of ECSN candidates. 
% Section~\ref{sec:LCmodeling} describes the models and methods. 
Section~\ref{sec:LCmodeling} describes the light-curve models and methods to compare the models with observations. 
Section~\ref{sec:res} presents the selection of ECSN candidates, their observational properties, and comparisons to theoretical light-curve models. 
Section~\ref{sec:disc} discusses the progenitor and explosion properties implied by the comparison to models as well as the inferred occurrence rate of ECSNe. 
Section~\ref{sec:concl} provides a summary of the main results and future prospects.

% This paper is structured as follows.
% We introduce the models and methods in Section~\ref{sec:method}. We next introduce SN II sample in Section~\ref{sec:obsmethod}.
% We present the selection of ECSN candidates and their observational properties with the comparison to theoretical light-curve models in Section~\ref{sec:res}.
% We then give a discussion on the occurrence rate and initial mass range, adopting a volume correction to the sample obtained from ZTF public results, as well as on the properties of the ECSN candidates inferred from the light-curve comparison with theoretical models in Section~\ref{sec:disc}.
% Finally, we present caveats, future observational prospects, and conclusions in Section~\ref{sec:conclusion}. 

% We show the resulting light curves in Section~\ref{sec:results}: the representative light curves in Section~\ref{subsec:typicalLC} and light curves with a wide range of physical quantities in Section~\ref{subsec:allLC}. We discuss their robust characteristics and propose an ECSN diagnostic method in Section~\ref{sec:discussion}. Finally, we present our conclusions in Section~\ref{sec:conclusion}. 

\section{Sample \label{sec:obsmethod}}
\subsection{SN II sample compilation \label{subsec:sample}}
% We explore ECSN candidates among SNe~II in the past from literature and public results of an astronomical transient survey, Zwicky Transient Facility (ZTF; \citealt{Bellm2018-cn}).
% In this section, we introduce the basic information of the obtained data and processing of ZTF data.
We compile two complementary samples of SNe~II.
The first is a literature sample having dense multi-band photometric data obtained by prior compilation studies \citep{Faran2014-fc, Valenti2016-ao, Galbany2016-zp, Anderson2024-md} with selected individual SNe of particular relevance, 
% the promising ECSN candidate, SN~2018zd \citep{Hiramatsu2021-er} and an SN II with a blue color plateau, SN~2023axu \citep{Shrestha2024-ty}. 
the promising ECSN candidate, SN~2018zd \citep{Hiramatsu2021-er}, an SN II with a blue color plateau, SN~2023axu \citep{Shrestha2024-ty}, and the prototypical low-luminosity SN~2005cs \citep{Pastorello2009-ge}. 
% The first is a literature sample having dense multi-band photometric data obtained by prior compilation studies \citep{Faran2014-fc, Valenti2016-ao, Galbany2016-zp, Anderson2024-md} with selected individual SNe of particular relevance (SN~2018zd; \citealt{Hiramatsu2021-er}, and SN~2023axu; \citealt{Shrestha2024-ty}). 
% used primarily for candidate vetting and for constraining observables that require high cadence and wavelength coverage.
% used primarily for candidate vetting and for investigating physical properties.
% used primarily to investigate the physical properties.
% We collect well-observed SN II absolute-magnitude light curves, including the ECSN candidate, SN~2018zd, from literature \citep{Faran2014-fc, Valenti2016-ao, Galbany2016-zp, Hiramatsu2021-er, Shrestha2024-ty}. 
% The SNe~II from \citet{Galbany2016-zp, Anderson2024-md} are corrected for the Milky-Way extinction but not for the host-galaxy extinction while the SNe~II from the other literature are corrected for both. 
The SNe~II from \citet{Pastorello2009-ge, Faran2014-fc, Valenti2016-ao, Hiramatsu2021-er, Shrestha2024-ty} are corrected for both Milky-Way and host-galaxy extinctions, whereas those from \citet{Galbany2016-zp, Anderson2024-md} are corrected only for the Milky-Way extinction.
% While the host-galaxy extinction is significant for some SNe~II, it does not appear to dominate the observed color diversity of SNe~II and may be modest in many cases \citep{deJaeger2018}. 
While host-galaxy extinction is not negligible for some SNe~II (\eg \citealt{Leonard2002-zx}), it does not appear to dominate the observed color diversity of SNe~II and may be modest in many cases \citep{De_Jaeger2018-vj,Das2025-fu}.
% Therefore, we do not apply additional extinction corrections for them, while evaluating its effect in the occurrence-rate analysis.  
% While the SNe~II from \citet{Pastorello2009-ge, Faran2014-fc, Valenti2016-ao, Hiramatsu2021-er, Shrestha2024-ty} are corrected for both of the Milky-Way and host-galaxy extinctions, those from \citet{Galbany2016-zp, Anderson2024-md} are corrected only for the Milky-Way extinction, but the host-galaxy extinction is expected to be negligible in most SNe~II \citep{De_Jaeger2018-vj,Anderson2024-md}. 
% The observational properties of the samples, the estimated explosion epoch and $\tPT$, are shown in Table~\ref{}. 
% They are observed with various filters, such as $u$-, $g$-, $r$-, $U$-, $B$- and $V$-bands and $UVW2$-, $UVM2$, $UVW1$-, $U$-, $B$-, and $V$-bands of the Swift-UVOT.
These SNe~II are observed in some or all of the filters of 
% $g$, $r$, and $i$ bands of SDSS, $U$, $B$, and $V$ bands of Johnson-Cousins system, $UVW2$, $UVM1$, $UVW1$, $U$, $B$, and $V$ bands of Swift/UVOT, and unfiltered observation of T.~Noguchi.
$g$, $r$, and $i$ bands of Sloan Digital Sky Survey system (SDSS, \citealt{Fukugita1996-ss}), $U$, $B$, and $V$ bands of Johnson-Cousins system \citep{Bessell2005-aw}, $UVW2$, $UVM1$, $UVW1$, $U$, $B$, and $V$ bands of Swift UltraViolet and Optical Telescope (UVOT, \citealt{Poole2007-sj}) and unfiltered observations of T.~Noguchi.
% $UVW2$, $UVM2$, $UVW1$, $U$, $B$, and $V$ bands of Swift/UVOT, $U$, $B$, and $V$ bands of Johnson-Cousins system \citep{Bessell2005-aw}, and $g$, $r$, and $i$ bands of the Sloan Digital Sky Survey system (\citep{Fukugita1996-ss}) and unfiltered photometries obtained by K. Itagaki (only for SN~2018zd).
% $u$, $g$, $r$, and $i$ bands of the SDSS system, $U$, $B$ and $V$ bands of the Johnson-Cousins, and $UVW2$, $UVM2$, $UVW1$, $U$, $B$, and $V$ bands of the Swift-UVOT.
% We adopted the explosion epoch ($\texp$) of the SNe~II from literature.
We adopt the literature value for the explosion epoch, $\texp$, and set $\texp$ as the origin of time through this paper.

% From the SNe~II obtained from literature, we discard those having a larger gap than 10~days between the $\texp$ and first detection, as $\texp$ estimation is somewhat uncertain for such objects.
From the SNe~II obtained from the literature, we discard those having a larger gap than 10~days between the $\texp$ and the first detection to select SNe~II with well-constrained $\texp$ estimation.
% From the SNe~II obtained from literature, we discard those having a larger gap than 10~days between the explosion epoch, $\texp$ estimated in the literature and the first detection as $\texp$ estimation is somewhat uncertain for such objects.
% Next we discard the objects that are not observed by the end of the plateau, for which we adopt $t_{end}$ defined by \citet{Anderson2014-fe} as the epoch at which the object gets $0.1$~mag fainter than the extrapolation of the straight line of the light-curve plateau.
Next we discard the objects that are not observed by the end of the plateau, $t_{\rm end}$.
We derive $t_{\rm end}$ following \citet{Anderson2014-fe} as the epoch at which the object gets $0.1$~mag fainter than the extrapolation of the straight line of the light-curve plateau.
We then discard the objects that do not show a light-curve plateau by visual inspection and finally obtain $36$ SNe~II from the literature.

% The second is a ZTF survey sample \citep{Bellm2018-cn,Masci2019-xo}, which--despite fewer bands--offers a uniform selection function
% and volumetric control, enabling population-level inferences such as the number of ECSN candidates and an estimate of the volumetric rate.
% The second is the ZTF survey sample \citep{Bellm2018-cn,Masci2019-xo}, which offers a systematic sample, enabling population inferences such as an estimate of the volumetric rate although the observation is limited to $g$, $r$, and $i$ bands of ZTF system.
The second is the ZTF survey sample \citep{Bellm2018-cn,Masci2019-xo}, which offers a systematic sample, enabling an estimate of the volumetric rate although the observation is limited to $g$, $r$, and $i$ bands of ZTF system.
% For comparison with a systematic, survey-mode sample, we further add SNe~II discovered and followed by the Zwicky Transient Facility (ZTF; \citealt{Bellm2018-cn,Masci2019-xo}), adopting the public data releases and standard calibrations where available.
% We also collect SNe~II from public results of ZTF to obtain large and homogeneous samples although their observation is limited to $g$-, $r$-, and $i$-bands and low cadence, compared to the well-observed SNe~II in Section~\ref{subsec:obslite}. We here mean 'homogeneous' that they are not selected arbitrarily. Thus, they still have observational biases, such that intrinsically faint SNe~II rarely observed, compared to luminous ones.
% We also collect SNe~II from the public results of ZTF to obtain a large and systematic sample although their observation is limited to $g$-, $r$-, and $i$-bands and to low cadence, compared to the well-observed SNe~II from literature. 
We follow a similar way to \citet{Silva-Farfan2024-ww} to select SNe~II observed by ZTF.
We first adopt the Automatic Learning for the Rapid Classification of Events (ALeRCE) broker light-curve classifier \citep{Forster2021-rz, Sanchez-Saez2021-jy} to select all the objects discovered before 2024-12-31 and classified as an SN~II with a probability higher than $0.4$. 
Then, we compare them with the Transient Name Server (TNS) database and select 683 objects that are spectroscopically classified as SNe~II. 
% Then, we compare them with Transient Name Server (TNS) database and select the objects that are spectroscopically classified as SNe~II. 
% We next select the objects that have a smaller gap than $10$~days between last non-detection and the first detection. 
% We 

We obtain forced-photometry data from the ZTF forced-photometry service \citep{Masci2019-xo} for all SNe~II.
We first supply the coordinates of the SNe~II from the TNS database and the time range from a sufficiently earlier (greater than $50$~days) epoch to a sufficiently later (greater than $400$~days) epoch than the first detection to obtain better baseline. 
% Then, we correct the baseline, rescale photometric uncertainties, and obtain apparent-magnitude forced-photometry light curves, following \citet{Masci2019-xo}.
% Then, we subtract the baseline from the differential-flux measurements, rescale the flux uncertainties using the PSF-fit reduced-$\chi^2$ and the scatter of the baseline S/N, and obtain apparent-magnitude forced-photometry light curves, following \citet{Masci2019-xo}.
Then, we subtract the baseline from the differential-flux measurements, rescale the flux uncertainties using the PSF-fit reduced-$\chi^2$ and the scatter of the baseline S/N, and obtain apparent-magnitude forced-photometry light curves, following \citet{Masci2019-xo}, with $3 \sigma$ detections, \ie $\mathrm{S/N} > 3$, and $5 \sigma$ upper limits for non-detections.

To produce absolute-magnitude light curves, we adopt redshifts from the TNS database and a standard $\rm{\Lambda CDM}$ model with the Hubble constant, $\hc = 70.4~{\rm km~Mpc^{-1}~s^{-1}}$ and the density parameter, $\omegac = 0.272$ \citep{Komatsu2011-oz}.
We also correct for dust extinction, adopting the Milky-Way extinction \citep{Schlegel1998-yn} at a coordinate where the forced photometry is obtained and assuming $R_V = 3.1$. 
% We do not correct for the host-galaxy extinction as it is expected to be negligible in most SNe~II \citep{De_Jaeger2018-vj,Anderson2024-md}.
We do not correct for the host-galaxy extinction similarly to the literature SNe~II from \citet{Galbany2016-zp,Anderson2024-md}.

% We adopt the literature value for the explosion epoch, $\texp$ of the SNe~II from literature while we consider the middle point between the last non-detection and earlier of the first detection and discovery as $\texp$ for the ZTF survey sample.
We consider the middle point between the last non-detection and earlier of the first detection and discovery as $\texp$ for the SNe~II obtained from ZTF.
% We set $\texp$ as the origin of time through this paper.
% $\texp$ obtained here is different from the $t_{0}$ defined in Section~\ref{subsec:diagnostic}, but the difference is negligible.
% 
% From the SNe~II, we discard the objects that have a gap greater than $10$~days between each observation, \ie detection and non-detection.
We then discard the objects that have a gap greater than $10$~days between each observation, including the non-detection.
% Since the epoch at which a light curve transitions from plateau to tail ($\tPT$, \citealt{Anderson2014-fe}) is required, we next discard those are not observed around by end of the plateau, for instance due to the Sun constraints.
% We discard the object that are not observed at $t_{end}$ defined by \citet{Anderson2014-fe} as the epoch at which the object gets $0.1$~mag fainter than the extrapolation of the straight line of light-curve plateau.
Next we discard the objects that are not observed by the end of the plateau in the same way as done for the literature sample and finally obtain $62$ SNe~II from the ZTF survey, named the ZTF survey sample.
Integrating the literature and ZTF survey samples, we finally obtain $98$ SNe~II in total.

\begin{figure}[ht!]
    \includegraphics[width=85truemm]{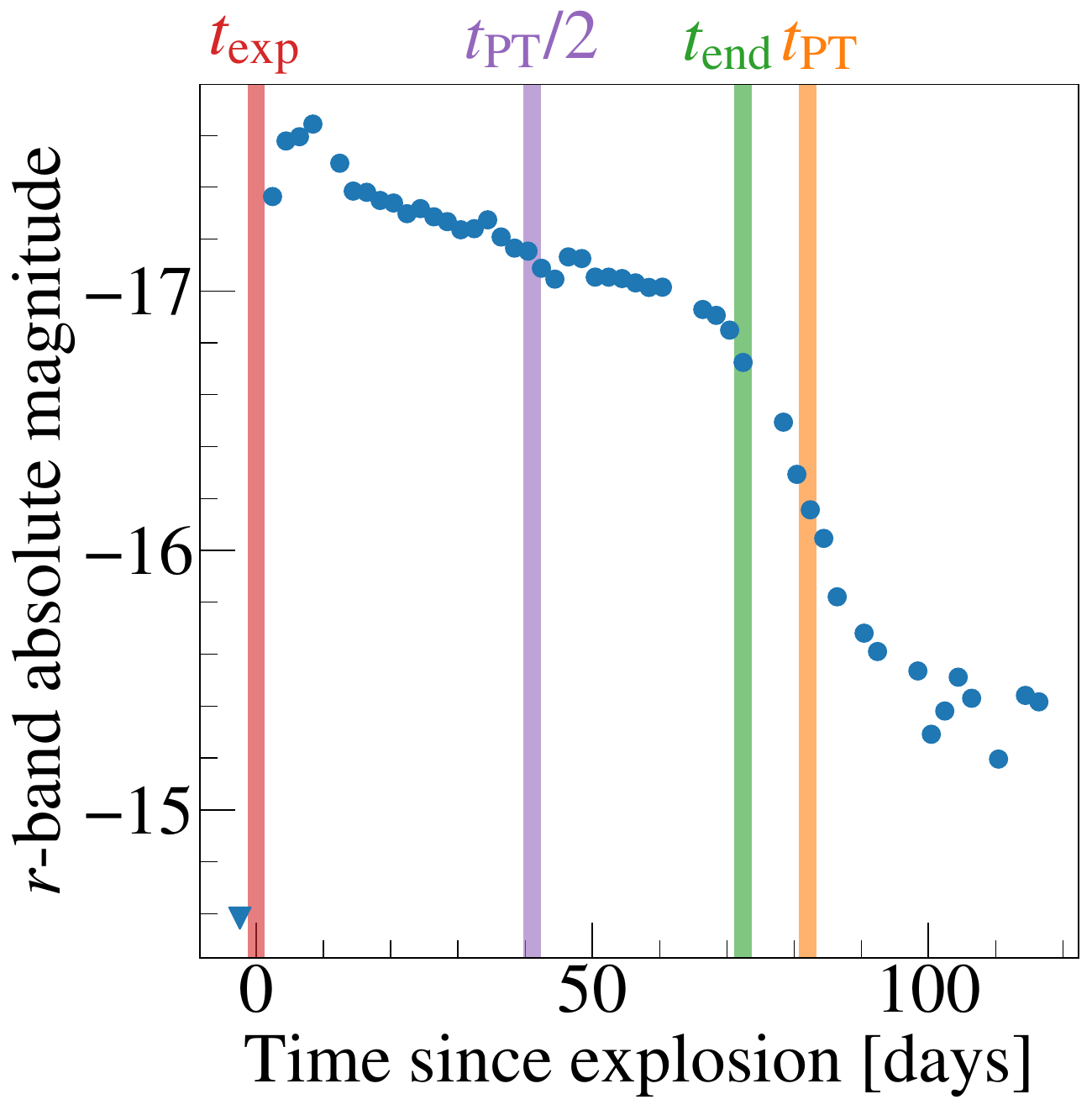}
\caption{
Example $r$-band light curve of a ZTF survey sample object, SN~2022jps.
The blue circles indicate the detections after 1-day binning for visual clarity and the blue downward triangle indicates the last non-detection.
% The detections after 1-day binning is shown in blue circles and the last non-detection is shown in blue downward triangle.
The derived $\texp$, $\tPT/2$, $t_{\rm end}$, and $\tPT$ are also indicated by the vertical red, purple, green, and orange lines, respectively.
% Example $r$-band light curve of a ZTF survey sample object, SN~2022jps (blue points), together with the derived $\texp$, $\tPT/2$, $t_{\rm end}$, and $\tPT$, indicated by vertical red, purple, green, and orange lines, respectively.
% Example $r$-band light curve of a ZTF survey sample object, SN~2019iex (blue points), together with the derived $\texp$, $\tPT/2$, $t_{\rm end}$, and $\tPT$, indicated by vertical red, purple, green, and orange lines, respectively.
We set $\texp$ as the origin of time in this work. $\tPT/2$ is the middle point between $\texp$ and $\tPT$, at which the color diagnostic is evaluated. 
% For visual clarity, the photometric points are shown after 1-day binning.
\label{fig:LCex}}
\end{figure}

% \subsection{Observational Quantities}
We next derive $\tPT$, the mid-point of the plateau end and tail onset, and the color index $g-r$ or $B-V$ at $\tPT/2$, $\grmid$ or $\BVmid$.
% We next derive $\tPT$, the the mid-point of the plateau end and tail onset, and the color index $g-r$ or $B-V$ at $\tPT/2$, $\grmid$ or $\BVmid$ to utilize the diagnostic proposed by \citet{Sato2024-kt} in the selection of ECSN candidates.
We adopt $\tPT$ from literature if available.
Otherwise, we estimate $\tPT$ following the same method as \citet{Felipe2010-tx, Anderson2014-fe}, except that we adopt the $r$ band in the case $V$ band is unavailable.
With the $\tPT$, we derive $\grmid$ and/or $\BVmid$, depending on the available photometry, linearly interpolating the light curves in the corresponding band.
% Table~\ref{table:obsqs} lists the $\texp$, $\tPT$, $\grmid$, and $\BVmid$ obtained either from literature or with our method.
The $\texp$, $\tPT$, $\grmid$, and $\BVmid$ of the SNe~II are listed in Table~\ref{table:obsprop_SNeII} in Appendix~\ref{app:obsprop_SNeII} with the reference from which we obtain the photometry data.
Figure~\ref{fig:LCex} shows an example of the derived $\texp$, $t_{\rm end}$, $\tPT$, and $\tPT/2$ for a ZTF survey sample object, SN~2022jps, along with its $r$-band light curve.
% Figure~\ref{fig:LCex} shows an example of the derived $\texp$, $t_{\rm end}$, $\tPT$, and $\tPT/2$ for a ZTF survey sample object, SN~2019iex, along with its $r$-band light curve.

% \subsection{ECSN diagnostic}
\subsection{Selection of ECSN candidates \label{subsec:selmethod}}
We adopt the color-based ECSN diagnostic method proposed by \citet{Sato2024-kt}.
% This method was developed based on one-dimensional radiation-hydrodynamical light-curve calculations of ECSNe and low-mass FeCCSNe, 
% and utilizes the fact that ECSNe exhibit systematically bluer colors during the plateau phase owing to their low-density, extended hydrogen-rich envelopes, which place the photosphere deeper than the H-recombination front during the plateau. 
% The diagnostic compares the color index at $\tPT/2$ for a given $\tPT$.  
% They performed light-curve calculations of 736 ECSN and 1266 FeCCSN models widely changing physical quantities (\eg the explosion energy and the CSM density profile) and find that the ECSNe exhibit bluer color index at $\tPT/2$ than the FeCCSN for a given $\tPT$ in the absence of the strong CSM interaction at $\tPT/2$.
% Performing one-dimensional radiation-hydrodynamical light-curve calculations of 736 ECSN and 1266 FeCCSN models widely changing physical quantities (\eg the explosion energy and the CSM density profile), they find that ECSNe exhibit systematically bluer colors during the plateau phase, if the plateau is not dominated by a strong CSM interaction, owing to their low-density, extended hydrogen-rich envelopes, which place the photosphere deeper than the H-recombination front during the plateau. 
Performing one-dimensional radiation-hydrodynamical light-curve calculations for 736 ECSN and 1266 low-mass FeCCSN models spanning a wide range of physical parameters (\eg explosion energy and CSM density profile), they show that the ECSNe exhibit systematically bluer plateaus than the FeCCSNe in the absence of strong CSM interaction, reflecting the low-density, extended hydrogen-rich envelopes of the ECSN progenitors, which place the photosphere deeper than the H-recombination front during the plateau.
% Using the models, they also find that the ECSNe exhibit bluer color index at $\tPT/2$ than the FeCCSN for a given $\tPT$ in the absence of the strong CSM interaction at $\tPT/2$ and propose the empirical linear boundary between them as ECSN diagnostic.
They also show that, for a given $\tPT$, the ECSNe have bluer color indices at $\tPT/2$ than the FeCCSNe in the absence of strong CSM interaction at $\tPT/2$, and propose empirical linear boundaries that separate the ECSNe and the FeCCSNe as an ECSN diagnostic.
% Specifically, ECSNe are expected to satisfy
Specifically, ECSNe are expected to satisfy
\begin{equation}
(g-r)_{\tPT/2} < 0.008 \times \tPT - 0.4
\label{eq:g-r(tPT/2)threshold}
\end{equation}
or
\begin{equation}
(B-V)_{\tPT/2} < 0.0089 \times \tPT - 0.36.
\label{eq:B-V(tPT/2)threshold}
\end{equation}
% reflecting their blue-color plateau in the absence of a strong CSM interaction at $\tPT/2$.
% reflecting their blue-color plateau.
% reflecting their blue-color plateau, where Equations~(\ref{eq:g-r(tPT/2)threshold}) and (\ref{eq:B-V(tPT/2)threshold}) correspond to the Equations~(3) and (C1) in \citet{Sato2024-kt}, respectively.
which correspond to Equations~(3) and (C1) of \citet{Sato2024-kt}, respectively.
% We apply the diagnostic to the available color index ($g-r$ and/or $B-V$).
% When both colors are available, we do not require an object to satisfy both criteria simultaneously; we classify an event as a photometric ECSN candidate if it satisfies either Equation~(1) or Equation~(2).
Considering the uncertainties in the photometry and its analysis (\eg extinction corrections), we select photometric ECSN candidates as events that satisfy either the $g-r$ criterion (Equation~\ref{eq:g-r(tPT/2)threshold}) or the $B-V$ criterion (Equation~\ref{eq:B-V(tPT/2)threshold}), using the available color information.
% Considering the uncertainties in the photometry and its analysis (\eg extinction corrections), we select photometric ECSN candidates as events that satisfy either the $g-r$ criterion (Equation~\ref{eq:g-r(tPT/2)threshold}) or the $B-V$ criterion (Equation~\ref{eq:B-V(tPT/2)threshold}), using the available color information.
% Considering the uncertainties in the photometry (\eg extinction corrections), we select photometric candidates if either Equation~(1) or Equation~(2) is satisfied, depending on the available color.

This diagnostic effectively separates ECSN and low-mass FeCCSN models
and successfully identified SN~2018zd as an ECSN candidate.  
% We note that the method assumes the absence of strong CSM interaction at $\tPT/2$, as the presence of dense and extended CSM can also result in blue colors through CSM interaction.  
We note that the method assumes the absence of strong CSM interaction at $\tPT/2$, as an FeCCSN can also be blue in the presence of a strong CSM interaction.  
% Therefore, we further inspect spectroscopic properties for each candidate, if available, to ensure that the blue color is not caused by CSM interaction.  
% Therefore, we assess the presence or absence of strong CSM interaction with spectra obtained around $\tPT/2$ if available, with particularly focusing on narrow lines, a presence of which indicate a presence of a dense CSM. 
% Therefore, when available, we assess the presence or absence of strong CSM interaction using spectra obtained around $\tPT/2$, focusing in particular on narrow lines, whose presence indicates the presence of dense CSM \citep{Yaron2017-lq,Bruch2021-mg,Bruch2023-nl}.
% Therefore, when available, we assess the presence or absence of strong CSM interaction using spectra closest to $\tPT/2$, practically within approximately $\pm 10$ days, focusing in particular on narrow lines, whose presence indicates the presence of dense CSM \citep{Yaron2017-lq,Bruch2021-mg,Bruch2023-nl}.
% Therefore, when available, we assess the presence or absence of strong CSM interaction using spectra obtained around $\tPT/2$, practically within about $\pm 10$ days, focusing in particular on narrow lines, whose presence indicates the presence of dense CSM \citep{Yaron2017-lq,Bruch2021-mg,Bruch2023-nl}.
Therefore, when available, we assess the presence or absence of strong CSM interaction using spectra obtained around $\tPT/2$, in practice within about $\pm 10$ days, focusing in particular on narrow lines, whose presence indicates the presence of dense CSM \citep{Yaron2017-lq,Bruch2021-mg,Bruch2023-nl}.
Although CSM interaction enhances the luminosity as well, it is difficult to discern it from the intrinsic SN luminosity or from enhancements due to other mechanisms. 
% We therefore do not rule out one showing possibility of CSM interaction only in photometric properties as ECSN candidates.
% We therefore rely on spectroscopic properties alone and not on photometric properties to rule out an SN interacting with a dense CSM as an ECSN candidate.
We therefore rely solely on spectroscopic, rather than photometric, properties 
to exclude SNe that interact with a dense CSM from ECSN candidates.

For the selected ECSN candidates, we adopt $\Mni$ inferred from the tail luminosity.
Tail-based $\Mni$ estimates are typically obtained by converting a pseudo-bolometric luminosity to $\Mni$ assuming full deposition of the radioactive decay power \citep{Arnett1980-dg,Nadyozhin1994-dx}, or scaling a pseudo-bolometric tail luminosity to that of SN~1987A evaluated over a comparable filter set \citep{Spiro2014-cr}.
% Tail-based $\Mni$ estimates are often obtained either by converting a pseudo-bolometric luminosity to $\Mni$ assuming full deposition of the radioactive decay power \citep{Suntzeff1990-ov}, or by scaling a pseudo-bolometric (or, if band coverage is limited, single-band) tail luminosity to SN~1987A \citep{Spiro2014-cr}.
We adopt the latter approach similarly to that adopted for SN~2018zd by \citet{Hiramatsu2021-er}.
% We adopt the latter approach to place our sample on the same scale as SN~2018zd \citep{Hiramatsu2021-er}.
We compute pseudo-bolometric luminosity following \citet{Valenti2007-yt} using multi-band photometry between $\tPT+15$ and $\tPT+60$~days since explosion, and calculate $\Mni$ following \citet{Spiro2014-cr}.
When late-time band coverage is insufficient, we derive an upper limit on $\Mni$ using the available photometric upper limits.
% Because ECSNe may have lower ejecta masses than SN~1987A, $\gamma$-ray escape fraction may be higher than in SN~1987A and bias $\Mni$ low, but over the early nebular phase this effect is expected to be modest.
Since the ejecta masses of ECSNe are lower than that of SN~1987A ($\sim 14~\Msun$; \citealt{Shigeyama1990-qq}), the $\gamma$-ray optical depths of ECSNe may be lower than that of SN~1987A, possibly resulting in an underestimate of $\Mni$.
However, since $\gamma$-rays are expected to be almost fully deposited in the early tail phase, the difference in the ejecta mass does not significantly affect the $\Mni$ estimates.
Note that the tail may also be powered by CSM interaction \citep{Moriya2014-oh} and/or pulsar spin-down \citep{Tominaga2013}.
In that case, the true $\Mni$ may be lower than the values inferred from the tail luminosity.
% Note that the tail may also be powered by CSM interaction and/or pulsar spin-down; in that case, the true $\Mni$ may be lower than the values inferred from the tail luminosity.

We also derive $\ha$ and \feii~$\lambda5169$ line velocity, $\vha$ and $\vfe$, of ECSN candidates from their optical spectral sequences if available.
We measure these velocities from the blue-shifted absorption trough of the P-Cygni profile, after correcting for redshift and extinction and subtracting the continuum.
% Following Eastman & Kirshner (1989), we adopt v_ph ≲ v_Hα/2 as an upper limit on the photospheric velocity,
% and we later compare these values with the model photospheric velocities in Section 4.3.

% % We also estimate the $\ha$ line velocity, $\vha$, evolution from the blue-shifted absorption trough of the line profile in the redshift-corrected optical spectral sequence of ASASSN-14ha and SN~2018zd.
% We also estimate the $\ha$ line velocity, $\vha$, evolution from the blue-shifted absorption trough of the line profile in the optical spectral sequence of ASASSN-14ha and SN~2018zd after the corrections of redshift and reddening and subtraction of continuum.
% % We also estimate the $\ha$ line velocity, $\vha$, evolution from the blue-shifted absorption trough of the line profile in the redshift-corrected optical spectral sequence of ASASSN-14ha and SN~2018zd, obtained from WISeREP.
% From the line velocity, we estimate the upper limit of the photospheric velocity as $\vph \lesssim \vha/2$ \citep{Eastman1989-es}.

\section{Model \label{sec:LCmodeling}}

% \subsection{Modeling Approach \label{subsec:modelingway}}
% We adopt a two-step approach to model the observed multicolor light curves of the candidates.
% In the Step 1, we calculate a broad parameter range and find the better-fitted models (\ie the models providing small $\chi ^2$) to the observed light curve of each candidate.
% In the Step 2, we calculate a fine parameter range around the parameter space found in Step 1.
% For the candidates other than ASASSN-14ha or SN~2018zd, we simply find the model providing the smallest $\chi^2$.
% For ASASSN-14ha and SN~2018zd, we find the best-fitting model among those having a consistent photospheric-velocity evolution to the photospheric velocity evolution estimated from observed $\ha$ absorption minima.
% % For ASASSN-14ha and SN~2018zd, we find the among best-fitting model that has a consistent photospheric-velocity evolution to the photospheric velocity evolution estimated from observed $\ha$ absorption minima.

We adopt a two-step approach to model the observed multicolor light curves of ECSN candidates.
Step~1 explores a broad parameter space with theoretically expected $\Mni$ and identifies a set of top-ranked models (\ie those with lower $\chi^2$) for each candidate.
% Step~2 refines the search with a denser local grid around the Step~1 region.
% We require the photospheric-velocity evolution of the model to be consistent with that inferred from the observed $\ha$ absorption minima, and we select the best-fitting model among those satisfy the criterion if the spectral time-series is available for the object.
% Otherwise, the best-fitting model is simply taken as the one with the minimum $\chi^2$.
Step~2 refines the search with a denser local exploration around the top-ranked region from Step~1, setting $\Mni$ to the tail-based estimates (Section~\ref{subsec:selmethod}), and the best-fitting model is simply taken as the one with the lowest $\chi^2$.
% Step~2 refines the search with a denser local grid around the best-fit parameters derived in the Step~1, adjusting $\Mni$ to estimates in literature if available, and the best-fitting model is simply taken as the one with the lowest $\chi^2$.
% Step~2 refines the search with a denser local grid around the Step~1 region, and the best-fitting model is simply taken as the one with the minimum $\chi^2$.
% For all candidates except ASASSN-14ha and SN~2018zd, the best-fitting model is taken as the one with the minimum $\chi^2$.
% For ASASSN-14ha and SN~2018zd, we additionally require that the photospheric-velocity evolution of the model to be consistent with that inferred from the observed $\ha$ absorption minima, and we select the best-fitting model among those satisfy the criterion.
% For both steps, we calculate $\chi^2$ between $\texp$ and $\tPT$ because modeling multicolor light curves in the tail phase requires solving the balance between radioactive heating and numerous line cooling and it is beyond the scope of this study.
% For both steps, we calculate and simply sum $\chi^2$ over all bands and epochs between $\texp$ and $\tPT$ because modeling multicolor light curves in the tail phase requires solving the balance between radioactive heating and numerous line cooling and it is beyond the scope of this study.
For both steps, we calculate $\chi^2$ by simply summing residuals over all bands and epochs between $\texp$ and $\tPT$ because modeling multicolor light curves in the tail phase requires solving the balance between radioactive heating and numerous line cooling, which is beyond the scope of this study.
% Here, we simply sum $\chi^2$ over all bands (with weights given only by the photometric uncertainties) and do not apply any additional cadence-based weighting.
% For both steps, we calculate $\chi^2$ between the estimated explosion epoch, $\texp$ and the the mid-point of the plateau end and tail onset, $\tPT$, because modeling multicolor light curves in the tail phase requires solving the balance between radioactive heating and numerous line cooling and it is beyond the scope of this study.
Although we do not include the tail phase in the $\chi^2$ calculation, we adjust $\Mni$ because radioactive heating extends the plateau \citep{Kasen2009-qk,Kozyreva2019-cw}.

\subsection{Progenitors \label{subsec:prog}}

We adopt the super-AGB progenitor models from \citet{Tominaga2013} to calculate ECSN light curves.
The hydrostatic and thermal equilibrium envelopes are attached to the $1.377~\Msun$ ONeMg cores of \citet{Nomoto1982-zj, Nomoto1984-au, Nomoto1987-fr}.
Since super-AGB stars experience uncertain mass loss and third dredge-up from thermal pulses, we consider various envelope masses $\Menv$ and H abundances in the envelope $\XH$.
% In Step~1, we adopt the super-AGB progenitor models from \citet{Tominaga2013} to calculate ECSN light curves. 
% The hydrostatic and thermal equilibrium envelopes are attached to the $1.377~\Msun$ ONeMg cores of \citet{Nomoto1982-zj, Nomoto1984-au, Nomoto1987-fr}.
% Since super-AGB stars experience uncertain mass loss and third dredge-up from thermal pulses,
% we adopt various envelope masses $\Menv$ and H abundances in the envelope $\XH$. 

We adopt the nucleosynthesis yields from \citet{Wanajo2009-yo}.
% We adopt the nucleosynthesis yields from \citet{Wanajo2009-yo} in Step~1.
% In Step~2, we set $\Mni$ to the value estimated from the tail luminosity, and scale the other elements correspondingly.
% In Step~2, we set $\Mni$ to the value estimated in literature, if available, and scale the other elements correspondingly.
% Otherwise, we retain the abundance distribution of \citet{Wanajo2009-yo} without modification.
% Nucleosynthesis is adopted from \citet{Wanajo2009-yo} in Step~1.
% % In Step~2, we modify $\Mni$ to the value proposed in literature and scale the other elements correspondingly, for ASASSN-14ha, SNe~2018zd, and 2023axu \citep{Valenti2016-ao, Hiramatsu2021-er, Shrestha2024-ty} while we do not modify it for the other candidates as there are no estimation of $\Mni$ for them.
% In Step~2, we modify $\Mni$ to the literature value, if estimated, and scale the other elements correspondingly.
% Otherwise, we adopt the abandance distribution of \citep{Wanajo2009-yo} without modification.
% For the other candidates, we consider a range of $\Mni$, $10^{-3}-10^{-2}~\Msun$ as there are no estimation of $\Mni$ for them and we do not attempt to estimate $\Mni$ from the limited $g$- and $r$- band photometry.
We adopt an inner mass cut at $1.363-1.373~\Msun$ to reduce the calculation cost, although it is estimated as $1.362~\Msun$ in the first-principles simulation \citep{Kitaura2006-ia}. 
This small difference does not substantially affect the light curve properties, especially on the plateau phase. 
% We adopted $\Mni$ from literature if available and otherwise $\Mni=10^{-3}-10^{-2}~\Msun$ to take the observational indication from SN~2018zd ($(8.6 \pm 0.5) \times 10^{-3}~\Msun$, \citealt{Hiramatsu2021-er}) into account.

While the first-principles simulations \citep{Kitaura2006-ia,Janka2008-ai} predict the explosion energy of an ECSN as $\Eexp \sim 10^{50}$~erg, we explore a wider range to account for possible diversity among candidates.
% While the first-principles simulation \citep{Kitaura2006-ia} predicts the explosion energy of an ECSN as $\Eexp = 1.5\times10^{50}$~erg, we explore a wider range to account for possible diversity among candidates.
% While the first-principles simulation \citep{Kitaura2006-ia} predicts the explosion energy of an ECSN as $\Eexp = 1.5\times10^{50}$~erg, we calculate light-curve models with wide parameter ranges.
% as listed in Table~\ref{table:modelparams_EC}. 

On top of the progenitor models, 
we attach a CSM structure following \citet{Moriya2018-ve}, with density
\begin{equation}
\rho_{CSM}(r) = \frac{\Mdot}{(4\pi\vw)r^{2}},
\label{eq:CSMdensity}
\end{equation}
% up to $r=\Rout$, where $r$ is the distance from the center of the star, $\vw$ is the wind velocity, and $\Rout$ is the CSM radius. 
up to $r=\Rout$, where $r$ is the distance from the center of the star, $\vw$ is the wind velocity, and $\Rout$ is the CSM radius though the mechanism driving the wind is yet to be revealed. 
For the wind velocity, acceleration was considered using a simple $\beta$ velocity law:
\begin{equation}
\vw(r) = \vi + (\vf-\vi)\Big(1-\frac{R}{r}\Big)^\beta ,
\label{eq:windvel}
\end{equation}
where $\vi$ is the initial velocity of the wind, $\vf$ is the terminal velocity of the wind, and $R$ is the progenitor radius.
Since the mass loss of super-AGB stars just before the explosion is uncertain, we consider wide parameter ranges for the CSM profile as listed in table~\ref{table:modelparams_EC}.

In Step~1, we explore a broad parameter space as listed in Table~\ref{table:modelparams_EC}.
In total, we consider 2080 models in Step~1.

In Step~2, we perform a local refinement around the top-ranked region from Step~1 and set $\Mni$ to the tail-based estimate (Section~\ref{subsec:selmethod}).
We placed \Nifs~at the bottom of the ejecta.
We do not consider the \Nifs~mixing because 3D simulations of ECSN explosions predict small mixing \citep{Stockinger2020-jx} and it has little impact on plateau duration and luminosity \citep{Nakar2016-xe,Moriya2015-rl,Kozyreva2019-cw}.
% We rescale the nucleosynthesis yields accordingly while keeping the relative abundance pattern from \citet{Wanajo2009-yo}.

\begin{deluxetable*}{ccccccccccc}
\tablewidth{0pt}
\tablecaption{Parameter sets for the light-curve modeling of ECSN candidates (Step~1)}
\label{table:modelparams_EC}
\tablehead{
\colhead{$\Menv$} & \colhead{$\XH$} & \colhead{R} & \colhead{Mass cut} & \colhead{$\Eexp$} & \colhead{$\Mni$}  & \colhead{$\Mdot$} & \colhead{$\Rcsm$} & \colhead{$\vi$} & \colhead{$\vf$} & \colhead{$\beta$} \\
\colhead{[$\Msun$]} & \colhead{} & \colhead{[$10^{13}$ cm]} & \colhead{[$\Msun$]} & \colhead{[$10^{50}$ ergs]} & \colhead{[$\Msun$]} & \colhead{[$\Myr$]} & \colhead{[cm]} & \colhead{[\kms]} & \colhead{[\kms]} & \colhead{}
}
\startdata
% \begin{table*}[t]
%  \caption{Parameter sets for the light-curve modeling of ECSN candidates}
%  \label{table:modelparams_EC}
%  \centering
%   \begin{tabular}{ccccccccccc}
%    \hline
   % \(\Menv\) & \(\XH\) & $R$ & Mass cut & \(\Eexp\) & $\Mni$  & \(\Mdot\) & $\Rcsm$ & \(\vi\) & \(\vf\) & \(\beta\) \\ \relax
   % [$\Msun$] & & [$10^{13}$ cm] & [$\Msun$] & [\(10^{50}\) ergs] & [$\Msun$] & [$\Myr$] & [cm] & [km/s] & [km/s] & \\
   % \hline
   % 2.0 & 0.5 & 6.5 & $1.363-1.366$ & $0.6-4.1$ & $0.002-0.003$ & \(10^{-4}-10^{-2}\) & \(10^{14}-10^{16}\) & $1-5$ & 10 & $1$ \\
   2.0 & 0.7 & 6.5 & $1.366-1.370$ & $0.2-14.2$ & $0.002-0.003$ & \(10^{-6}-10^{-2}\) & \(10^{14}-10^{16}\) & $1-5$ & 10 & $1-5$ \\
   3.0 & 0.2 & 7.0 & $1.364-1.370$ & $0.2-13.3$ & $0.002-0.003$ & \(10^{-6}-10^{-2}\) & \(10^{14}-10^{16}\) & $1-5$ & 10 & $1-5$ \\
   3.0 & 0.5 & 7.0 & $1.364-1.370$ & $0.2-13.9$ & $0.002-0.003$ & \(10^{-4}-10^{-2}\) & \(10^{14}-10^{16}\) & $1-5$ & 10 & $1-5$ \\
   3.0 & 0.7 & 7.1 & $1.364-1.373$ & $0.1-13.2$ & $0.002-0.003$ & \(10^{-6}-10^{-2}\) & \(10^{14}-10^{16}\) & $1-5$ & 10 & $1-5$ \\
   4.7 & 0.5 & 7.3 & $1.363-1.370$ & $0.2-13.7$ & $0.002-0.003$ & \(10^{-6}- 3\times 10^{-2}\) & \(10^{14}-10^{16}\) & $1-5$ & 10 & $1-5$ \\
   4.7 & 0.7 & 7.2 & $1.363-1.373$ & $0.2-13.7$ & $0.002-0.003$ & \(10^{-6}- 3\times 10^{-1}\) & \(10^{14}-10^{16}\) & $1-5$ & 10 & $1-5$ \\
  %  \hline
  % \end{tabular}
\enddata
\tablecomments{
% \\
% \(\Menv\) and \(\XH\) are the envelope mass and hydrogen abundance of the progenitor, $R$ is the progenitor radius, \(\Eexp\) is the explosion energy, \(\Mdot\) is the mass-loss rate, $\Rcsm$ is the CSM radius, \(\vi\) is the initial velocity of the wind, \(\vf\) is the terminal velocity of the wind, and \(\beta\) is the acceleration parameter of the wind. 
% The ejected \Nifs~mass, $\Mni$, is set to $0.002-0.003~\Msun$ \citep{Wanajo2009-yo} in Step~1 ($\Mni$ is set to the tail-based estimates in Step~2).
\(\Menv\) and \(\XH\) are the envelope mass and hydrogen abundance of the progenitor, $R$ is the progenitor radius, \(\Eexp\) is the explosion energy, $\Mni$ is the ejected \Nifs~mass, \(\Mdot\) is the mass-loss rate, $\Rcsm$ is the CSM radius, \(\vi\) is the initial velocity of the wind, \(\vf\) is the terminal velocity of the wind, and \(\beta\) is the acceleration parameter of the wind. Step~1 explores 2080 models in total.
}
% \end{table*}
\end{deluxetable*}

% Nevertheless, since to estimate $\Mni$, either of derivations of bolometric light curves of tail phase using multicolor photometries with wide wavelength coverages or multicolor light-curve modeling of tail phase with detailed radiation hydrodynamical calculations is required and it is beyond the scope of this dissertation, which focuses on light-curve plateaus, we do not estimate $\Mni$.
% LCモデリングでは、基本的にLC計算の値を採用しつつ、表のとおり必要に応じて範囲を拡張した。
% MNiについては、先行研究で明らかになっているものはその値を採用し、そうでないものについては18zdの示唆を考慮して、1e-3~1e-2に設定する。
% しかし、MNiの推定には、bolometric LCを作るか、詳細なradiative transferに基づく多色光度曲線が必要であり、今回のプラトーにフォーカスする研究の範疇を超えているため、MNiは推定しない。

\subsection{Light-curve Calculations \label{subsec:stella}}
We adopt the one-dimensional multi-group radiation hydrodynamics code $\stella$ to calculate light curves because it solves radiation hydrodynamics with non-equilibrium prescription between gas and radiation \citep{Blinnikov1993-vi, Blinnikov1998-ye, Blinnikov2000-xq}, which is required in the low-density regions, \ie the super-AGB envelope and CSM. 
$\stella$ first injects thermal energy at the center of the progenitor and solves time-dependent equations about the angular moments of averaged intensity over up to 100 fixed-frequency bins to follow the evolution, with providing a spectral energy distribution (SED)  at each timestep.
We adopt the standard 100 frequency bins ranging over $1-50,000$ \AA.
Bolometric light curves are produced, integrating the SED at each time step.
Light curves with any filter could be calculated by convolving the SED with the corresponding response function.
% In this paper, we present light curves in the $g$, $r$, and $i$ bands of SDSS system, $U$, $B$, and $V$ bands of Johnson-Cousins system, $UVW2$, $UVM1$, $UVW1$, $U$, $B$, and $V$ bands of Swift/UVOT and unfiltered observation of T.~Noguchi.
In this paper, we present light curves in the 
% $g$, $r$, and $i$ bands of Sloan Digital Sky Survey system (SDSS, \citealt{Fukugita1996-ss}), $U$, $B$, and $V$ bands of Johnson-Cousins system \citep{Bessell2005-aw}, $UVW2$, $UVM1$, $UVW1$, $U$, $B$, and $V$ bands of Swift UltraViolet and Optical Telescope (UVOT, \citealt{Poole2007-sj}) and unfiltered observation of T.~Noguchi.
% $g$, $r$, and $i$ bands of SDSS, $U$, $B$, and $V$ bands of Johnson-Cousins system, $UVW2$, $UVM1$, $UVW1$, $U$, $B$, and $V$ bands of Swift/UVOT and unfiltered observation of T.~Noguchi.
$g$, $r$, and $i$ bands of SDSS, $U$, $B$, and $V$ bands of Johnson-Cousins system, $UVW2$, $UVM1$, $UVW1$, $U$, $B$, and $V$ bands of Swift/UVOT and unfiltered observation of T.~Noguchi
\footnote{
% For the observations by T.~Noguchi, we adopt a Gaussian in wavelength,
% with central wavelength of $6500~\AA$ \citep{Hiramatsu2021-er}, and width of $3\sigma=2500~\AA$.
% The width does not appreciably affect the light curves if it is $3\sigma=2000-3000~\AA$.
% We verified that varying the width within $3\sigma=2000$--$3000~\AA$ does not appreciably affect the light curves.
For the observations by T.~Noguchi, we adopt the quantum-efficiency curve of the \href{https://cdn.prod.website-files.com/6787ccd15f899cb85836f2b9/680bbef3ef428b7330b3c074_MicroLine\%20Discontinued\%20Models.pdf}{MicroLine ML0261E}.
}.
While the ZTF $gri$ filters are similar but not identical to the SDSS system \citep{Bellm2018-cn},
given the small differences in the transmission curves, we adopt the SDSS filter functions in the comparison with ZTF photometry.

% Given the small differences in the transmission curves, we adopt the SDSS filter functions for the light-curve calculations of ZTF photometry.
% In this paper, we compare light curves in the $g$, $r$, and $i$ bands of Sloan Digital Sky Survey \citep{Fukugita1996-ss}, $U$, $B$, and $V$ bands of Johnson-Cousins \citep{Bessell2005-aw}, $UVW2$, $UVM1$, $UVW1$, $U$, $B$, and $V$ bands of Swift UltraViolet and Optical Telescope (UVOT, \citealt{Poole2007-sj}) and unfiltered observation of T.~Noguchi.

We obtain photospheric velocity ($\vph$) evolution for each light-curve model.
The photosphere is defined as the location where the Rosseland mean optical depth becomes 2/3.

\section{Results \label{sec:res}}
\subsection{ECSN candidates \label{subsec:ECSNselection}}
% we explore ECSN candidates among the SNe~II with the light-curve diagnostic proposed by \citet{Sato2024-kt}, in which $\tPT$, the the mid-point of the plateau end and tail onset and the color index $g-r$ or $B-V$ at $\tPT/2$, $\grmid$ or $\BVmid$ are required.
We explore ECSN candidates among the SN II sample with the light-curve diagnostic proposed by \citet{Sato2024-kt}, adopting the $\tPT$, $\grmid$, and $\BVmid$ presented in Section~\ref{sec:obsmethod}.

\begin{figure*}[ht!]
\centering
\includegraphics[width=180truemm]{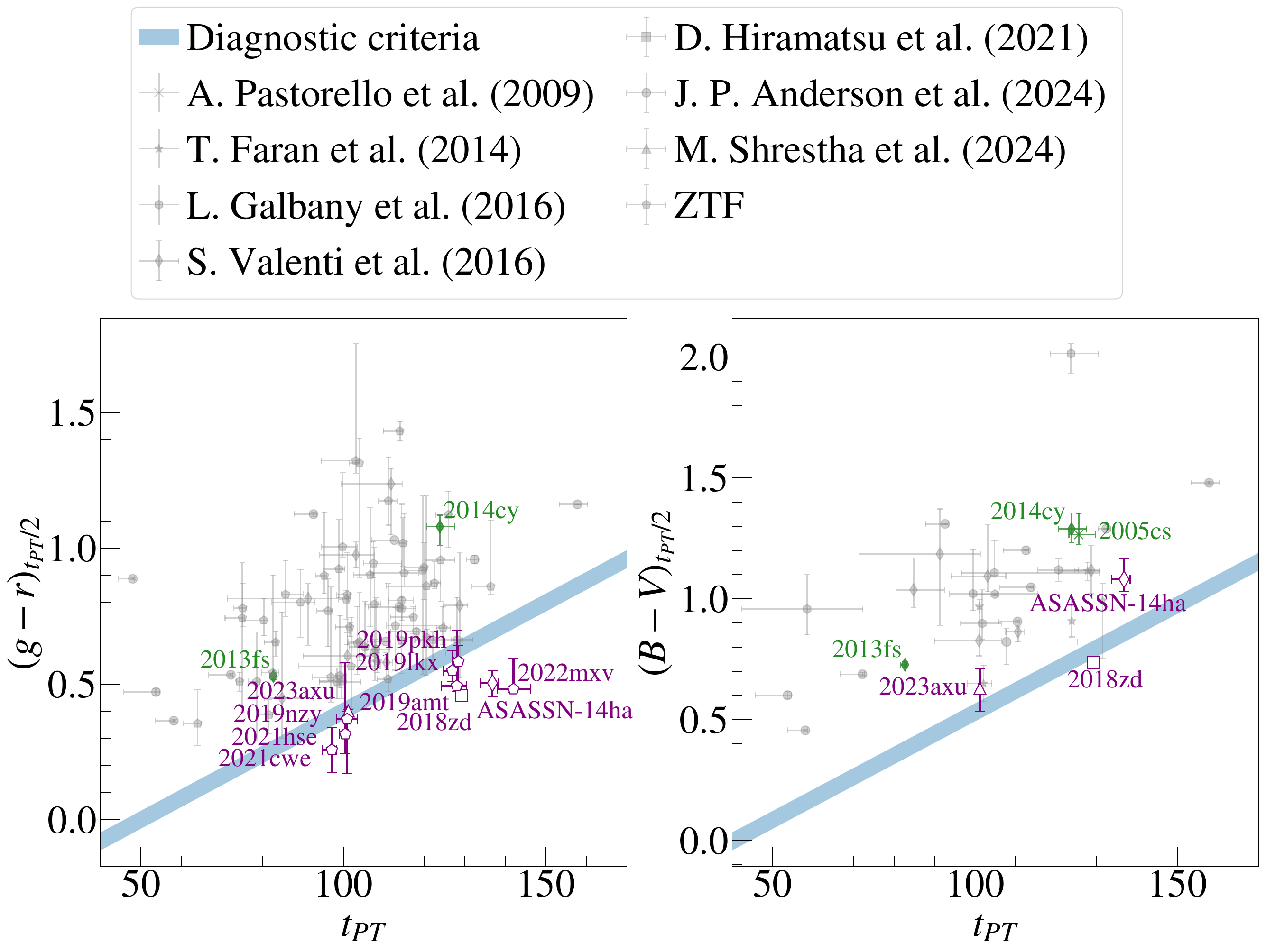}
\caption{
The diagnostic of the SNe~II with the colors $g-r$ and $B-V$ in the left and right panels respectively.
Purple points indicate the photometric ECSN candidates, \ie objects bluer than the selection criteria in either $g-r$ or $B-V$ \citep{Sato2024-kt}.
Gray points indicate objects that are not bluer than the criteria in either color.
The reference SNe are highlighted in green.
\label{fig:diag}}
\end{figure*}

Figure \ref{fig:diag} shows the diagnostic of the SNe~II with the color indices $g-r$ and $B-V$.
We identify ASASSN-14ha, SNe~2018zd, 2019amt, 2019lkx, 2019nzy, 2019pkh, 2021cwe, 2021hse, 2022mxv, and 2023axu as photometric ECSN candidates.
% For ASASSN-14ha and SN~2023axu, $g-r$ satisfies Equation~(\ref{eq:g-r(tPT/2)threshold}) whereas $B-V$ does not satisfy Equation~(\ref{eq:B-V(tPT/2)threshold}), but we retain them as photometric candidates because we consider objects satisfying either color criterion as photometric candidates, as described in Section~\ref{subsec:selmethod}.
For ASASSN-14ha and SN~2023axu, $g-r$ satisfies Equation~(\ref{eq:g-r(tPT/2)threshold}) whereas $B-V$ does not satisfy Equation~(\ref{eq:B-V(tPT/2)threshold}), but we retain them as we regard objects satisfying either color criterion as photometric candidates.
% For ASASSN-14ha and SN~2023axu, $g-r$ satisfies Equation~(\ref{eq:g-r(tPT/2)threshold}) whereas $B-V$ does not satisfy Equation~(\ref{eq:B-V(tPT/2)threshold}), but we retain them because, as described in Section~\ref{subsec:selmethod}, we regard objects satisfying either color criterion as photometric candidates.
% For ASASSN-14ha and SN~2023axu, $g-r$ satisfies Equation~(\ref{eq:g-r(tPT/2)threshold}) whereas $B-V$ does not satisfy Equation~(\ref{eq:B-V(tPT/2)threshold}), but we retain them as photometric candidates, as described in Section~\ref{subsec:selmethod}.
% Among the normal SNe~II, we select nearby and well-observed reference SNe, sampling the plateau luminosity range from the prototypical low-luminosity SN~2005cs \citep{Pastorello2009-ge} through the normal SN~2014cy \citep{Valenti2016-ao} and to the relatively bright (but still normal) SN~2013fs \citep{Yaron2017-lq}, all of which exhibit plateau durations typical of SNe~II.
Among the normal SNe~II, we highlight three well-observed events (SNe~2005cs, 2013fs, and 2014cy) that span the plateau-luminosity range, as reference objects for comparison.

\begin{figure*}[ht!]
\includegraphics[width=180truemm]{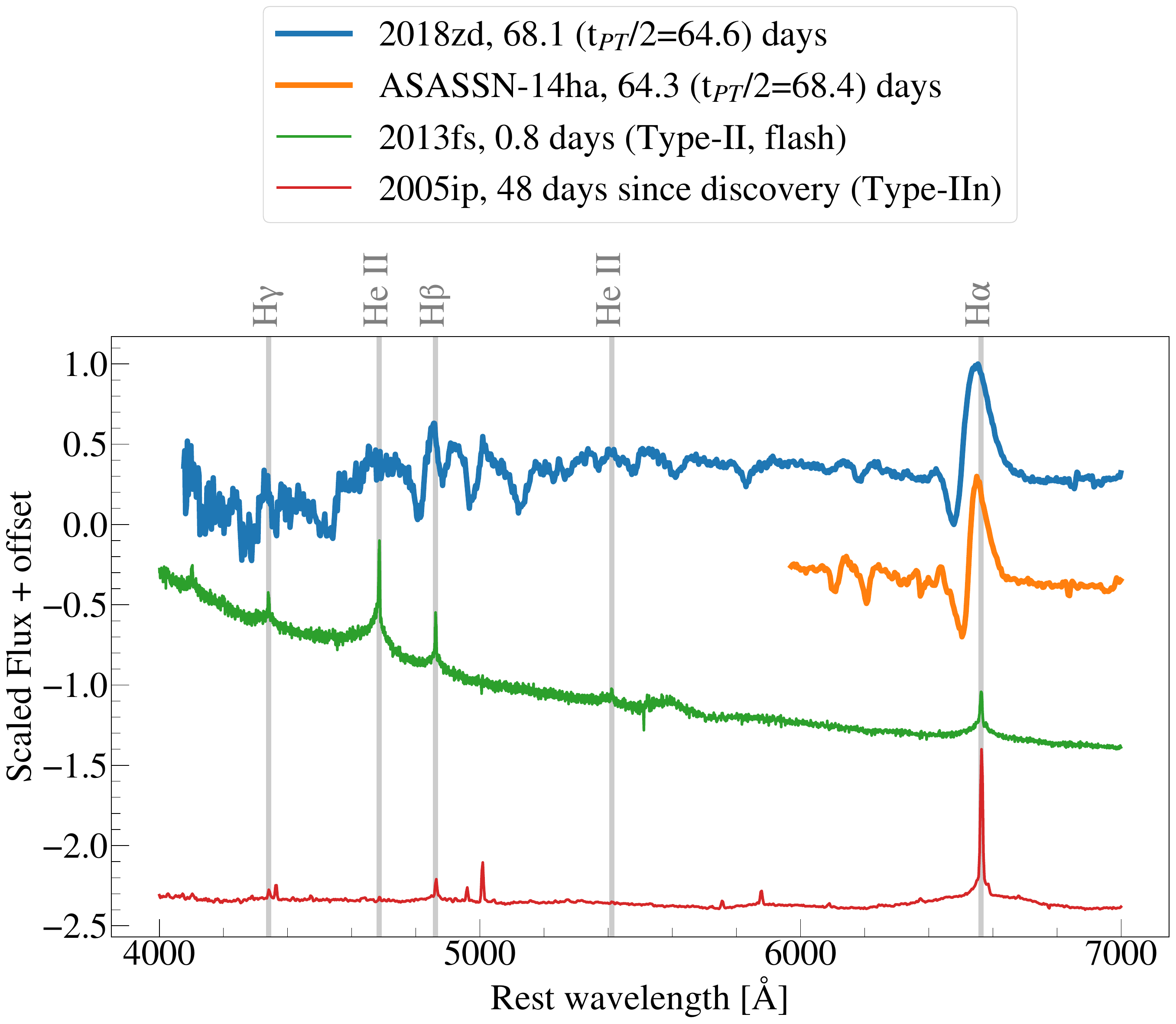}
\caption{
Spectra of SN~2018zd (thick blue line) and ASASSN-14ha (thick orange line) around $\tPT/2$.
The flash spectrum of Type-II SN~2013fs (thin green line) and a spectrum of the Type-IIn SN~2005ip (thin red line) are also shown as representatives of spectra with narrow lines due to CSM interaction.
All spectra are obtained from WISeREP.
% The spectra are scaled but not corrected for extinction.
\label{fig:spec.tPT2}}
\end{figure*}

We next examine the spectra of the photometric ECSN candidates around $\tPT/2$ to search for signatures of strong CSM interaction. 
% Figure~\ref{fig:spec.tPT2} shows optical spectra of SNe~2004er, 2018zd, and ASASSN-14ha at epochs closest to $\tPT/2$ ($79.7$, $68.1$, and $64.3$~days, respectively), without extinction correction. 
% The spectrum of SN~2004er was obtained from the CSP, while those of SN~2018zd and ASASSN-14ha were taken from WISeREP \citep{Yaron2012-aw}. 
% Figure~\ref{fig:spec.tPT2} shows optical spectra of SN~2018zd and ASASSN-14ha at epochs closest to $\tPT/2$ ($68.1$ and $64.3$~days, respectively), without extinction correction. 
% Figure~\ref{fig:spec.tPT2} shows optical spectra of SN~2018zd \citep{Zhang2020-dz} and ASASSN-14ha at epochs closest to $\tPT/2$ ($68.1$ and $64.3$~days, respectively). 
Figure~\ref{fig:spec.tPT2} shows optical spectra of SN~2018zd \citep{Zhang2020-dz} and ASASSN-14ha at epochs closest to $\tPT/2$ ($68.1$ and $64.3$~days, respectively), together with two representative spectra showing narrow lines: the flash spectrum of the Type-II SN~2013fs at $\sim 0.8$~days since explosion \citep{Yaron2017-lq} and a spectrum of the Type-IIn SN~2005ip at $\sim 48$~days since discovery \citep{Stritzinger2012-ik}.
Since SN~2005ip likely exploded $8-10$~days before discovery \citep{Smith2009-jo}, the spectrum approximately corresponds to $\sim 60$~days after explosion.
These spectra were taken from WISeREP \citep{Yaron2012-aw}. 
For SN~2023axu, no spectra around $\tPT/2$ are available in WISeREP, so we refer to \citet{Shrestha2024-ty}. 
% SN~2004er exhibits prominent narrow $\ha$ and H$\beta$ emission lines; although they may originate from the host galaxy, we conservatively discard this object.
ASASSN-14ha, SNe~2018zd, and 2023axu show no narrow lines around $\tPT/2$, indicating the absence of strong CSM interaction, and we designate them as the \textit{gold candidates}. 
% In contrast, ASASSN-14ha and SNe~2018zd and 2023axu show no such narrow lines around $\tPT/2$, indicating the absence of strong CSM interaction, and we designate them as the \textit{gold candidates}. 
The remaining photometric ECSN candidates without spectra are treated as the \textit{silver candidates}.

The estimated $\Mni$ for ASASSN-14ha, SNe~2018zd, and 2023axu in the literature are $0.0014 \pm 0.0002$, $0.0086 \pm 0.0005$, and $0.058 \pm 0.017~\Msun$, respectively \citep{Valenti2016-ao, Hiramatsu2021-er, Shrestha2024-ty}. 
% Since no previous $\Mni$ estimates are available for the silver candidates, we estimate $\Mni$ from their tail luminosities (Section~\ref{subsec:selmethod}), obtaining $\Mni = 0.2$, $0.1$, $0.3$, $0.09$, $0.1$, $0.2$, and $0.03~\Msun$ (for SNe~2019amt, 2019lkx, 2019nzy, 2019pkh, 2021cwe, 2021hse, and 2022mxv, respectively).
Since no previous $\Mni$ estimates are available for the silver candidates, we estimate $\Mni$ from their tail $g$ and $r$ photometry (Section~\ref{subsec:selmethod}).
% We obtain $\Mni = 0.2 \pm 0.02$, $0.1 \pm 0.006$, $0.09 \pm 0.003$, and $0.1 \pm 0.005 ~\Msun$ for SNe~2019amt, 2019lkx, 2019pkh, and 2021cwe.
We obtain $\Mni = 0.2 \pm 0.02$, $0.1 \pm 0.01$, $0.09 \pm 0.003$, and $0.1 \pm 0.01 ~\Msun$ for SNe~2019amt, 2019lkx, 2019pkh, and 2021cwe, with the uncertainties reflecting the photometric errors.
% We obtain $\Mni = 0.2 \pm 0.02$, $0.1 \pm 0.006$, $0.09 \pm 0.003$, and $0.1 \pm 0.005 ~\Msun$ for SNe~2019amt, 2019lkx, 2019pkh, and 2021cwe, with the uncertainties reflecting the photometric errors.
For SNe 2019nzy, 2021hse, and 2022mxv, we instead derive upper limits, $\Mni \lesssim 0.3$, $0.2$, and $0.03~\Msun$, respectively as there is no epoch with detections in both $g$ and $r$ after $\tPT+15$~days.

\subsection{Photometric properties}
\subsubsection{Gold candidates}

\begin{figure*}[ht!]
\centering
\includegraphics[width=180truemm]{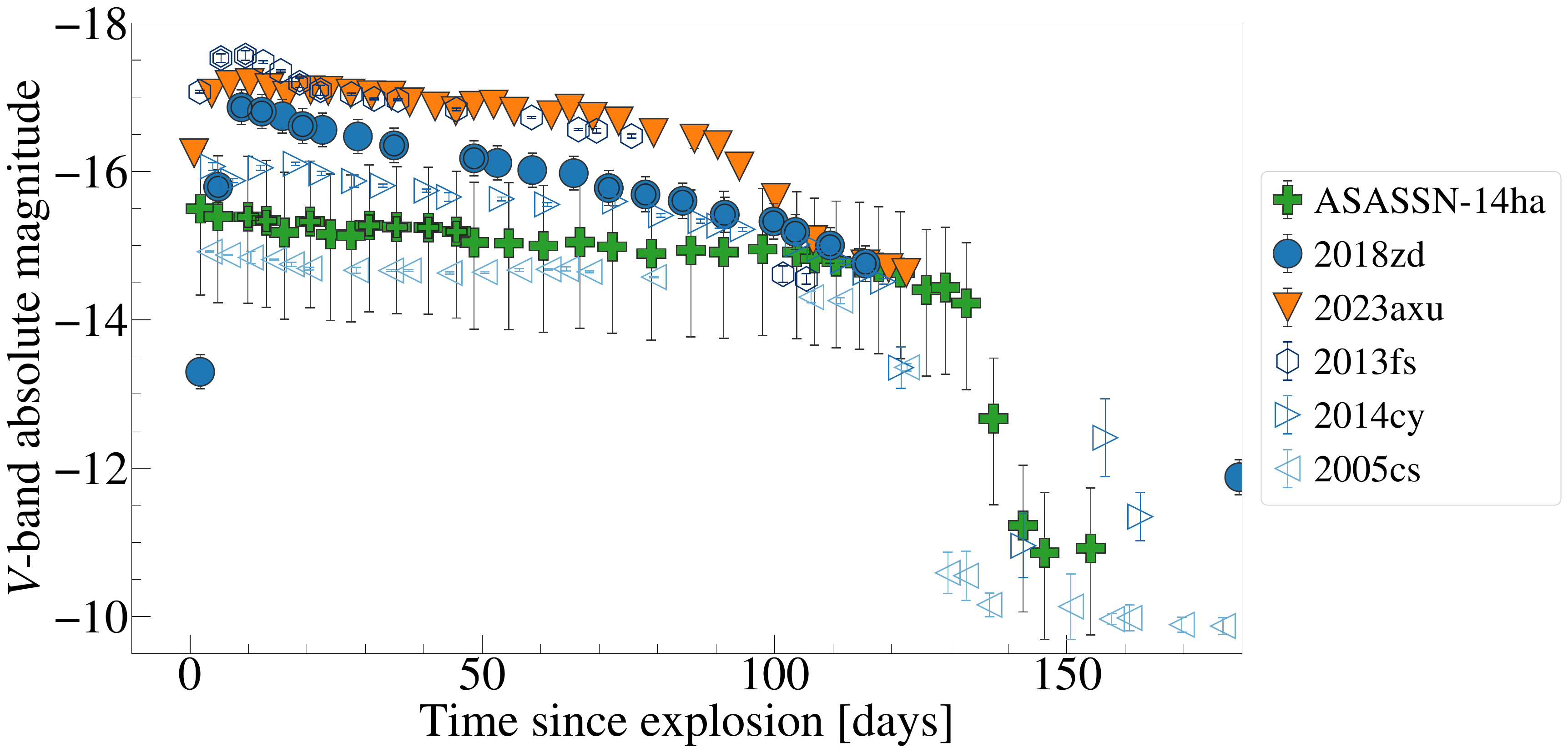}
% \centering
% \includegraphics[width=180truemm]{99_fig_compLC.r.pdf}
\includegraphics[width=180truemm]{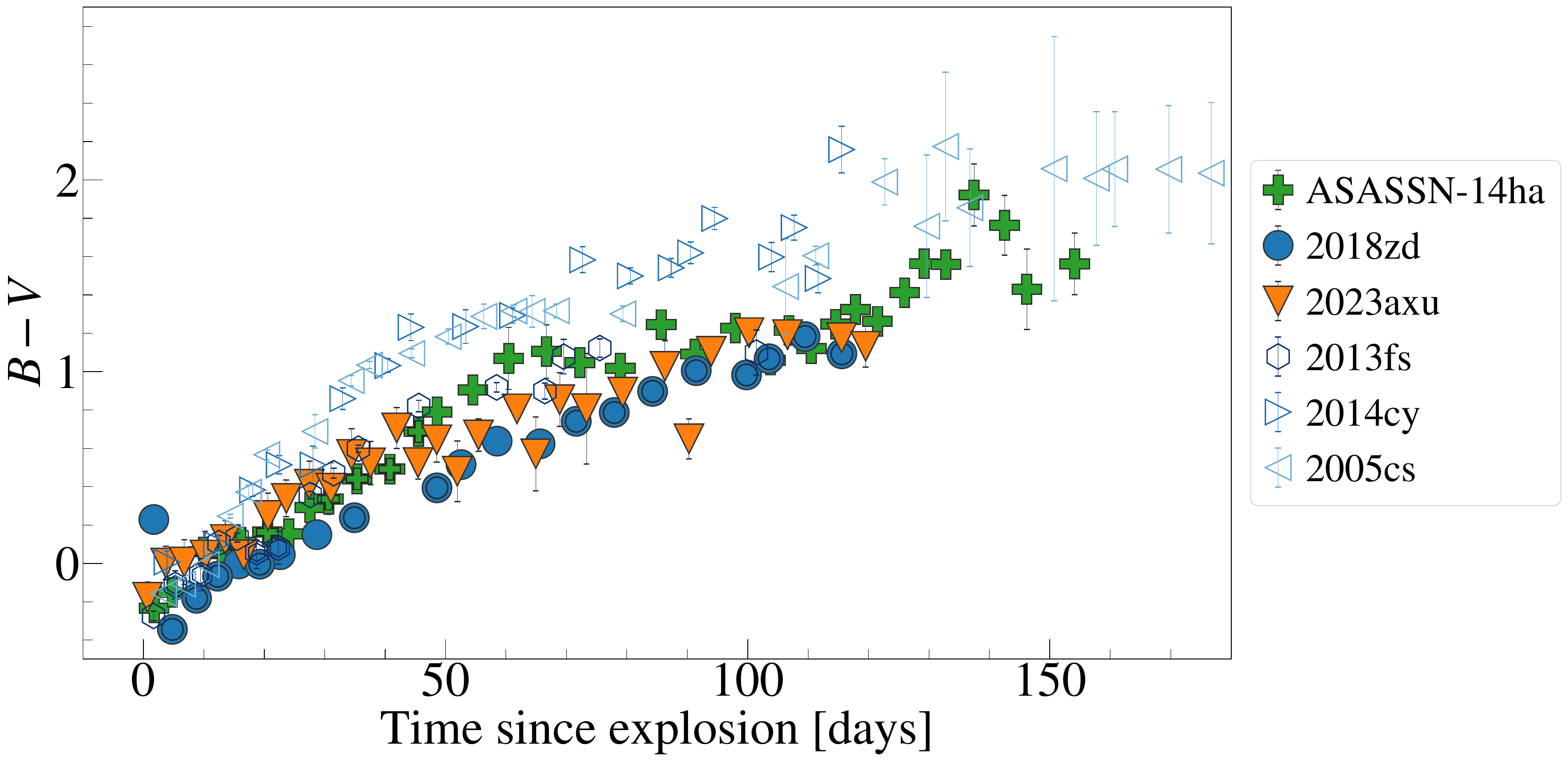}
\caption{
% The observed $r$- and $V$-band absolute-magnitude light curves of the ECSN candidates.
The observed $V$-band absolute-magnitude light curves (\textit{top}) and $B-V$ evolutions (\textit{bottom}) of the \textit{gold candidates}, ASASSN-14ha, SNe~2018zd, and 2023axu, together with the reference SNe~2005cs, 2013fs, and 2014cy.
Ground-based observations are indicated by single marker edges, whereas Swift observations are indicated by double marker edges.
Markers are shown only for epochs separated by $\geq 3$~days to reduce crowding.
\label{fig:obsLC_gold}}
\end{figure*}

% Figure~\ref{fig:obsLC_gold} shows the observed $V$-band absolute-magnitude light curves and $B-V$ evolutions of the \textit{gold candidates}, compared with those of a low-luminosity SN~2005cs \citep{Pastorello2009-ge} and normal SNe~2013fs \citep{Yaron2017-lq} and 2014cy \citep{Valenti2016-ao}.
% Figure~\ref{fig:obsLC_gold} shows the observed $V$-band absolute-magnitude light curves and $B-V$ evolutions of the \textit{gold candidates}, compared with those of the reference SNe.
Figure~\ref{fig:obsLC_gold} shows the observed $V$-band absolute-magnitude light curves and $B-V$ evolutions of the \textit{gold candidates}, compared with those of the reference SNe, including Swift observations.
% Figure~\ref{fig:obsLC_gold} shows the observed $V$-band absolute-magnitude light curves and $B-V$ evolutions of the \textit{gold candidates}, compared with those of nearby and well-observed reference SNe.
% The reference SNe sample the plateau luminosity range from the prototypical low-luminosity SN~2005cs \citep{Pastorello2009-ge} through the normal SN~2014cy \citep{Valenti2016-ao} and to the relatively bright (but still normal) SN~2013fs \citep{Yaron2017-lq}, all of which exhibit plateau durations typical of SNe~II.
For visual clarity, markers are plotted only for epochs separated by $\geq 3$~days.
% The plateau of ASASSN-14ha is around $140$~days, longer than SNe~2005cs, 2013fs, and 2014cy.
The large errors of ASASSN-14ha are due to the distance uncertainty toward the host galaxy, with the distance modulus ranging from $29.53$ \citep{Valenti2016-ao} to $30.86$ \citep{Rodriguez2022-ck}.

While the plateau length of ASASSN-14ha is around $140$~days, significantly longer than those of SNe~2005cs ($\sim 100-130$~days), 2013fs ($\sim 80-100$~days), and 2014cy ($\sim 120$~days), those of SNe~2018zd and 2023axu are around $120$ and $100$~days, respectively, comparable to the reference SNe.
% The plateau lengths of ASASSN-14ha, SNe~2018zd, and 2023axu are around $140$, $120$, $100$~days, respectively.
% ASASSN-14ha shows a long plateau for $\sim 140$~days.

The plateau of ASASSN-14ha is faint, $V \sim -15$~mag, slightly brighter than that of the low-luminosity SN~2005cs ($V \sim -14.5$~mag) but still fainter than that of the normal SNe~2013fs ($V \sim -17$~mag) and 2014cy ($V \sim -15.5$~mag).
The plateau brightness of SNe~2018zd and 2023axu is $V \sim -16$ and $-17$~mag, respectively, similar to that of the normal SNe~2013fs and 2014cy.

The tail of ASASSN-14ha is faint, $V \sim -11$~mag around $150$~days, brighter than that of SN~2005cs ($V \sim -10$~mag around $150$~days) but fainter than that of SNe~2013fs ($V\sim -14.5$~mag around $100$~days) and 2014cy ($V\sim -12$~mag around $160$~days).
The tail of SN~2018zd ($V \sim -11.5$ mag at $\sim$180 days) is similar to that of SN~2014cy, while the tail of SN~2023axu ($V \sim -15$ mag at $\sim$100 days) is comparable to that of SN~2013fs.
% The estimated $\Mni$ for ASASSN-14ha, SNe~2018zd, 2023axu, 2005cs, 2013fs, and 2014cy are $0.0014 \pm 0.0002$, $0.0086 \pm 0.0005$, $0.058 \pm 0.017$, $0.0021 \pm 0.0002$, $0.0545 \pm 0.0003$, and $0.0037 \pm 0.0038~\Msun$, respectively \citep{Valenti2016-ao, Hiramatsu2021-er, Shrestha2024-ty}, which are broadly consistent with the $V$-band absolute magnitudes observed during the tails. 
The estimated $\Mni$ for the reference SNe~2005cs, 2013fs, and 2014cy are $0.0021 \pm 0.0002$, $0.0545 \pm 0.0003$, and $0.0037 \pm 0.0038~\Msun$, respectively.
Together with the $\Mni$ estimates for ASASSN-14ha, SNe~2018zd, and 2023axu in Section~\ref{subsec:ECSNselection}, these estimates are broadly consistent with the $V$-band absolute magnitudes observed during the tails.

% While ECSNe are theoretically predicted to produce a small amount of \Nifs, $0.002-0.004~\Msun$ \citep{Kitaura2006-ia,Janka2008-ai,Wanajo2009-yo}, SN~2023axu exhibits a relatively bright tail, from which a comparatively large $\Mni$ is inferred. 
While ECSNe are theoretically predicted to produce a small amount of \Nifs, $0.002-0.004~\Msun$ \citep{Kitaura2006-ia,Janka2008-ai,Wanajo2009-yo}, the tail-based $\Mni$ estimate for SN~2023axu is comparatively large. 
% However, such luminous tails can also be powered by additional energy sources such as CSM interaction or pulsar spin-down, as suggested for SN~1054 by \citet{Moriya2014-oh,Tominaga2013}. 
However, additional energy sources such as CSM interaction or pulsar spin-down can also power a luminous tail without necessarily affecting the plateau, as suggested for SN~1054 \citep{Moriya2014-oh,Tominaga2013}. 
In such cases, the true $\Mni$ may be lower than that inferred under the assumption that the tail is powered solely by radioactive decay. 
We therefore do not rule out SN~2023axu from the candidates.

The $B-V$ evolutions of the \textit{gold candidates} are similar to those of the reference SNe before $\sim 20$~days after explosion.
During the plateau phase thereafter, the \textit{gold candidates} are persistently bluer than SNe~2005cs and 2014cy on average.
While the \textit{gold candidates} show similar $B-V$ evolutions to SN~2013fs, their plateau durations are longer than that of SN~2013fs, suggesting that both the color evolution and plateau length are important for distinguishing ECSN candidates from normal SNe~II.

% Overall, the \textit{gold candidates} span a wide range in plateau and tail luminosities, from the relatively faint ASASSN-14ha to the brighter SN~2023axu, illustrating the diversity among potential ECSN events.

\subsubsection{Silver candidates}

\begin{figure*}[ht!]
\centering
\includegraphics[width=180truemm]{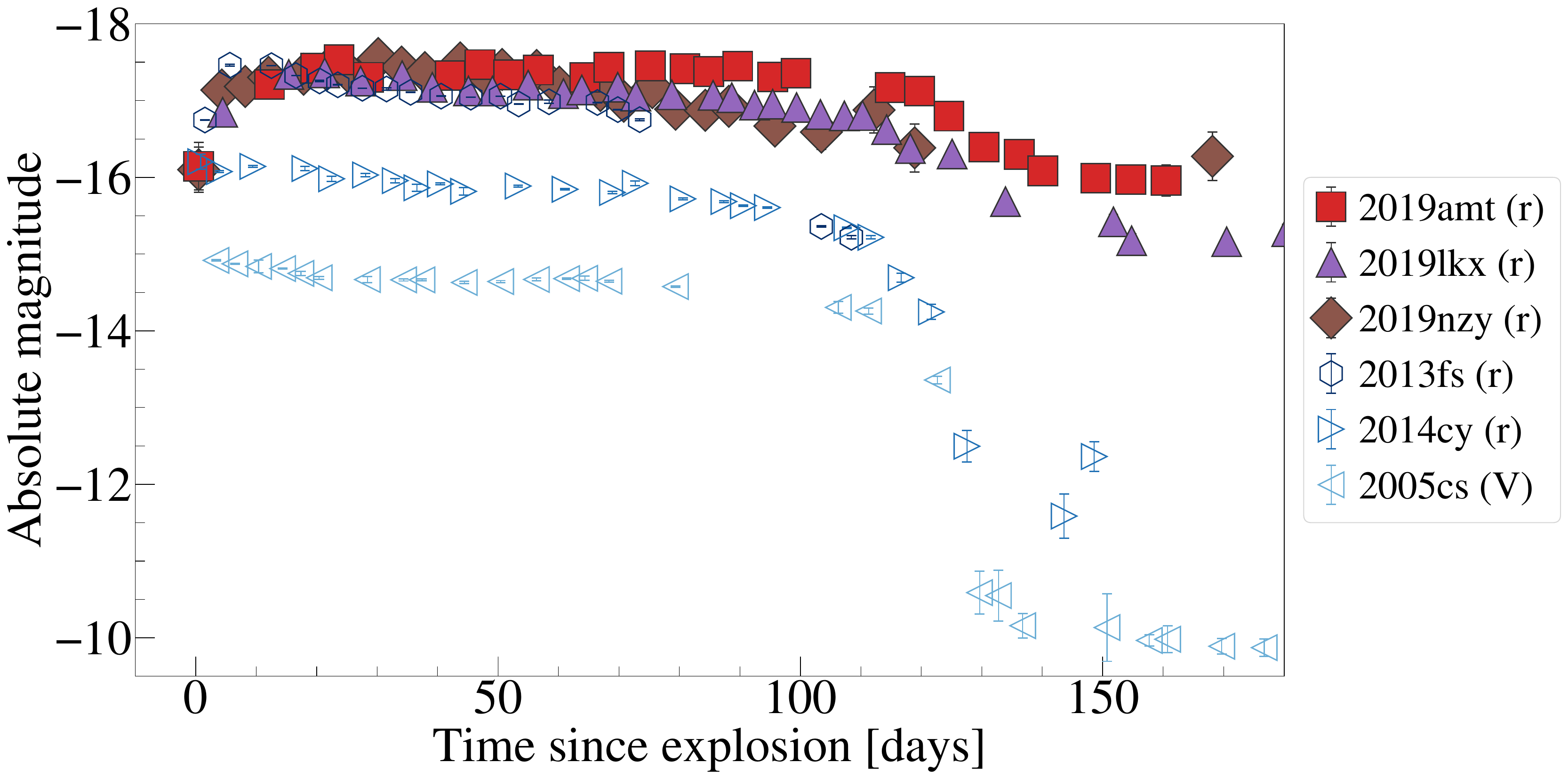}
\centering
\includegraphics[width=180truemm]{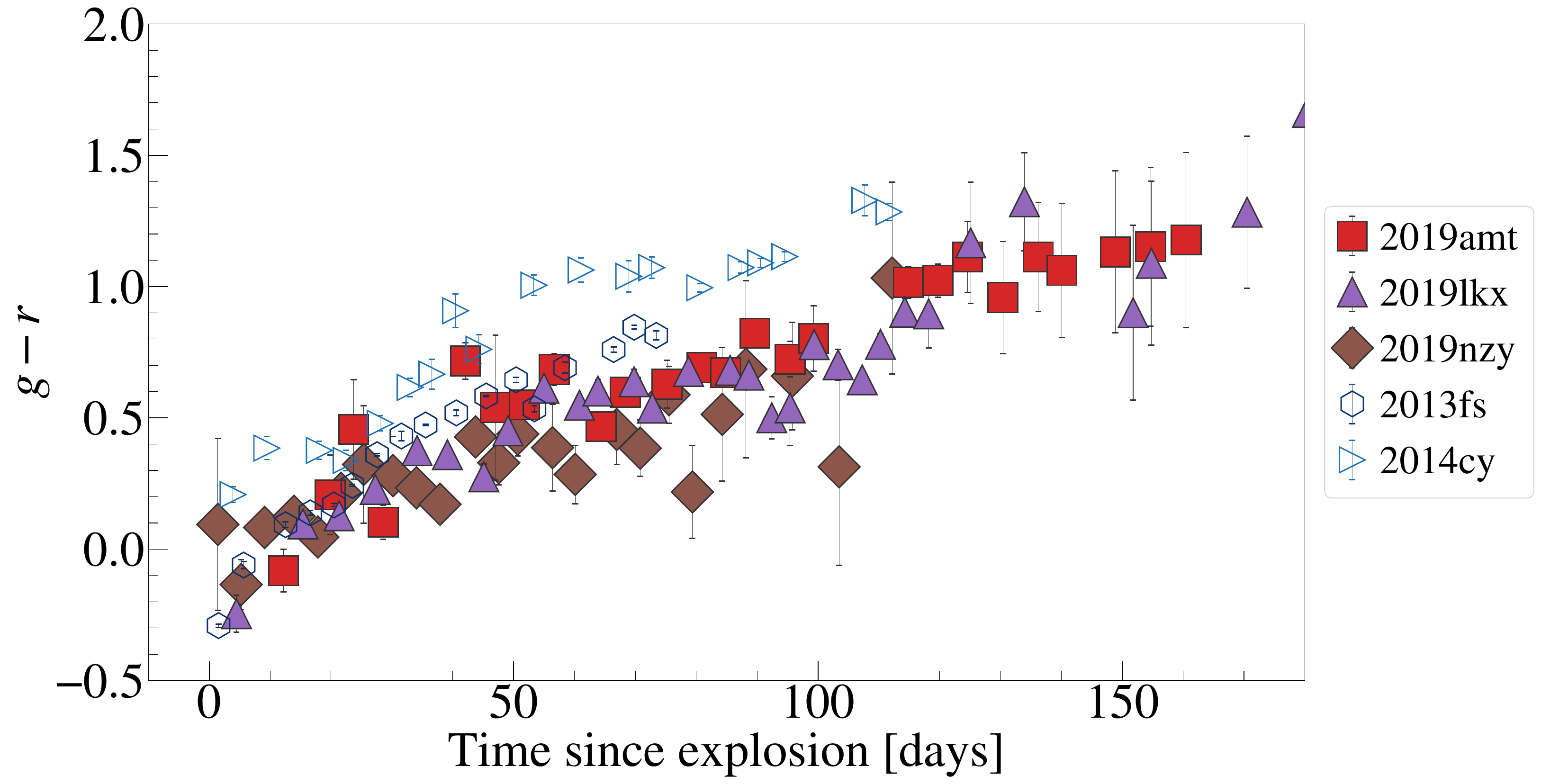}
\caption{
% The observed $r$- and $V$-band absolute-magnitude light curves of the ECSN candidates.
The observed $r$-band absolute-magnitude light curves (\textit{top}) and $g-r$ evolutions (\textit{bottom}) of the \textit{silver candidates}, SNe~2019amt, 2019lkx, and 2019nzy, together with those of the reference SNe~2013fs and 2014cy. 
For comparison, the observed $V$-band absolute magnitude light-curve of the reference SN~2005cs is also shown in the \textit{top} panel.
Markers are shown only for epochs separated by $\geq 3$~days to reduce crowding.
\label{fig:obsLC_silver1}}
\end{figure*}

\begin{figure*}[ht!]
\centering
\includegraphics[width=180truemm]{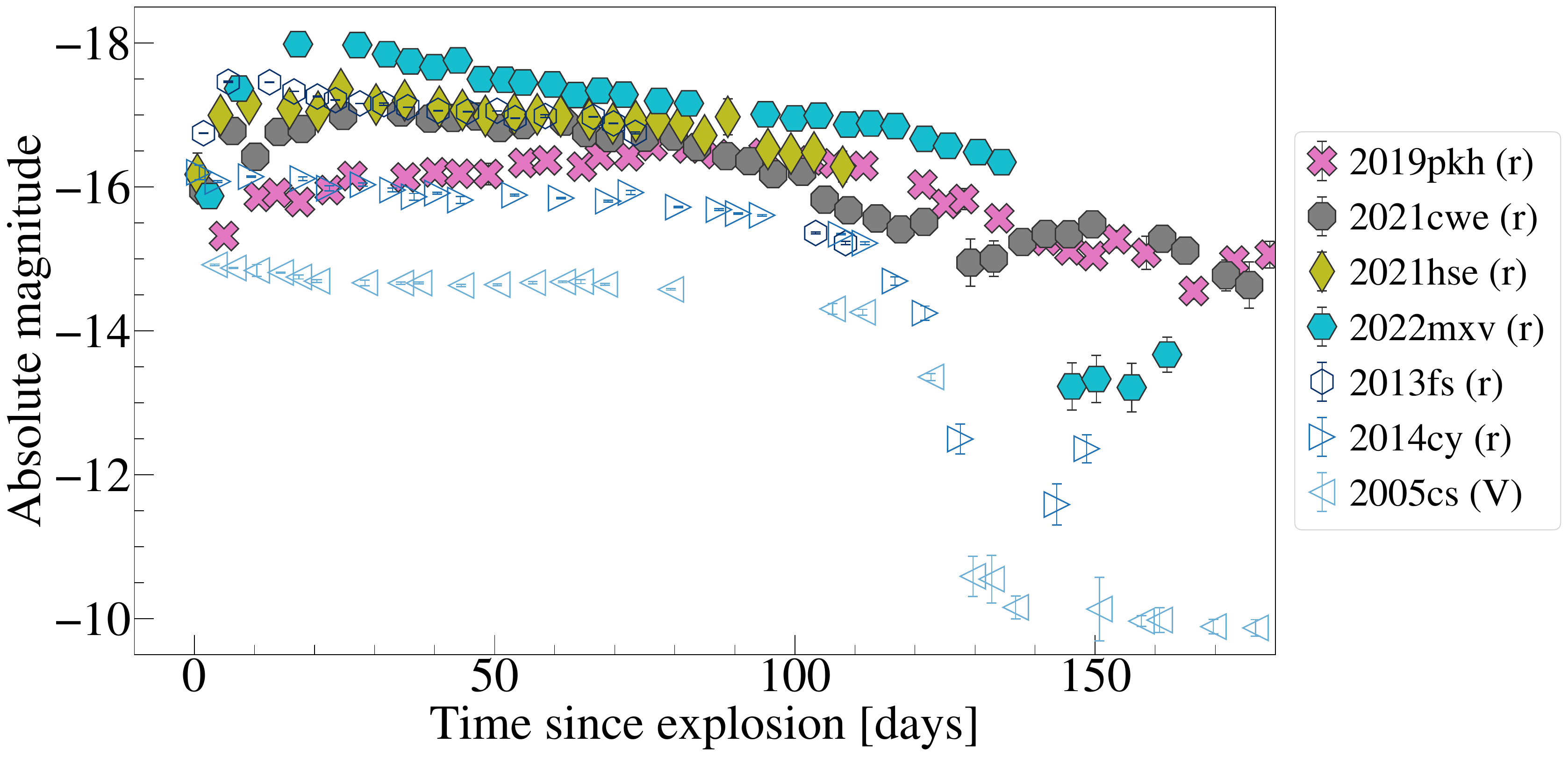}
\centering
\includegraphics[width=180truemm]{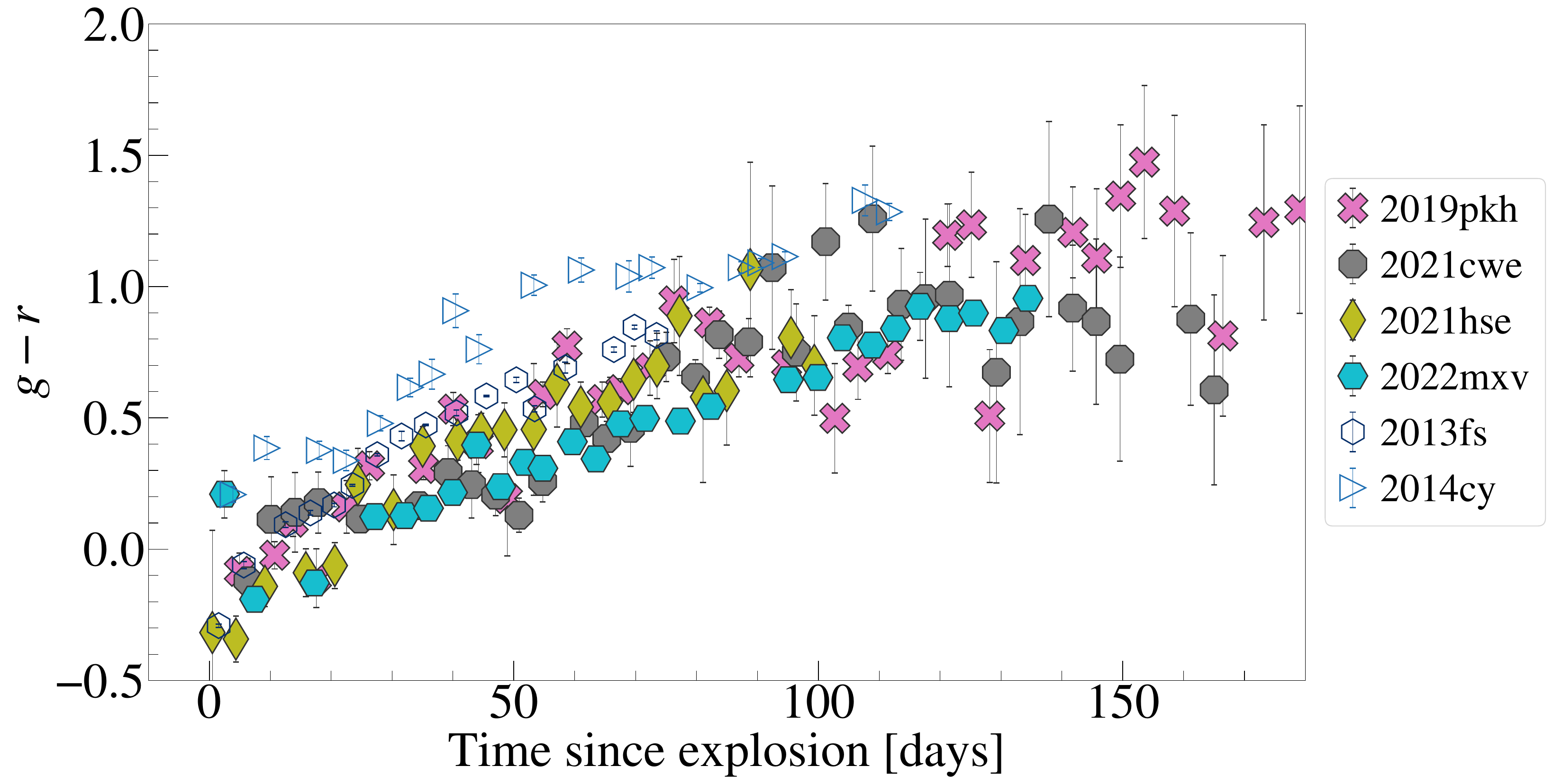}
\caption{
% The observed $r$- and $V$-band absolute-magnitude light curves of the ECSN candidates.
Same as figure~\ref{fig:obsLC_silver1}, but for the \textit{silver candidates}, SNe~2019pkh, 2021cwe, 2021hse, and 2022mxv.
\label{fig:obsLC_silver2}}
\end{figure*}

Figures~\ref{fig:obsLC_silver1} and \ref{fig:obsLC_silver2} show the observed $r$-band absolute-magnitude light curves of the \textit{silver candidates}, compared with those of SNe~2013fs and 2014cy and the $V$-band light curve of SN~2005cs.
For visual clarity, markers are plotted only for epochs separated by $\geq 3$~days.

% \begin{figure*}[ht!]
% \centering
% \includegraphics[width=180truemm]{99_fig_compLC.r-1.pdf}
% \centering
% \includegraphics[width=180truemm]{99_fig_compLC.r-2.pdf}
% \caption{
% % The observed $r$- and $V$-band absolute-magnitude light curves of the ECSN candidates.
% The observed $r$-band absolute-magnitude light curves of the \textit{silver candidates}, SNe~2019amt, 2019lkx, 2019nzy, 2019pkh, 2021cwe, 2021hse, and 2022mxv, together with those of SNe~2013fs and 2014cy. 
% For comparison, the observed $V$-band absolute magnitude light-curve of SN~2005cs is also shown.
% Markers are shown only for epochs separated by $\geq 3$~days to reduce crowding.
% \label{fig:obsLC_silver}}
% \end{figure*}

The plateau lengths of SNe~2019amt, 2019nzy, 2019pkh, 2021cwe and 2021hse are around $100-120$~days, longer than SN~2013fs and consistent with SNe~2005cs and 2014cy.
In contrast, the plateau lengths of SNe~2019lkx and 2022mxv are around $130$ and $140$~days, respectively, significantly longer than the reference SNe.

The plateaus of SNe~2019lkx, 2019nzy, 2021cwe, and 2021hse are of similar brightness ($r\sim -17$~mag) to those of SN~2013fs and brighter than those of SNe~2005cs and 2014cy.
Those of SNe~2019amt and 2022mxv are brighter ($r\sim -17.5$~mag) than those of the reference SNe, suggesting that they belong to the brighter end of the SN~II population.
% The brightness evolution of SN~2019pkh in the plateau is somewhat peculiar, faint in the early phase ($r \sim -15.5$~mag around $5$~days) and gradually increasing toward the end of the plateau ($r \sim -16.5$~mag around $110$~days).
The brightness evolution of SN~2019pkh in the plateau is somewhat peculiar, faint in the early phase ($r \sim -15.5$~mag around $5$~days) and gradually increasing toward the end of the plateau ($r \sim -16.5$~mag around $110$~days),
% as seen in some low-luminosity SNe \citep{Valerin2022-ow}.
possibly due to the increase of the photospheric radius, as seen in some low-luminosity SNe \citep{Valerin2022-ow}.

The tails of SNe~2019amt, 2019lkx, 2019pkh, and 2021cwe are $r \sim -16$~mag around $130$~days, 
likely similar to SN~2013fs and brighter than SNe~2005cs and 2014cy, 
% with SN~2021cwe showing a $\sim 1$~mag bump around $150$~days.
with SNe~2019pkh and 2021cwe showing bumpy evolution around $150$~days, possibly due to interaction with a CSM of complex structure.
The tail of SN~2019nzy is $r \sim -17$~mag around $170$~days, brighter than those of the reference SNe.
The tail of SN~2022mxv is $r\sim-13.5$~mag around $140$~days, brighter than SNe~2005cs and 2014cy and fainter than SN~2013fs.
These bright tails of the \textit{silver candidates} result in the $\Mni$ estimates similar to or higher than the $\Mni$ estimates for the reference SNe.
% The tails of the \textit{silver candidates} are similar to or brighter than those of the reference normal SNe 2013fs and 2014cy, resulting in the $\Mni$ estimates similar to or higher than the $\Mni$ estimates for the reference SNe.
% The $\Mni$ estimates for the \textit{silver candidates} are similar to or higher than those for the reference SNe.
The tail-based $\Mni$ estimates are also higher than the theoretical predictions \citep{Kitaura2006-ia,Janka2008-ai,Wanajo2009-yo}, possibly indicating that they are not ECSNe or that their tails are powered by additional sources, most plausibly CSM interaction and/or a central engine. 
Because spectra around $\tPT/2$ are unavailable for the silver candidates, we cannot distinguish these possibilities. 
Also, the tail brightness may be enhanced by a moderate CSM interaction that does not necessarily affect the plateau \citep{Moriya2014-oh}.
% We therefore retain the silver candidates not as robust ECSN identifications, but as photometrically selected candidates that may include false positives and are used solely to derive an upper limit on the ECSN occurrence ratio.
We therefore retain the silver candidates not as robust ECSN identifications, but as photometrically selected candidates that may include false positives, so as not to exclude potential ECSN candidates.
% We do not, however, rule out the \textit{silver candidates}, as in the case of the \textit{gold candidate} SN~2023axu.

The $g-r$ evolutions of the \textit{silver candidates} during the plateau phase are, on average, bluer than that of SN~2014cy .
Among them, SNe~2019amt and 2019pkh, show $g-r$ evolutions comparable to SN~2013fs but with longer plateau durations.
% Among them, SNe~2019amt and 2019pkh, show $g-r$ evolutions comparable to SN~2013fs but with longer plateau durations, suggesting the importance of both color evolution and plateau duration in ECSN candidate selection, as also seen for the \textit{gold candidates}.
% Among them, SNe~2019amt and 2019pkh, show $g-r$ evolutions comparable to SN~2013fs but with longer plateau durations, suggesting that both color evolution and plateau duration are important for distinguishing ECSN candidates from normal SNe~II, as also seen for the \textit{gold candidates}.
% , while the other \textit{silver candidates} are bluer than SN~2013fs.
The remaining \textit{silver candidates} are generally bluer than SN~2013fs.

% Taken together, the \textit{silver candidates} tend to show somewhat longer and brighter plateaus, and relatively luminous tails compared to the reference SNe, though with considerable diversity among individual events.

\subsection{Comparison with Light-Curve Models}

\begin{figure*}[ht!]
  % \begin{minipage}[b]{0.32\linewidth}
  %   \centering
  %   \includegraphics[width=50truemm]{99_fig_2018zd_ONeMgM4.7X0.70247a2012totinc.WanajoST.mni1e-2_m3d-2vf1d1vi5d0r3d14z400s2b2_STELLA2023r_M1.364hE0.06BQ0Teps3d-3bq1d0_110213cfca.eps}
  % \end{minipage}
  % \begin{minipage}[b]{0.32\linewidth}
  %   \centering
  %   \includegraphics[width=50truemm]{99_fig_ASASSN-14ha_ONeMgM4.7X0.70247a2012totinc.WanajoST.mni1e-2_m1d-3vf1d1vi5d0r1d15z400s2b1_STELLA2023r_M1.364hE0.04BQ0Teps3d-3bq1d0_110213cfca.eps}
  % \end{minipage}
  % \begin{minipage}[b]{0.32\linewidth}
  %   \centering
  %   \includegraphics[width=50truemm]{99_fig_2023axu_ONeMgM4.7X0.70247a2012totinc.WanajoST.mniuni1e-2_m3d-2vf1d1vi5d0r6d14z400s2b1_STELLA2023r_M1.373hE0.09BQ0Teps3d-3bq1d0_110213cfca.eps}
  % \end{minipage}
  % \plotone{99_fig_fit.multicolor.pdf}
  \includegraphics[width=180truemm]{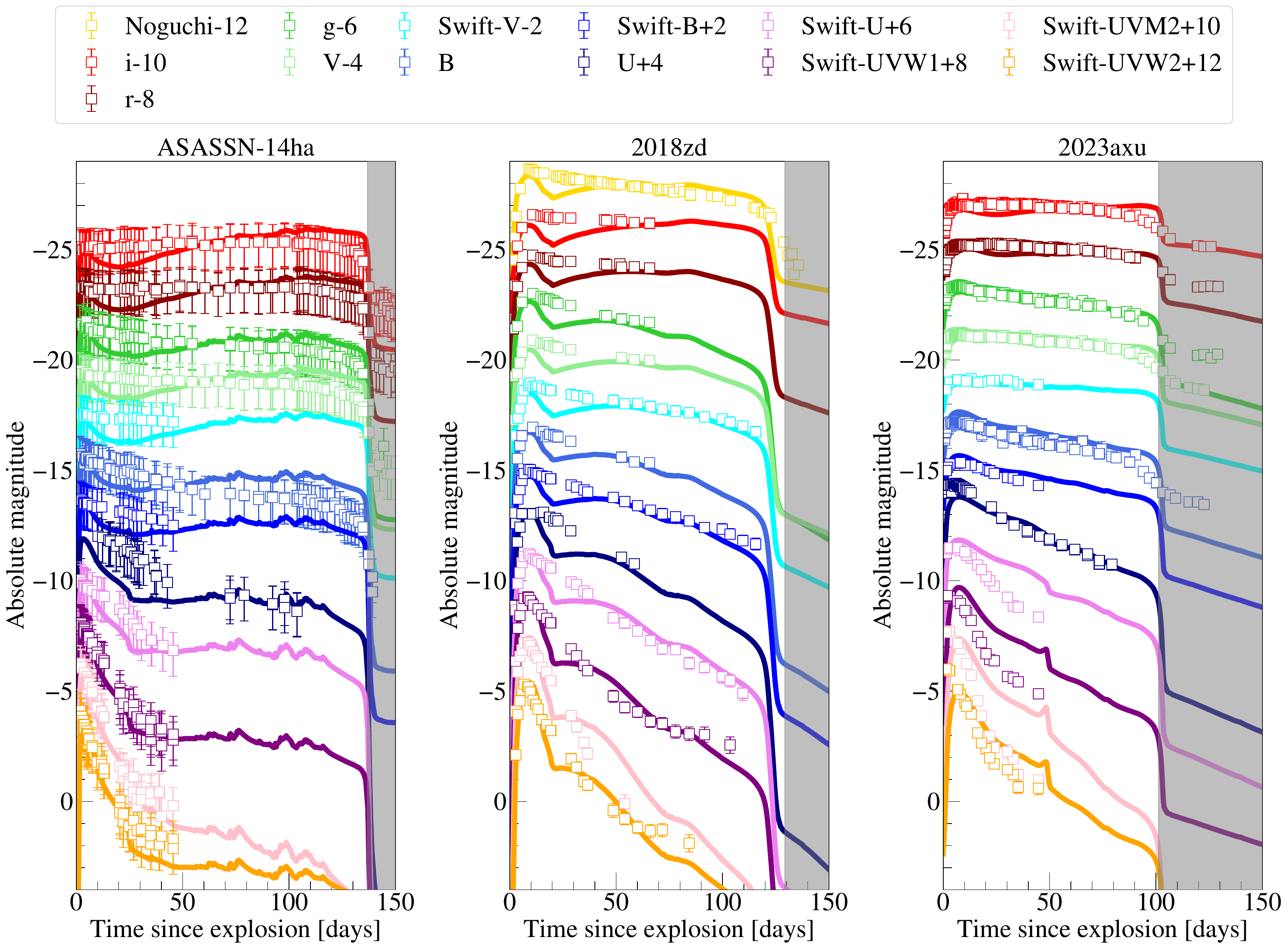}
\caption{
% \textit{Top panels}: 
Comparison of the observed light curves (points) of the \textit{gold candidates}, ASASSN-14ha (\textit{left}), SNe~2018zd (\textit{center}) and 2023axu (\textit{right}) and the best-fitting ECSN multicolor light-curve models (lines).
The gray-shaded region corresponds to epochs later than $\tPT$ and is excluded from the $\chi^2$ calculation.
% \textit{Bottom panels}: The photospheric velocity estimates ($\vph$) from the light-curve models (lines), compared with half of the velocity derived from the $\ha$ absorption minima ($\vha/2$; points).
% \textit{Bottom panels}: The upper limits of the photospheric velocity (points) estimated from the $\ha$ absorption minima and the photospheric velocity estimates from the light-curve models (lines).
\label{fig:LCreproduction1}}
\end{figure*}

\begin{figure*}[ht!]
  % \begin{minipage}[b]{0.32\linewidth}
  %   \centering
  %   \includegraphics[width=50truemm]{99_fig_2019lkx_ONeMgM4.7X0.70247a2012totinc.WanajoST_m1d-2vf1d1vi1d0r3d15z400s2d0b5_STELLA2023r_M1.364hE0.09BQ0Teps3d-3bq1d0_110213cfca.eps}
  % \end{minipage}
  % \begin{minipage}[b]{0.32\linewidth}
  %   \centering
  %   \includegraphics[width=50truemm]{99_fig_2019lkx_ONeMgM4.7X0.70247a2012totinc.WanajoST_m1d-2vf1d1vi1d0r3d15z400s2d0b5_STELLA2023r_M1.364hE0.09BQ0Teps3d-3bq1d0_110213cfca.eps}
  % \end{minipage}
  % \begin{minipage}[b]{0.32\linewidth}
  %   \centering
  %   \includegraphics[width=50truemm]{99_fig_2019lkx_ONeMgM4.7X0.70247a2012totinc.WanajoST_m1d-2vf1d1vi1d0r3d15z400s2d0b5_STELLA2023r_M1.364hE0.09BQ0Teps3d-3bq1d0_110213cfca.eps}
  % \end{minipage}
  % \plotone{99_fig_fit.ztfgri.pdf}
  % \includegraphics[width=180truemm]{99_fig_fit.ztfgri.pdf}
  \includegraphics[width=180truemm]{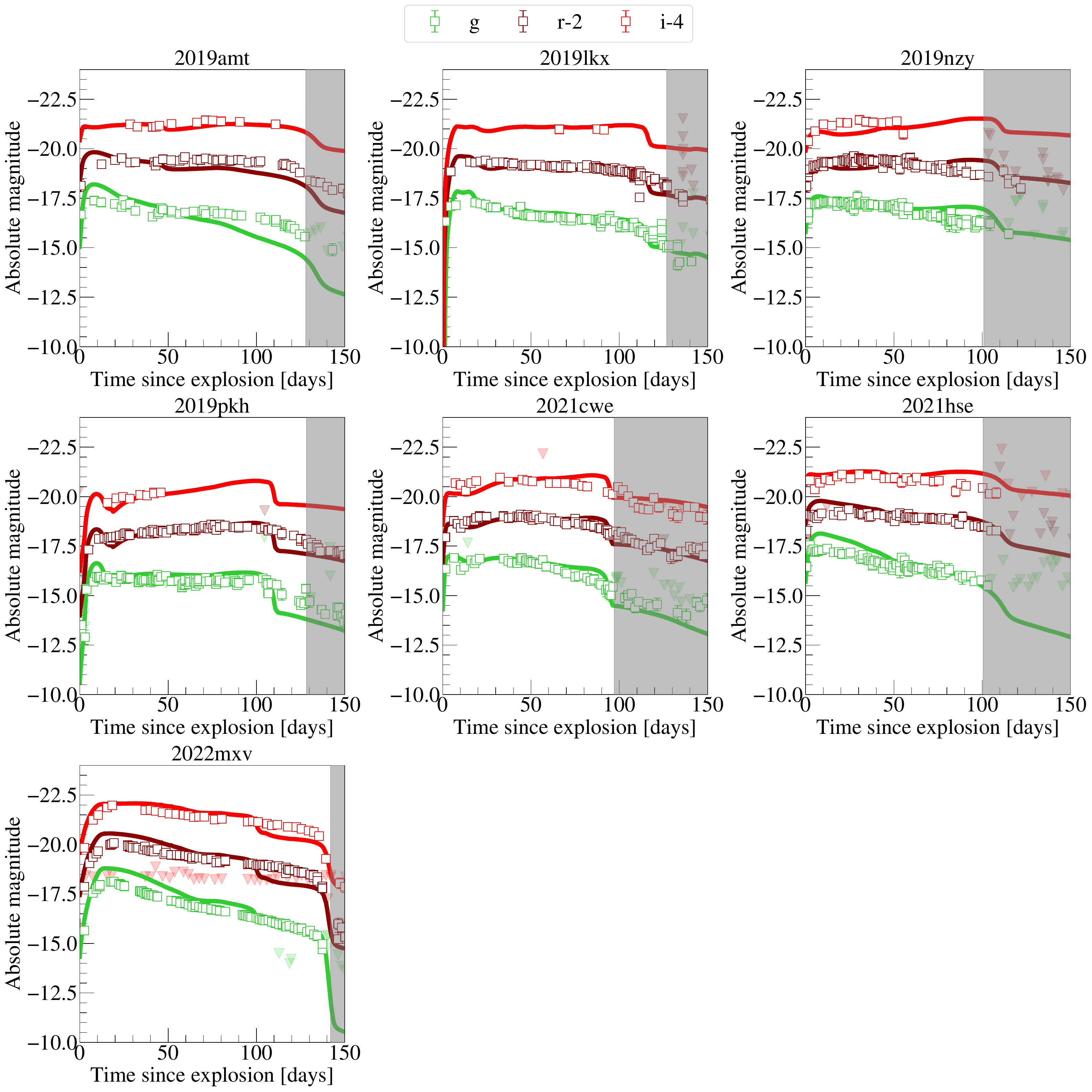}
\caption{
Comparison of the observed light curves (points) of the \textit{silver candidates}, SNe 2019amt, 2019lkx, 2019nzy, 2019pkh, 2021cwe, 2021hse, and 2022mxv and the best-fitting ECSN multicolor light-curve models (lines).
The gray-shaded region corresponds to epochs later than $\tPT$ and is excluded from the $\chi^2$ calculation.
\label{fig:LCreproduction2}}
\end{figure*}

Figures~\ref{fig:LCreproduction1} and \ref{fig:LCreproduction2} compare the best-fitting ECSN light-curve models and the observed light curves of the \textit{gold} and \textit{silver candidates}, respectively.
Table~\ref{table:properties_ECSNe} summarizes the explosion and progenitor properties inferred from the best-fitting models. 

% \begin{table*}[t]
%  \caption{Properties of the ECSN candidates, inferred from multicolor light-curve modelings}
%  \label{table:properties_ECSNe}
%  \centering
%   \makebox[1 \textwidth][l]{
%     % \resizebox{0.9 \textwidth}{!}{   % 必要ならコメントアウトを外してリサイズする
%   \begin{tabular}{ccccccccccc}
%    \hline
%    SN Name & \(\Menv\) & \(\XH\) & \(\Eexp\) & \(\Mdot\) & $\Rcsm$ & \(\vi\) & \(\vf\) & \(\beta\) \\ \relax
%    & [$\Msun$] & & [\(10^{50}\) ergs] & [$\Myr$] & [$10^{14}$cm] & [km/s] & [km/s] & \\
%    \hline
%    ASASSN-14ha & $3.0$ & $0.7$ & $0.1$ & $3 \times 10^{-3}$ & $3$ & $1-5$ & $10$ & $1-5$ \\
%    2018zd & $\leq 4.7$ & $0.7$ & $\leq 0.8$ & $(1-3) \times 10^{-2}$ & $3$ & $1$ & $30$ & $3-5$ \\
%    2019rqe & $3.0$ & $0.7$ & $1.0$ & $10^{-2}$ & $6-10$ & $1-5$ & $10$ & $3-5$ \\
%    2020abtf & $\leq 4.7$ & $0.7$ & $\leq 0.9$ & $(0.3-1) \times 10^{-2}$ & $10$ & $1-5$ & $10$ & $3-5$ \\
%    2021ucg & $4.7$ & $0.7$ & $1.8$ & $10^{-2}$ & $10$ & $1-5$ & $10$ & $3-5$ \\
%    2023axu & $3.0$ & $0.7$ & $1.3$ & $(1-3) \times 10^{-2}$ & $6$ & $1-5$ & $10$ & $3-5$ \\
%    \hline
%   \end{tabular}
%   }
%   % }
% \end{table*}

\begin{deluxetable*}{cccccccccc}
\tablewidth{0pt}
\tablecaption{Properties of the ECSN candidates, inferred from the multicolor light-curve modelings}
\label{table:properties_ECSNe}
\tablehead{
% \colhead{Number} & \colhead{Units} & \colhead{Label} & \colhead{Explanation}
\colhead{SN Name} & \colhead{\(\Menv\)} & \colhead{\(\XH\)} & \colhead{\(\Eexp\)} & \colhead{\(\Mdot\)} & \colhead{$\Rcsm$} & \colhead{\(\vi\)} & \colhead{\(\vf\)} & \colhead{\(\beta\)} & \colhead{CSM mass} \\
\colhead{} & \colhead{[$\Msun$]} & \colhead{} & \colhead{[\(10^{50}\) ergs]} & \colhead{[$\Myr$]} & \colhead{[$10^{14}$cm]} & \colhead{[km/s]} & \colhead{[km/s]} & \colhead{} & \colhead{[$\Msun$]}
}
\startdata
   ASASSN-14ha & 4.7 & 0.7 & 0.4 & $3 \times 10^{-3}$ & $3$ & 5 & 10 & 5 & 0.04 \\
   2018zd & 4.7 & 0.7 & 0.8 & $3 \times 10^{-2}$ & $3$ & 5 & 10 & 5 & 0.40 \\
   2023axu & 4.7 & 0.7 & 1.7 & $10^{-2}$ & $10$ & 5 & 10 & 2 & 0.37 \\
   2019amt & 4.7 & 0.7 & 1.8 & $10^{-2}$ & $10$ & 5 & 10 & 1 & 0.34 \\
   2019lkx & 4.7 & 0.7 & 1.7 & $3 \times 10^{-3}$ & $30$ & 5 & 10 & 1 & 0.30 \\
   2019nzy & 3.0 & 0.7 & 1.4 & $10^{-2}$ & $10$ & 5 & 10 & 3 & 0.40 \\
   2019pkh & 3.0 & 0.7 & 0.8 & $3 \times 10^{-2}$ & $3$ & 5 & 10 & 3 & 0.38 \\
   2021cwe & 3.0 & 0.7 & 1.4 & $3 \times 10^{-3}$ & $10$ & 5 & 10 & 3 & 0.12 \\
   2021hse & 3.0 & 0.7 & 1.3 & $10^{-2}$ & $10$ & 5 & 10 & 1 & 0.34 \\
   2022mxv & 4.7 & 0.7 & 2.7 & $10^{-2}$ & $30$ & 5 & 10 & 3 & 1.07 \\
\enddata
% \tablecomments{Table 2 is published in its entirety in the electronic 
% edition of the {\it Astrophysical Journal}.  A portion is shown here 
% for guidance regarding its form and content. The {\tt\string \digitalasset}\ command highlights the Table title to visually indicate to the reader that there is data associated with this table.}
\end{deluxetable*}

While the models show good agreement with the observations for the \textit{gold candidates}, they are more (SN~2023axu) or less (ASASSN-14ha and SN~2018zd) luminous than the observations just after the peak.
This may indicate the existence of more complicated CSM profiles (\eg aspherical profile as indicated for SN~2023ixf; \citealt{Shrestha2025-wz,Smith2023-gc,Singh2024-hp}) than considered here.
On the other hand, since the peak luminosity reflects the overall density profile of the surrounding CSM, the agreement at the peak lends support to the inferred $\Mdot$ and $\vi$ although they are degenerate. 
% On the other hand, since the peak luminosity reflects the overall density profile of the surrounding CSM, the agreement at the peak supports the inferred $\Mdot$ and $\vi$ in a 1D simplified picture although they are degenerate. 
% Since the peak luminosity reflects the overall density profile of the surrounding CSM, this lends support to the inferred $\Mdot$ and $\vi$ although they are degenerated. 
Also, since the duration and later-phase luminosity of plateaus reflect the properties of the H-rich envelopes of the progenitors and explosion energies \citep{Litvinova1985-ty, Popov1993-cu, Eastman1994-hs}, 
the similarities in the late plateaus lend support to the $\Menv$ and $\Eexp$ inferred here.
The observed $\vha$ during the plateau phase of both ASASSN-14ha and SN~2018zd are around $2-3$ times larger than the $\vph$ evolutions of the best-fitting model, while the detailed line formation cannot be directly derived from \stella~calculations.
% Although the line-forming region cannot be directly derived from \stella~calculations, it is likely extended to outside from the recombination front as the $\ha$ line opacity is significantly higher than the electron-scattering opacity, which dominates the continuum (\eg $\vha / \vph \sim 1-2$ is indicated for normal SNe~II and SN~1987A \citep{Eastman1989-es,Dessart2005-sb}).
% Although the line-forming region cannot be directly derived from \stella~calculations, it is likely extended to outside from the recombination front as the $\ha$ line opacity is significantly higher than the electron-scattering opacity, which dominates the continuum \citep{Eastman1989-es,Utrobin2005-iw}.
% Although the line-forming region cannot be directly derived from \stella~calculations, it is likely extended to outside from the continuum photosphere as the $\ha$ line opacity is significantly higher than the electron-scattering opacity, which dominates the continuum \citep{Eastman1989-es,Utrobin2005-iw}.
% The relation between $\vha$ and $\vph$ has not been investigated for ECSNe, whereas $\vha / \vph \sim 1-2$ is indicated for normal SNe~II and SN~1987A \citep{Eastman1989-es,Dessart2005-sb}.
The relation between line velocities and $\vph$ has not been investigated for ECSNe, whereas $\vha / \vph \sim 1-2$ is indicated for normal SNe~II and SN~1987A 
% because the $\ha$ can remain optically thick above the electron-scattering photosphere, likely forming around the H recombination front.
% due to the large $\ha$ optical depth above the electron-scattering photosphere \citep{Eastman1989-es,Utrobin2005-iw,Dessart2005-sb}, likely forming around the H recombination front.
due to the large $\ha$ optical depth above the electron-scattering photosphere \citep{Eastman1989-es,Utrobin2005-iw,Dessart2005-sb}, which likely lies around the H recombination front.
While $\vfe/\vph \sim 0.9-1.1$ is indicated for normal SNe~II \citep{Dessart2005-sb}, the observed $\vfe$ for ASASSN-14ha and SN~2018zd are also significantly higher than the model $\vph$ ($\vfe/\vph \sim 2-3$ and $\sim 1.5-2.5$, respectively).
In ECSNe, the photospheric radius may be smaller than the radius of the H recombination front due to the low ejecta density, while both radii tend to be closer in FeCCSNe \citep{Sato2024-kt}.
% In ECSNe, the photospheric radius may be smaller than the H recombination front by a factor of $\lesssim 2$ because the ejecta is significantly low even in the hot H-ionized region, while, in FeCCSNe, both radii are almost the same because the density is sufficiently high \citep{Sato2024-kt}.
% Therefore, if the $\ha$ line-forming region extends around and above the recombination front as in normal SNe~II, it may be further outside the photosphere in ECSNe, which may lead to $\vha / \vph \sim 2-3$ for ASASSN-14ha and SN~2018zd.
% Therefore, if the $\ha$ line-forming region extends around and above the recombination front, $\vha/\vph$ may be higher in ECSNe than in normal SNe II.
% Therefore, if the $\ha$ and Fe~{\sc ii} line-forming regions extend around and above the recombination front, these velocity ratios may be higher in ECSNe than in normal SNe~II.
% Therefore, if the optical depth in $\ha$ and Fe~{\sc ii} becomes large around and above the recombination front, these velocity ratios may be higher in ECSNe than in normal SNe~II.
Therefore, if the line absorption in $\ha$ and Fe~{\sc ii} is dominated at and above the H recombination front, these velocity ratios may be higher in ECSNe than in normal SNe~II.
% In particular for $\ha$, the absorption may be further biased to at and above the recombination front because the highly ionized layers below it can strongly suppress the $\ha$ optical depth.
% In particular for $\ha$, the optical depth may be strongly suppressed in the highly ionized layers below the recombination front and can rise rapidly once a substantial neutral fraction is present, further enhancing this effect.
In particular for $\ha$, the optical depth may be suppressed in the highly ionized layers below the H recombination front, 
% enhancing the bias toward absorption at and above the recombination front.
% making the absorption more prominent at and above the recombination front.
so that the absorption minimum can be set at larger radii, as it traces the region where the line optical depth becomes significant \citep{Kasen2001-si,Dessart2005-sb}.

For the \textit{silver candidates}, the models are broadly consistent with the observations.
% Figure~\ref{fig:LCreproduction2} compares the observed $g$-, $r$-, and $i$-band light curves of the \textit{silver candidates} with the best-fitting ECSN light-curve models. 
% Figure~\ref{fig:LCreproduction2} shows the comparisons for the \textit{silver candidates}. 
% For SNe~2019amt, 2019lkx, 2019nzy, and 2022mxv, although the models are fit only to the plateaus (from $\texp$ to $\tPT$), they tend to favor extended CSM distributions ($\Rcsm = 3 \times 10^{15}-10^{16}$~cm) that would continue to affect the luminosity into the tails. 
For SNe~2019lkx and 2022mxv, although the models are fit only to the plateaus (from $\texp$ to $\tPT$), they tend to favor extended CSM distributions ($\Rcsm = 3 \times 10^{15}$~cm) that would continue to affect the luminosity into the tails. 
This may indicate that the blue colors observed during the plateau are partly influenced by CSM interaction rather than being intrinsic to the SN ejecta.
% This may indicate that the blue colors observed during the plateau are partly influenced by CSM interaction rather than being intrinsic to the explosion.
While we do not exclude them from the \textit{silver candidates}, this possibility should be considered when interpreting their properties. 
% In particular, SN~2019amt is best fit by a model with a tenuous but extended CSM ($\Mdot = 10^{-6}~\Myr$, $\Rcsm = 10^{16}$~cm), leaving it ambiguous whether its blue plateau reflects intrinsic ECSN-like properties or mild CSM interaction. 
% By contrast, the other objects show models in which the CSM interaction ceases before the onset of the tail, and their late plateaus show overall good agreements with the observations without invoking strong CSM effects. 
By contrast, the other objects favor the models with less extended CSM ($\Rcsm \leq 10^{15}$~cm), for which the interaction ceases before the onset of the tail, and their late plateaus show overall good agreements with the observations without invoking strong CSM effects. 
% These cases provide physically consistent fits that remain compatible with an ECSN interpretation, although the overall diversity highlights that CSM properties can significantly complicate the identification of ECSNe based solely on light-curve morphology.

\section{Discussion \label{sec:disc}}

\subsection{Properties of the ECSN Candidates \label{subsec:ECSNprop}}

The $\Menv$ inferred for the \textit{gold candidates} are $4.7~\Msun$ (or $3.0–4.7~\Msun$ when the \textit{silver candidates} are included; Table~\ref{table:properties_ECSNe}), 
% indicating a lack of both high- and low-mass envelope progenitors. 
% indicating a lack of progenitors with both massive and less-massive envelope. 
indicating a lack of progenitors with either very massive or very low-mass envelopes.
Given the progenitor mass range of super-AGB stars ($\sim 8-10~\Msun$), this may suggest that their mass loss is extensive enough to strip a large fraction of the envelope, preventing the presence of progenitors with high envelope masses. 
% Conversely, if the mass loss is strong enough to leave only a low-mass envelope, the progenitor is likely to be embedded in dense and extended CSM, and the resulting explosion would more likely be classified as a Type IIn SN, a subclass of Type II SN with persistent narrow emission lines caused by interaction with dense and extended CSM, rather than an ordinary SN II, which may explain the apparent absence of systems with low envelope masses in our sample.
% Conversely, if the mass loss is strong enough to leave only a low-mass envelope, the progenitor is likely to be embedded in a dense and extended CSM, and the resulting explosion would more likely be classified as a Type IIn SN rather than an ordinary SN II, which may explain the apparent absence of systems with low envelope masses in our sample.
Conversely, if the mass loss is strong enough to leave only a low-mass envelope, the progenitor is likely to be embedded in a dense and extended CSM, and the resulting explosion would more likely be classified as a Type IIn SN, as discussed in the context of the faint-fast subclass by \citet{Hiramatsu2024-to}, rather than an ordinary SN II, which may explain the apparent absence of progenitors with low envelope masses in our sample.

\begin{figure}[ht!]
    \includegraphics[width=85truemm]{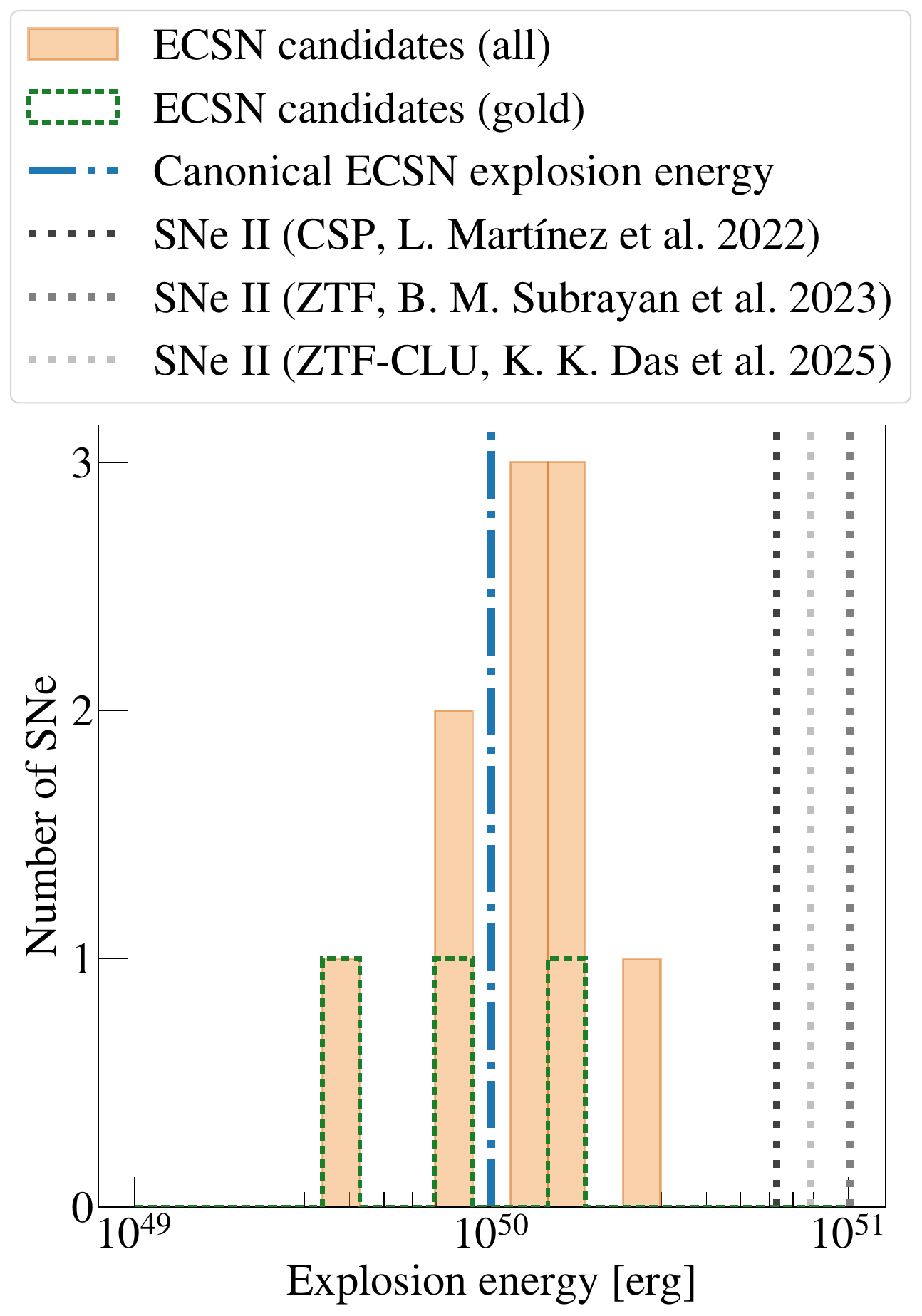}
\caption{
The distribution of the $\Eexp$ inferred for the ECSN candidates (the \textit{gold candidates} in green dotted line and the \textit{silver candidates} in orange), together with the median values of the $\Eexp$ of normal SNe~II (gray dotted lines) from literature \citep{Martinez2022-lv,Subrayan2023-md,Das2025-wj} and the canonical ECSN explosion energy of $10^{50}$~erg (blue dash-dotted line; \citealt{Kitaura2006-ia,Janka2008-ai}).
\label{fig:histE}}
\end{figure}

% Figure~\ref{fig:histE} shows the distribution of the $\Eexp$ inferred for the ECSN candidates, together with those of SNe~II in \citet{Das2025-wj}.
% The explosion energies of the SNe II are inferred based on the same code, $\stella$ as this work.
Figure~\ref{fig:histE} shows the distribution of the $\Eexp$ inferred for the ECSN candidates, together with the median values inferred for normal SNe~II in \citet{Martinez2022-lv,Subrayan2023-md,Das2025-wj} and the canonical ECSN explosion energy of $10^{50}$~erg \citep{Kitaura2006-ia,Janka2008-ai}.
% The $\Eexp$ inferred for the ECSN candidates are broadly consistent with that predicted by the first-principles simulations ($\sim 10^{50}$~erg; \citealt{Kitaura2006-ia, Janka2008-ai}) and less energetic than the median values inferred for normal SNe~II.
The $\Eexp$ inferred for the ECSN candidates are $(0.4-1.7) \times 10^{50}$~erg for the \textit{gold candidates} and $(0.4-2.7) \times 10^{50}$~erg including the \textit{silver candidates}, less energetic than the median values inferred for normal SNe~II.
% The $\Eexp$ inferred for the ECSN candidates are $\sim 10^{50}$~erg, less energetic than the median values inferred for normal SNe~II.
% The ECSN candidates are generally less energetic than the SNe~II.
% possibly because the shock revival occurs with less neutrino heating in the ECSN than in the FeCCSN, due to the steep density gradient at the super-AGB core surface as shown by explosion simulations.
% This trend is consistent with the explosion simulations where the shock revival occurs with less neutrino heating in the ECSN than in the FeCCSN, due to the steep density gradient at the super-AGB core surface.
% This is broadly consistent with the prediction of the first-principles simulations where the shock revival occurs with less neutrino heating in the ECSN than in the FeCCSN, due to the steep density gradient at the super-AGB core surface, resulting in the low explosion energy of $\sim 10^{50}$~erg \citep{Kitaura2006-ia,Janka2008-ai}.
This is broadly consistent with first-principles simulations, which find that ECSNe can achieve shock revival with less neutrino heating than FeCCSNe owing to the steep density gradient at the super-AGB core surface, leading to low explosion energies ($\sim 10^{50}$~erg; \citealt{Kitaura2006-ia,Janka2008-ai}).

% The $\Rcsm$ and $\Mdot$ range $(3-10) \times 10^{14}$~cm and $3 \times (10^{-3}-10^{-2})~\Myr$ for the \textit{gold candidates}, assuming $\vf=10$~\kms, indicating such strong mass loss lasts $\sim 10-20$ years before the explosions.
% The $\Mdot$ is much greater than that at the beginning of the thermal pulse phase, $\sim (2-3) \times 10^{-5}~ \Myr$, predicted by stellar evolution calculations \citep{Limongi2023-db}.
% This elevation may be because the electron degeneracy pressure increases as the core approaches the Chandrasekhar mass.
The inferred $\Rcsm$ and $\Mdot$ ranges are $(3-10) \times 10^{14}$~cm and $3 \times 10^{-3} - 3 \times 10^{-2}~\Myr$, respectively, for the \textit{gold candidates}, assuming $\vf=10$~\kms.
The $\Mdot$ is much greater than that at the beginning of the thermal pulse phase, $\sim (2-3) \times 10^{-5}~ \Myr$, predicted by stellar evolution calculations \citep{Limongi2023-db}.
The compact $\Rcsm$ indicates that such strong mass loss lasts $\sim 10-20$ years before the explosions and thus occurred during the very final stage in the super-AGB phase, lasting typically $\sim 10^4 - 10^5$~years \citep{Doherty2017-vu}.
% This enhancement of $\Mdot$ may reflect the increase of the electron degeneracy pressure as the core approaches the Chandrasekhar mass.
% This enhancement of $\Mdot$ may reflect the increase of electron degeneracy pressure as the core approaches the Chandrasekhar mass, which could make the envelope more extended and thus more prone to enhanced mass loss, potentially facilitated by thermal pulses \citep{Doherty2017-vu}.
% This enhancement of $\Mdot$ may reflect the increasing electron degeneracy pressure as the core approaches the Chandrasekhar mass, which steepens the pressure gradient at the core-envelope boundary and naturally leads to a more extended envelope, thereby making the star more susceptible to enhanced mass loss, potentially facilitated by thermal pulses \citep{Doherty2017-vu}.
This enhancement of $\Mdot$ is possibly related to the decreasing interpulse period as the core approaches the Chandrasekhar mass \citep{Poelarends2008-hq}.

Given that the tail-based $\Mni$ may be overestimated, we test the sensitivity of our results by using the models in Step~1, which adopt a theoretically expected $\Mni$ \citep{Wanajo2009-yo}.
The inferred properties are listed in Table~\ref{table:properties_ECSNe_lowMNi} in Appendix~\ref{app:param_lowMNi}.
The $\Menv$ are unchanged or increased from $3.0$ to $4.7~\Msun$ and the $\Eexp$ are unchanged or decreased by up to factor of two.
% These increased $\Menv$ and decreased $\Eexp$ are required to maintain plateau length and brightness with less \Nifs~heating.
These $\Menv$ and $\Eexp$ changes are required to maintain plateau length and brightness with less \Nifs~heating.
% The increased $\Menv$ and decreased $\Eexp$ are required to maintain plateau length and brightness with less \Nifs~heating.
% The $\Mdot$ are unchanged or changed by factor of $3$ and $\Rcsm$ are similarly unchanged or changed by factor of 3.
The $\Mdot$ and $\Rcsm$ are unchanged or changed by factors of three.
% We find that the inferred $\Menv$ and $\Mdot$ are unchanged for SNe~2018zd, 2019nzy, 2021hse, 2022mxv, and 2023axu, while the best-fitting $\Eexp$ decreases modestly, from $0.8$ to $0.6$, from $1.4$ to $1.0$, from $1.7$ to $1.4$ $\times 10^{50}$~erg for SN~2018zd, 2019nzy, and 2023axu and remains the same for SNe~2021hse and 2022mxv.
% The inferred $\Menv$ is unchanged for ASASSN-14ha, SNe~2018zd, 2023axu, 2019amt, 2019lkx, and 2023axu, while the best-fitting $\Eexp$ decreases modestly, from $0.8$ to $0.6$, from $1.4$ to $1.0$, from $1.7$ to $1.4$ $\times 10^{50}$~erg for SN~2018zd, 2019nzy, and 2023axu and remains the same for SNe~2021hse and 2022mxv.
% This indicates that our discussion based on the inferred $\Eexp$, $\Menv$, and $\Mdot$ is essentially unchanged within plausible uncertainties in $\Mni$.
These unchanged or modestly changed properties indicate that our discussion based on the inferred $\Menv$, $\Eexp$, $\Mdot$, and $\Rcsm$ is essentially unchanged within plausible uncertainties in $\Mni$.

\subsection{Occurrence Rate \label{subsec:rate}}
With the use of the obtained ECSN candidates, we estimate a volumetric rate for ECSNe.
In this work, the \textit{silver candidates} are drawn exclusively from the ZTF survey sample, whereas the \textit{gold candidates} are drawn exclusively from the literature sample.
While the ZTF survey sample provides a systematic data set, strong CSM interaction cannot be excluded for the \textit{silver candidates}, and they may include false positives.
We therefore use the \textit{silver candidates} to derive an upper limit on the ECSN occurrence ratio among SNe~II.
In contrast, while the literature sample is heterogeneous, it contains more robust \textit{gold candidates}.
Thus, we use the \textit{gold candidates} to derive a lower limit on the ECSN occurrence ratio among SNe~II.

We exclude SNe~2005cs, 2018zd, and 2023axu from the literature sample in the rate discussion to obtain a more uniform sample, since they were selected based on prior interest.
% We exclude two \textit{gold candidates}, SN~2023axu and SN~2018zd from the rate discussion because they were selected based on prior interest.
% as ECSN candidates or their blue colors.
After this exclusion, the literature sample contains 33 SNe~II, among which we identify one \textit{gold candidate}, ASASSN-14ha.
% This gives an ECSN occurrence ratio among SNe~II, $\recii$, of $3.0^{+10.6}_{-2.9}~\%$ (central 90\% confidence interval), which we interpret as a lower-limit constraint.
This gives an ECSN occurrence ratio among SNe~II, $\recii$, of $3.0^{+10.6}_{-2.9}~\%$ with the uncertainties corresponding to the central 90\% confidence interval for a binomial proportion.
We interpret this as a lower-limit constraint.
% Treating this as a binomial proportion and using the exact Clopper--Pearson interval, we obtain a central 90\% confidence interval for the ECSN occurrence ratio among SNe~II of $\recii \gtrsim 3.0^{+10.6}_{-2.9}~\%$.
% Since a systematic sample is required to derive the volumetric rate, we simply take the lowest possible limit of $> 0~\%$ with the identifications of gold candidates.
% We simply take the lowest possible limit of $> 0~\%$ with the identifications of the \textit{gold candidates}, similarly to \citet{Hiramatsu2021-er}, as the quantitative discussion is difficult from the heterogeneous literature sample.
% Since a systematic sample is required to derive the volumetric rate, 
% % Since a systematic sample is required to derive the luminosity function and volumetric rate, 
% we limit our consideration to the ZTF survey sample.
% Since the candidates in the ZTF survey sample, SNe~2019amt, 2019lkx, 2019nzy, 2019pkh, 2021cwe, 2021hse, and 2022mxv, are \textit{silver candidates} (\ie the CSM interaction is not spectroscopically excluded), the volumetric rate derived here should be regarded as an upper limit.

% We utilize the \textit{silver candidates} found in the systematic ZTF survey sample to estimate the upper limit as the CSM interaction is not spectroscopically excluded for them.
% We use the $r$-band photometry to derive the volumetric rate. 
% A volume correction is applied using the $1/V_{\rm max}$ method \citep{Schmidt1968-lc} as done by \citet{Das2025-fu, Tanaka2016-ia, Tominaga2019-bq, Toshikage2024-oo}.
For the ZTF survey sample, we use the $r$-band photometry and apply a volume correction using the $1/V_{\rm max}$ method \citep{Schmidt1968-lc} as done by \citet{Das2025-fu, Tanaka2016-ia, Tominaga2019-bq, Toshikage2024-oo}. 
% We use the $r$-band photometry and apply a volume correction using the $1/V_{\rm max}$ method \citep{Schmidt1968-lc} as done by \citet{Das2025-fu, Tanaka2016-ia, Tominaga2019-bq, Toshikage2024-oo}. 
Each object is weighted by $1/V_{\rm max} = 1/D_{\rm max}^3$, where $D_{\rm max}$ is the furthest distance at which the transient can be detected given the limiting magnitude, 
% which we adopt $20.6$ mag \citep{Masci2019-xo}. 
which is $r=20.6$ mag for ZTF \citep{Masci2019-xo}. 
% Since our sample is required to be detected around $\tPT$, we adopt $1/V_{\rm max}$ method using the photometry at $\tPT$.
Since our sample is required to be detected around $\tPT$, we adopt $1/V_{\rm max}$ method using the brightness at $\tPT$ estimated from the linear interpolation or extrapolation of the observed photometry.
While \citet{Das2025-fu} correct for the spectroscopic completeness and the ZTF pipeline recovery efficiency factor as well, we do not as it is difficult to evaluate them for our complicated selection criteria.
% After the volume correction, $\recii$ is estimated as $15.7^{+17.3}_{-12.7}~\%$ (central 90\% confidence interval) from the \textit{silver candidates}, which we interpret as an upper-limit constraint.
After the volume correction, $\recii$ is estimated as $15.7^{+17.3}_{-12.7}~\%$ from the \textit{silver candidates}, 
% where the uncertainties are the central 90\% interval, estimated via bootstrap resampling of the $1/V_{\rm max}$-weighted sample.
where the uncertainties correspond to the central 90\% confidence interval from bootstrap resampling of the $1/V_{\rm max}$-weighted sample.
We interpret this as an upper-limit constraint.
% Combined with the estimate from the \textit{gold candidate}, we interpret the allowed range as $0.16\le \recii \le 33.1~\%$.

% Adopting the representative $\recii$ range as $3.0-15.7~\%$ from the \textit{gold} and \textit{silver candidates}, 
With the representative $\recii$ range, $3.0-15.7~\%$ from the \textit{gold} and \textit{silver candidates}, 
we next discuss the ECSN ratio among CCSNe, $\reccc$, the occurrence rate of ECSNe, $\recsn$, and the initial-mass window for stars that explode as ECSNe, $\dmec$.
% After the volume correction, the fraction of \textit{silver candidates} is $15.7$~\% of the entire SN II population.
% With the estimate of lower limit from the \textit{gold candidates}, the \textit{silver candidates} gives a ratio of the ECSNe among SNe~II of $0< \recii \lesssim15.7$~\% after the volume correction.
% Combining the lower limit from the \textit{gold candidates} with the volume-corrected contribution of the \textit{silver candidates}, we obtain an ECSN ratio among SNe~II of $0< \recii \lesssim15.7$~\%.
% With the SN II ratio among CCSNe from the LOSS survey ($\sim 58.3$~\%, \citealt{Li2011-bx}) and the estimate of lower limit from the \textit{gold candidates}, it gives a ratio of the ECSNe among CCSNe as $0< \reccc \lesssim9.2$~\%.
% If we assume the SN II ratio among CCSNe from the LOSS survey ($\sim 58.3$~\%, \citealt{Li2011-bx}), this results in a ratio of the ECSNe among CCSNe as $0< \reccc \lesssim9.2$~\%.
% Adopting the fraction of SNe II among CCSNe from the LOSS survey ($\sim 58.3$~\%, \citealt{Li2011-bx}), the $\recii$ estimates from the \textit{gold} and \textit{silver} candidates correspond to ECSN ratios among CCSNe, $\reccc$, of $1.8^{+6.2}_{-1.7}$ and $9.2^{+10.1}_{-7.4}~\%$, respectively; the quoted uncertainties include only the statistical uncertainty from our samples.
Adopting the fraction of SNe II among CCSNe from the LOSS survey ($\sim 58.3$~\%, \citealt{Li2011-bx}), 
% this corresponds to an ECSN ratio among CCSNe of $0.1< \reccc \lesssim9.2$~\%.
this corresponds to $\reccc = 1.8-9.2$~\%.
% Adopting the fraction of SNe II among CCSNe from the LOSS survey ($\sim 58.3$~\%, \citealt{Li2011-bx}), this corresponds to an ECSN ratio among CCSNe of $0< \reccc \lesssim9.2$~\%.
% With the SN II ratio among CCSNe from the LOSS survey ($\sim 58.3$~\%, \citealt{Li2011-bx}) and the estimate of lower limit from the \textit{gold candidates}, it gives a ratio of the ECSNe among CCSNe as $>0$ and $\lesssim9.2$~\%.
This is broadly consistent with 
% These broadly agree with
the estimate from a comparison of $^{86}$Kr abundance between the ECSN nucleosynthetic calculation and solar abundance ($\leq 8.5$~\%, \citealt{Wanajo2018-qu}).
% This is consistent with the ECSN ratio among CCSNe estimated from a comparison of $^{86}$Kr abundance between the ECSN nucleosynthetic calculation and solar abundance ($8.5$~\%, \citealt{Wanajo2018-qu}).
% This is consistent with the ECSN ratio ($8.5$~\%) estimated assuming ECSNe as the dominant production site of $^{86}$Kr \citep{Wanajo2018-qu}.

Combined with the volumetric rate of SNe~II from the ZTF-CLU survey ($3.9 \times 10^4~\gpcyr$, \citealt{Das2025-fu}), the $\recii$ range gives a $\recsn$ range of $(1.2-6.1) \times 10^3 ~\gpcyr$.
With the estimated $\reccc$, we discuss the initial-mass window for stars that explode as ECSNe, $\dmec$. 
% With the estimated ECSN ratios among CCSNe, we discuss the initial-mass window for stars that explode as ECSNe, $\dmec$. 
% With the estimated ECSN ratio among CCSNe, we discuss the initial-mass window for stars that explode as ECSNe, $\dmec$. 
% Assuming the Salpeter initial-mass function (IMF; \citealt{Salpeter1955-fe}), the ratios from \textit{gold} and \textit{silver candidates} correspond to progenitor mass windows of $\dmec \lesssim0.7~\Msun$, if the upper-mass limit for ECSNe lies between $8$ and $10~\Msun$.
% Assuming the Salpeter initial-mass function (IMF; \citealt{Salpeter1955-fe}), the $\reccc$ range corresponds to a progenitor mass window of $0.1 \lesssim \dmec \lesssim0.7~\Msun$, if the upper-mass limit for ECSNe lies between $8$ and $10~\Msun$.
% Assuming the Salpeter initial-mass function (IMF; \citealt{Salpeter1955-fe}), the $\reccc$ range corresponds to a progenitor mass window of $0.1 \lesssim \dmec \lesssim0.7~\Msun$, if the upper-mass limit for the stars to explode as ECSNe is whithin $8-10~\Msun$.
% Assuming the Salpeter initial-mass function (IMF; \citealt{Salpeter1955-fe}), the $\reccc$ range corresponds to a progenitor mass window of $0.1 \lesssim \dmec \lesssim0.7~\Msun$, if the mass boundary between ECSN and FeCCSN is in the range $8-10~\Msun$.
Assuming the Salpeter initial-mass function (IMF; \citealt{Salpeter1955-fe}), the $\reccc$ range corresponds to a progenitor mass window of $0.1 \lesssim \dmec \lesssim0.7~\Msun$, for an ECSN/FeCCSN boundary mass of $8-10~\Msun$.
% Assuming the Salpeter initial-mass function (IMF; \citealt{Salpeter1955-fe}), this ratio corresponds to a progenitor mass range of approximately $0.7~\Msun$, if the upper-mass limit for ECSNe is taken to be $9~\Msun$.
% This estimate is obtained by integrating the IMF over the corresponding mass range to reproduce the inferred ECSN fraction.
The derived mass window is broadly consistent with the estimate by \citet{Hiramatsu2021-er}, $0.06-0.69~\Msun$.

We next discuss the effect of host-galaxy extinction.
In this work, the literature sample from \citet{Anderson2014-fe,Galbany2016-zp} and the ZTF survey sample are not corrected for host-galaxy extinction.
% , possibly resulting in the underestimate of the ECSN rate.
% If host-galaxy extinction of $\ebv=0.1$~mag is assumed for them, number of the \textit{silver candidates} increases to $14$ whereas that of the \textit{gold candidates} remains same.
If a uniform host-galaxy extinction of $\ebv=0.1$~mag is assumed for these samples, the number of \textit{silver candidates} increases to $14$, whereas the number of \textit{gold candidates} does not change.
% This difference between the \textit{silver} and \textit{gold candidates} may reflect that, objects that appear blue because of strong CSM interaction may be classified as SNe~IIn and are less likely to remain in the literature sample as they are mostly nearby events and generally has multi-epoch spectroscopy, whereas similar objects with blue colors near the selection criteria may remained in the ZTF survey sample as the spectroscopic observation is typically less frequent and lower resolution for them.
% This difference between the \textit{silver} and \textit{gold candidates} is possibly due to the fact that nearby objects that appear blue due to strong CSM interaction tend to be classified as SNe~IIn with multi-epoch spectroscopy and are therefore less likely to remain in the literature sample, whereas similar blue objects near the selection criteria may remain in the ZTF survey sample, where spectroscopic observations are typically less frequent and of lower resolution.
This difference between the \textit{silver} and \textit{gold candidates} is possibly due to the fact that nearby, spectroscopically well-observed objects that appear blue due to strong CSM interaction may be classified as SNe~IIn, and are therefore less likely to remain in the literature SN~II sample, whereas similar blue objects may remain near the selection criterion in the ZTF survey sample, where spectroscopic observations are typically less frequent and of lower resolution.
% This difference between the \textit{silver} and \textit{gold candidates} is possibly due to the fact that nearby objects that appear blue due to strong CSM interaction are more likely to be spectroscopically classified as SNe~IIn, and therefore less likely to remain in the literature SN~II sample, which often has multi-epoch spectroscopy, whereas similar blue objects may remain near the selection criteria in the ZTF survey sample, where spectroscopic observations used for classification are typically less frequent and of lower resolution.
% This difference between the \textit{silver} and \textit{gold candidates} is possibly due to the fact that objects that appear blue due to strong CSM interaction may be classified as SNe~IIn and are therefore less likely to remain in the literature sample, which consists mostly of nearby events and generally has multi-epoch spectroscopy, whereas similar blue objects near the selection criteria may remain in the ZTF survey sample, where spectroscopic observations are typically less frequent and of lower resolution.
% This difference between the \textit{silver} and \textit{gold candidates} may indicate that objects that appear blue due to strong CSM interaction may be classified as SNe~IIn and are therefore less likely to remain in the literature sample, which consists mostly of nearby events and generally has multi-epoch spectroscopy, whereas similar blue objects near the selection criteria may remain in the ZTF survey sample, where spectroscopic observations are typically less frequent and of lower resolution.
The corresponding upper-side estimates are $\sim 25.6~\%$ for $\recii$, $\sim 14.9~\%$ for $\reccc$, $\sim1.0 \times 10^4~\gpcyr$ for $\recsn$, and $\sim 1.1~\Msun$ for $\dmec$, while the lower-side estimates remain unchanged. 
% This corresponds to the upper-side estimates of $\sim 25.6~\%$ for $\recii$, $\sim 14.9~\%$ for $\reccc$, $\sim1.0 \times 10^4~\gpcyr$ for $\recsn$, and $\sim 1.1~\Msun$ for $\dmec$, while the lower-side estimates remain unchanged. 
% Since this assumption applies the host-galaxy extinction, $\ebv=0.1$~mag, uniformly to all the SNe in the ZTF survey sample, the resulting rates and the corresponding $\dmec$ are likely overestimated, because host-galaxy extinction may be modest for most SNe~II \citep{De_Jaeger2018-vj,Das2025-fu}.
% Since this assumption applies the host-galaxy extinction, $\ebv=0.1$~mag, uniformly to all the SNe in the ZTF survey sample, these values are likely overestimated, because host-galaxy extinction may be modest for most SNe~II \citep{De_Jaeger2018-vj,Das2025-fu}.
If $\sim 10-20$~\% of SNe~II have host-galaxy extinction of $\ebv \ge 0.1$~mag (\eg \citealt{Das2025-fu}), this corresponds to $\sim 6-12$ of 62 SNe~II in the ZTF survey sample.
% Since this is comparable to the increase of the 7 \textit{silver candidates} under the uniform $\ebv=0.1$~mag assumption and , the resulting increase may be interpreted as a rough upper-side estimate, even if such non-negligible extinction is preferentially associated with the objects lying just above the ECSN selection criterion.
This is comparable to the increase of the 7 \textit{silver candidates} under the uniform $\ebv=0.1$~mag assumption, suggesting that the increase may not be much larger even if SNe~II close to the ECSN selection criterion preferentially had such non-negligible extinction.
% Since this assumption applies the host-galaxy extinction, $\ebv=0.1$~mag, uniformly to all the SNe in the ZTF survey sample, we interpret these values as upper-end estimates rather than as reliably corrected estimates, because host-galaxy extinction may be modest for most SNe~II \citep{De_Jaeger2018-vj,Das2025-fu}.
% While we assume the same host-galaxy extinction $\ebv=0.1$~mag for all the ZTF survey sample, it may be almost negligible for most SNe~II \citep{De_Jaeger2018-vj,Das2025-fu}, suggesting that the derived ECSN rates and the corresponding $\dmec$ are likely overestimated.
Also, the increase of the $7$ \textit{silver candidates} under this assumption suggests that correction of host-galaxy extinction, \eg with sufficiently high-resolution spectra or well-sampled multi-band photometry, is important for completely identifying possible ECSN candidates.
On the other hand, the upper-limit estimate can also be biased high by possible contamination from SNe~II with blue colors caused by strong CSM interaction among the \textit{silver candidates}.
% If SNe~2019amt, 2019lkx, 2019nzy, and 2022mxv are excluded, for which the light-curve models infer the existence of extended CSM, the $\recii$ from \textit{silver candidates} is $3.2^{+4.5}_{-2.7}$~\%.
% This gives a $\reccc$ of $\sim 1.9$~\%, a $\recsn$ of $\sim 1.2 \times 10^3 ~\gpcyr$, and a $\dmec$ of $\sim 0.1~\Msun$.
If SNe~2019amt, 2019lkx, 2019pkh, and 2021cwe are excluded, for which the bright tails may suggest high $\Mni$ or long-lasting CSM interaction, the $\recii$ from the \textit{silver candidates} is $11.6^{+16.9}_{-11.6}$~\%.
This gives a $\reccc$ range of $\sim 1.8-6.8$~\%, a $\recsn$ range of $\sim (1.2-4.5) \times 10^3 ~\gpcyr$, and a $\dmec$ range of $\sim 0.1-0.5\Msun$.
Also, if SNe~2019lkx and 2022mxv are excluded, for which the light-curve models infer the existence of extended CSM, the $\recii$ from the \textit{silver candidates} is $4.2^{+5.0}_{-3.0}$~\%.
This gives a $\reccc$ range of $\sim 1.8-2.5$~\%, a $\recsn$ range of $\sim (1.2-1.7) \times 10^3 ~\gpcyr$, and a $\dmec$ range of $\sim 0.1-0.2\Msun$.
% This gives a $\reccc$ of $\sim 1.9$~\%, a $\recsn$ of $\sim 1.2 \times 10^3 ~\gpcyr$, and a $\dmec$ of $\sim 0.1~\Msun$.
% If SNe~2019amt, 2019lkx, 2019nzy, and 2022mxv are excluded, for which the light-curve models infer the existence of extended CSM, the ratio of ECSNe among SNe~II is $0< \recii \lesssim3.2$~\%, that among CCSNe is $0< \reccc \lesssim1.9$~\%, the volumetric rate of ECSNe is $0< \recsn\lesssim 1.2 \times 10^3 ~\gpcyr$, and the progenitor mass window is $0< \dmec \lesssim0.1~\Msun$.
% If SNe~2019amt, 2019lkx, 2019nzy, and 2022mxv are excluded, for which the light-curve models infer the existence of extended CSM, the fraction of the \textit{gold candidates} is $3.2$~\% of the entire SN II population, the ratio of ECSNe among CCSNe is $0< \reccc \lesssim1.9$~\%, the volumetric rate of ECSNe is $0< \recsn\lesssim 1.2 \times 10^3 ~\gpcyr$, and the progenitor mass window is $0< \dmec \lesssim0.1~\Msun$.

% \section{Final Remarks and Conclusion \label{sec:summary}}
\section{Conclusion \label{sec:concl}}
% % In this work, we have searched for ECSN candidates in previously published samples and ZTF survey samples of SNe~II and investigated their multicolor light curves.
% In this work, we have searched for ECSN candidates in the past from previously published SN II samples and a ZTF survey SN II sample and investigated their multicolor light curves.
% % In this work, we have searched for ECSN candidates in the past and investigated their multicolor light curves.
% Finally, we summarize the main results and mention future prospects.

% \subsection{Summary of Main Results \label{subsec:summary}}

In this study, we have searched for ECSN candidates among previously reported SNe~II in the literature and within the ZTF public results, using the color-based diagnostic proposed by \citet{Sato2024-kt}. 
We selected ten candidates and investigated their photometric properties in detail. 
By comparing the multicolor light curves with radiation-hydrodynamical models, we inferred physical properties for the candidates, with particularly reliable estimates for the \textit{gold candidates}.
% For the \textit{gold candidates}, the inferred explosion energies are $(0.4–1.7)\times10^{50}$~erg ($(0.4-2.7)\times10^{50}$~erg including the \textit{silver candidates}), consistent with first-principles predictions \citep{Kitaura2006-ia,Janka2008-ai}, and the inferred mass-loss rates are $3\times(10^{-3}–10^{-2})~\Myr$ ($10^{-6}-3\times10^{-2}~\Myr$ including the \textit{silver candidates}), higher than those predicted for the early super-AGB phase \citep{Limongi2023-db}. 
% For the \textit{gold candidates}, the inferred explosion energies are $(0.4–1.7)\times10^{50}$~erg ($(0.4-2.7)\times10^{50}$~erg including the \textit{silver candidates}), consistent with first-principles predictions \citep{Kitaura2006-ia,Janka2008-ai}, and the inferred mass-loss rates are $3\times(10^{-3}–10^{-2})~\Myr$ (same if the \textit{silver candidates} are included), higher than those predicted for the early super-AGB phase \citep{Limongi2023-db}. 
The inferred explosion energies are $(0.4–1.7)\times10^{50}$~erg for the \textit{gold candidates} and $(0.4-2.7)\times10^{50}$~erg including the \textit{silver candidates}, consistent with first-principles predictions \citep{Kitaura2006-ia,Janka2008-ai}.
The inferred mass-loss rates are $3\times10^{-3}–3\times10^{-2}~\Myr$ for the \textit{gold candidates} and remain the same if the \textit{silver candidates} are included, higher than those predicted for the early super-AGB phase \citep{Limongi2023-db}. 
% For the \textit{gold candidates}, the inferred explosion energies are $(0.4–1.7)\times10^{50}$~erg ($(0.4–5.2)\times10^{50}$~erg including the \textit{silver candidates}), consistent with first-principles predictions \citep{Kitaura2006-ia,Janka2008-ai}, and the inferred mass-loss rates are $3\times(10^{-3}–10^{-2})~\Myr$ ($10^{-6}-3\times10^{-2}~\Myr$ including the \textit{silver candidates}), higher than those predicted for the early super-AGB phase \citep{Limongi2023-db}. 
% For the \textit{gold candidates}, the inferred explosion energies are $(0.4$–$1.7)\times10^{50}$~erg, consistent with first-principles predictions \citep{Kitaura2006-ia,Janka2008-ai}, and the inferred mass-loss rates are $3\times(10^{-3}$–$10^{-2})~\Myr$, higher than those predicted for the early super-AGB phase \citep{Limongi2023-db}, suggesting enhanced mass loss possibly driven by the increasing electron degeneracy pressure as the core mass approaches the Chandrasekhar limit. 
% For the ZTF survey sample, we applied a volume correction and derived the observational implication for the ECSN occurrence rate, $\leq 9.2~\%$ of CCSNe, corresponding to a volumetric rate of $\leq 6.1\times10^3~\gpcyr$, which implies a initial mass window of $\sim0.7~\Msun$ for their progenitors, assuming the standard IMF \citep{Salpeter1955-fe} and an upper-mass limit for ECSNe between $8$ and $10~\Msun$. 
% Utilizing the candidates, we derived the observational implication for the ECSN occurrence ratio among SNe~II, $0< r {\rm (\frac{ECSN}{SN~II})} \lesssim 15.7~\%$.
We derived the observational implication for the ECSN occurrence ratio among SNe~II, $\recii$, of $3.0^{+10.6}_{-2.9}$ and $15.7^{+17.3}_{-12.7}~\%$ from \textit{gold} and \textit{silver candidates}, respectively, which we interpret as the lower and upper limits, respectively.
% We derived the observational implication for the ECSN occurrence ratio among SNe~II, $\recii$, of $3.0^{+10.6}_{-2.9}$ and $15.7^{+17.3}_{-12.7}~\%$ from \textit{gold} and \textit{silver candidates}, respectively, which we interpret as the lower and upper limits, respectively, whereas the upper-side estimate from the \textit{silver candidates} can be affected by possible host-galaxy extinction and contamination by SNe~II showing blue color due to a strong CSM interaction.
The estimate from the \textit{silver candidates} may be affected by possible host-galaxy extinction and contamination by SNe~II showing blue color due to a strong CSM interaction.
% This corresponds to an ECSN occurrence ratio among CCSNe of $0< r {\rm (\frac{ECSN}{CCSN})} \lesssim 9.2~\%$, and a volumetric rate of $0 < \recsn \lesssim 6.1\times10^3~\gpcyr$, which implies an initial mass window of $0< \dmec \lesssim 0.7~\Msun$ for their progenitors, assuming the standard IMF \citep{Salpeter1955-fe} and an upper-mass limit for ECSNe between $8$ and $10~\Msun$. 
% The representative $\recii$ range, $3.0-15.7~\%$, corresponds to a range of ECSN occurrence ratio among CCSNe, $\reccc = 1.8-9.2~\%$, and a range of volumetric rate of ECSNe, $\recsn = (1.2-6.1)\times10^3~\gpcyr$, which implies an initial mass window of $0.1 \lesssim \dmec \lesssim 0.7~\Msun$ for their progenitors, assuming the standard IMF \citep{Salpeter1955-fe} and an upper-mass limit for ECSNe between $8$ and $10~\Msun$. 
The representative $\recii$ range, $3.0-15.7~\%$, corresponds to a range of ECSN occurrence ratio among CCSNe, $\reccc = 1.8-9.2~\%$, and a range of volumetric rate of ECSNe, $\recsn = (1.2-6.1)\times10^3~\gpcyr$, which implies an initial mass window of $0.1 \lesssim \dmec \lesssim 0.7~\Msun$ for their progenitors, assuming the standard IMF \citep{Salpeter1955-fe} and an ECSN/FeCCSN boundary mass of $8-10~\Msun$. 
% Utilizing the candidates, we derived the observational implication for the ECSN occurrence rate among CCSNe, $0< r {\rm (\frac{ECSN}{CCSN})} \lesssim 9.2~\%$, corresponding to a volumetric rate of $0 < \recsn \lesssim 6.1\times10^3~\gpcyr$, which implies an initial mass window of $0< \dmec \lesssim 0.7~\Msun$ for their progenitors, assuming the standard IMF \citep{Salpeter1955-fe} and an upper-mass limit for ECSNe between $8$ and $10~\Msun$. 
% Our findings give new empirical implications on ECSN progenitors and explosion physics, and provide an observational basis for future theoretical and time-domain studies of ECSNe.
% Our findings provide new empirical implications for ECSN progenitors and explosion physics, and an observational basis for future theoretical and time-domain studies of ECSNe.

% \subsection{Future Prospects \label{subsec:future}}

% We have searched for and investigated ECSN candidates in the past with a particular focus on the properties of their light-curve plateaus.
We finally note an outlook.
For the \textit{silver candidates}, we cannot rule out contamination of SNe that appear blue due to strong CSM interaction, as spectra around $\tPT/2$ are not available for these objects.
% However, for the \textit{silver candidates}, we cannot rule out contamination of SNe that appear blue due to strong CSM interaction, as spectra around $\tPT/2$ are not available for these objects.
Consequently, the allowed ranges of the ECSN occurrence rate and progenitor mass window remain relatively broad. 
This highlights the need for spectroscopic follow-up of future ECSN candidates, particularly around the middle of the plateau, to enable a more robust identification of ECSNe.

\begin{acknowledgements}
We thank the anonymous referee for helpful comments that improved the manuscript.
We thank Yudai Suwa, Kazumi Kashiyama, Keiichi Maeda, Ken’ichi Nomoto, Marco Limongi, Masaomi Tanaka, José Prieto, Régis Cartier, and Jared Goldberg for helpful discussions and input.
% This work is supported by the Japan Society for the Promotion of Science Open Partnership Bilateral Joint Research Project between Japan and Chile (JPJSBP120239901, 120259901) and the RSF grant 24-12-00141.
This work is supported by the Japan Society for the Promotion of Science Open Partnership Bilateral Joint Research Project between Japan and Chile (JPJSBP120239901, 120259901).
MP is supported by the RSF grant 24-12-00141 for spectral analysis.
Numerical computations were in part carried out on the general-purpose PC cluster at the Center for Computational Astrophysics, National Astronomical Observatory of Japan.
This research has made use of the NASA/IPAC Infrared Science Archive, which is funded by the National Aeronautics and Space Administration and operated by the California Institute of Technology. The relevant IPAC/IRSA service DOI is 10.26131/IRSA539.
\end{acknowledgements}

\appendix
\section{The Observational Properties of the SNe~II
\label{app:obsprop_SNeII}}
The observational properties of the SNe~II are summarized in Table~\ref{table:obsprop_SNeII}.

% \clearpage
\startlongtable
\begin{deluxetable}{ccccccccc}
\tablewidth{0pt}
\tablecaption{The Observational Properties of the SNe~II}
\label{table:obsprop_SNeII}
\tablehead{
\colhead{SN Name} & \colhead{$\texp$} & \colhead{$\tPT$} & \colhead{$\grmid$} & \colhead{$\BVmid$} & \colhead{Reference of photometry data} \\
\colhead{} & \colhead{[MJD]} & \colhead{[days]} & \colhead{[mag]} & \colhead{[mag]} & 
}
\startdata
   2005cs & $53548.50 \pm 0.25$  & $125.6~^{+1.9}_{-4.0}$ \tablenotemark{a} & -- & $1.27~^{+0.09}_{-0.04}$ & \citet{Pastorello2004-pz} \\
   1999gi & $51518.00 \pm 2.00$  & $127.8~^{+3.1}_{-3.1}$ \tablenotemark{a} & -- & $1.12~^{+0.04}_{-0.04}$ & \citet{Faran2014-fc} \\
   2000dj & $51788.00 \pm 3.50$  & $123.9$ \tablenotemark{d} & -- & $0.91~^{+0.09}_{-0.07}$ & \citet{Faran2014-fc} \\
   2002gd & $52552.00$  & $101.1$ \tablenotemark{d} & -- & $0.97~^{+0.03}_{-0.03}$ & \citet{Faran2014-fc} \\
   2003hl & $52867.00 \pm 2.50$  & $131.5$ \tablenotemark{d} & -- & $0.99~^{+0.07}_{-0.24}$ & \citet{Faran2014-fc} \\
   2003iq & $52920.00 \pm 1.00$  & $102.1~^{+2.0}_{-2.0}$ \tablenotemark{a} & -- & $0.65~^{+0.08}_{-0.08}$ & \citet{Faran2014-fc} \\
   1986l & $46708.00 \pm 1.50$ \tablenotemark{b} & $104.9~^{+2.3}_{-0.7}$ \tablenotemark{d} & -- & $1.02~^{+0.00}_{-0.01}$ & \citet{Galbany2016-zp} \\
   1992af & $48798.80 \pm 4.00$ \tablenotemark{b} & $58.5~^{+8.1}_{-13.8}$ \tablenotemark{d} & -- & $0.96~^{+0.14}_{-0.11}$ & \citet{Galbany2016-zp} \\
   1999cr & $51246.50 \pm 2.00$ \tablenotemark{b} & $101.8~^{+1.3}_{-4.7}$ \tablenotemark{d} & -- & $0.90~^{+0.14}_{-0.06}$ & \citet{Galbany2016-zp} \\
   1999em & $51476.50 \pm 2.50$ \tablenotemark{b} & $104.8~^{+3.2}_{-25.8}$ \tablenotemark{d} & -- & $1.11~^{+0.13}_{-0.01}$ & \citet{Galbany2016-zp} \\
   2003bn & $52694.50 \pm 1.50$ \tablenotemark{b} & $120.6~^{+2.1}_{-4.9}$ \tablenotemark{d} & -- & $1.12~^{+0.05}_{-0.05}$ & \citet{Galbany2016-zp} \\
   2003bl & $52696.50 \pm 2.00$ \tablenotemark{b} & $99.5~^{+1.8}_{-16.0}$ \tablenotemark{d} & -- & $1.02~^{+0.18}_{-0.07}$ & \citet{Galbany2016-zp} \\
   2003hg & $52865.50 \pm 2.50$ \tablenotemark{b} & $123.7~^{+0.1}_{-6.8}$ \tablenotemark{d} & -- & $2.01~^{+0.04}_{-0.08}$ & \citet{Galbany2016-zp} \\
   2013bu & $56399.30 \pm 2.25$  & $103.1~^{+4.5}_{-4.5}$  & $0.98~^{+0.05}_{-0.05}$ & $1.09~^{+0.21}_{-0.06}$ & \citet{Valenti2016-ao} \\
   2013fs & $56570.62 \pm 0.25$  & $82.7~^{+0.5}_{-0.5}$  & $0.53~^{+0.01}_{-0.01}$ & $0.73~^{+0.01}_{-0.01}$ & \citet{Valenti2016-ao} \\
   lsq13dpa & $56642.20 \pm 1.00$  & $128.7~^{+2.0}_{-2.0}$  & $0.79~^{+0.19}_{-0.27}$ & $1.12~^{+0.10}_{-0.10}$ & \citet{Valenti2016-ao} \\
   lsq14gv & $56674.30 \pm 1.00$  & $84.8~^{+2.4}_{-7.7}$  & $0.45~^{+0.32}_{-0.15}$ & $1.04~^{+0.13}_{-0.07}$ & \citet{Valenti2016-ao} \\
   ASASSN-14dq & $56841.00 \pm 2.75$  & $101.0~^{+5.5}_{-5.5}$  & $0.60~^{+0.08}_{-0.01}$ & $0.83~^{+0.14}_{-0.06}$ & \citet{Valenti2016-ao} \\
   2014cy & $56899.50 \pm 0.50$  & $123.8~^{+2.2}_{-3.7}$  & $1.08~^{+0.04}_{-0.07}$ & $1.29~^{+0.07}_{-0.05}$ & \citet{Valenti2016-ao} \\
   ASASSN-14gm & $56900.50 \pm 0.75$  & $110.6~^{+1.5}_{-1.5}$  & $0.58~^{+0.04}_{-0.06}$ & $0.86~^{+0.02}_{-0.04}$ & \citet{Valenti2016-ao} \\
   ASASSN-14ha & $56910.00 \pm 0.75$  & $136.8~^{+1.5}_{-1.5}$  & $0.50~^{+0.05}_{-0.05}$ & $1.08~^{+0.08}_{-0.05}$ & \citet{Valenti2016-ao} \\
   2014dw & $56957.50 \pm 5.00$  & $91.3~^{+10.0}_{-10.0}$  & $0.81~^{+0.06}_{-0.17}$ & $1.19~^{+0.19}_{-0.29}$ & \citet{Valenti2016-ao} \\
   2015W & $57024.50 \pm 5.00$  & $111.8~^{+2.3}_{-2.7}$  & $1.24~^{+0.06}_{-0.19}$ & -- & \citet{Valenti2016-ao} \\
   2018zd & $58178.41 \pm 0.05$  & $129.1~^{+0.5}_{-1.5}$ \tablenotemark{d} & $0.46~^{+0.00}_{-0.00}$ & $0.74~^{+0.00}_{-0.00}$ & \citet{Hiramatsu2021-er} \\
   2004dy & $53241.00 \pm 1.50$  & $48.0~^{+0.5}_{-0.5}$ \tablenotemark{c} & $0.89~^{+0.00}_{-0.00}$ & -- & \citet{Anderson2024-md} \\
   2004er & $53271.80 \pm 1.00$  & $157.8~^{+2.5}_{-2.5}$ \tablenotemark{c} & $1.16~^{+0.00}_{-0.00}$ & $1.48~^{+0.01}_{-0.01}$ & \citet{Anderson2024-md} \\
   2004fc & $53293.50 \pm 0.50$  & $132.4~^{+1.0}_{-1.0}$ \tablenotemark{c} & $0.96~^{+0.01}_{-0.01}$ & $1.29~^{+0.00}_{-0.00}$ & \citet{Anderson2024-md} \\
   2005J & $53382.80 \pm 3.50$  & $113.8~^{+0.7}_{-0.7}$ \tablenotemark{c} & $0.78~^{+0.00}_{-0.00}$ & $1.05~^{+0.00}_{-0.00}$ & \citet{Anderson2024-md} \\
   2005dz & $53619.50 \pm 2.00$  & $110.5~^{+0.6}_{-0.6}$ \tablenotemark{c} & $0.66~^{+0.00}_{-0.00}$ & $0.91~^{+0.00}_{-0.00}$ & \citet{Anderson2024-md} \\
   2006Y & $53766.50 \pm 2.00$  & $58.1~^{+0.5}_{-0.5}$ \tablenotemark{c} & $0.36~^{+0.01}_{-0.01}$ & $0.46~^{+0.00}_{-0.00}$ & \citet{Anderson2024-md} \\
   2006ai & $53781.80 \pm 2.50$  & $72.2~^{+0.5}_{-0.5}$ \tablenotemark{c} & $0.53~^{+0.01}_{-0.01}$ & $0.69~^{+0.01}_{-0.01}$ & \citet{Anderson2024-md} \\
   2007X & $54143.90 \pm 2.50$  & $112.6~^{+0.6}_{-0.6}$ \tablenotemark{c} & $1.03~^{+0.00}_{-0.00}$ & $1.20~^{+0.00}_{-0.00}$ & \citet{Anderson2024-md} \\
   2008K & $54477.70 \pm 2.00$  & $92.6~^{+0.8}_{-0.8}$ \tablenotemark{c} & $1.13~^{+0.01}_{-0.01}$ & $1.31~^{+0.00}_{-0.00}$ & \citet{Anderson2024-md} \\
   2008bu & $54566.80 \pm 3.50$  & $53.7~^{+1.0}_{-1.0}$ \tablenotemark{c} & $0.47~^{+0.01}_{-0.01}$ & $0.60~^{+0.01}_{-0.01}$ & \citet{Anderson2024-md} \\
   ASASSN-15bb & $57037.50$  & $107.8~^{+17.5}_{-17.5}$ \tablenotemark{c} & $0.61~^{+0.05}_{-0.10}$ & $0.82~^{+0.05}_{-0.09}$ & \citet{Anderson2024-md} \\
   2023axu & $59971.48 \pm 0.01$  & $101.2~^{+0.3}_{-0.3}$  & $0.40~^{+0.01}_{-0.01}$ & $0.63~^{+0.08}_{-0.10}$ & \citet{Shrestha2024-ty} \\
   2018inm & $58433.41 \pm 0.46$  & $75.1~^{+1.2}_{-0.9}$  & $0.78~^{+0.09}_{-0.07}$ & -- & This work \\
   2019va & $58497.61 \pm 1.51$  & $122.1~^{+1.4}_{-7.5}$  & $0.66~^{+0.03}_{-0.11}$ & -- & This work \\
   2019amt & $58513.93 \pm 0.24$  & $128.0~^{+3.4}_{-2.4}$  & $0.49~^{+0.20}_{-0.02}$ & -- & This work \\
   2019awk & $58524.48 \pm 0.95$  & $78.6~^{+0.2}_{-0.7}$  & $0.51~^{+0.18}_{-0.22}$ & -- & This work \\
   2019cec & $58561.86 \pm 0.23$  & $108.0~^{+2.3}_{-3.4}$  & $0.63~^{+0.19}_{-0.03}$ & -- & This work \\
   2019cct & $58563.29 \pm 0.87$  & $104.0~^{+0.7}_{-0.1}$  & $1.31~^{+0.09}_{-0.42}$ & -- & This work \\
   2019ceg & $58564.87 \pm 1.23$  & $124.0~^{+1.4}_{-3.6}$  & $0.96~^{+0.12}_{-0.15}$ & -- & This work \\
   2019cpo & $58565.29 \pm 1.43$  & $125.9~^{+0.3}_{-0.6}$  & $1.12~^{+0.09}_{-0.12}$ & -- & This work \\
   2019cvz & $58576.46 \pm 0.47$  & $118.0~^{+5.4}_{-0.5}$  & $0.69~^{+0.01}_{-0.02}$ & -- & This work \\
   2019dek & $58577.93 \pm 1.18$  & $98.9~^{+1.6}_{-1.4}$  & $0.92~^{+0.18}_{-0.11}$ & -- & This work \\
   2019fce & $58611.48 \pm 1.95$  & $106.6~^{+0.2}_{-0.7}$  & $0.90~^{+0.25}_{-0.19}$ & -- & This work \\
   2019fem & $58614.42 \pm 0.92$  & $95.3~^{+0.3}_{-0.5}$  & $0.90~^{+0.23}_{-0.07}$ & -- & This work \\
   2019gmh & $58634.28 \pm 0.45$  & $120.5~^{+9.1}_{-2.6}$  & $0.86~^{+0.33}_{-0.33}$ & -- & This work \\
   2019hrb & $58638.93 \pm 0.73$  & $103.1~^{+7.1}_{-0.8}$  & $1.32~^{+0.43}_{-0.05}$ & -- & This work \\
   2019hpv & $58644.96 \pm 0.72$  & $105.3~^{+2.5}_{-0.3}$  & $0.64~^{+0.17}_{-0.03}$ & -- & This work \\
   2019iex & $58660.45 \pm 0.00$  & $114.4~^{+0.1}_{-0.8}$  & $0.66~^{+0.03}_{-0.01}$ & -- & This work \\
   2019lkx & $58676.95 \pm 0.74$  & $126.9~^{+0.9}_{-0.1}$  & $0.55~^{+0.07}_{-0.04}$ & -- & This work \\
   2019mfu & $58691.97 \pm 0.71$  & $119.9~^{+4.1}_{-0.7}$  & $0.93~^{+0.09}_{-0.12}$ & -- & This work \\
   2019nzy & $58711.90 \pm 0.24$  & $100.9~^{+2.2}_{-2.5}$  & $0.37~^{+0.03}_{-0.20}$ & -- & This work \\
   2019odf & $58714.89 \pm 0.25$  & $128.1~^{+0.9}_{-3.9}$  & $0.66~^{+0.16}_{-0.02}$ & -- & This work \\
   2019ovq & $58718.99 \pm 2.21$  & $101.9~^{+8.5}_{-7.2}$  & $0.56~^{+0.18}_{-0.16}$ & -- & This work \\
   2019pkh & $58722.46 \pm 0.49$  & $128.4~^{+0.2}_{-0.7}$  & $0.58~^{+0.06}_{-0.09}$ & -- & This work \\
   2019roa & $58749.50 \pm 0.97$  & $107.6~^{+0.1}_{-1.7}$  & $0.79~^{+0.21}_{-0.24}$ & -- & This work \\
   2019rms & $58752.57 \pm 1.46$  & $117.3~^{+2.8}_{-0.1}$  & $0.75~^{+0.04}_{-0.06}$ & -- & This work \\
   2020gpe & $58947.82 \pm 1.62$  & $80.3~^{+1.9}_{-1.0}$  & $0.73~^{+0.08}_{-0.06}$ & -- & This work \\
   2020jww & $58982.37 \pm 0.51$  & $96.2~^{+1.7}_{-1.2}$  & $0.77~^{+0.12}_{-0.13}$ & -- & This work \\
   2020kbd & $58982.96 \pm 0.74$  & $111.0~^{+1.2}_{-0.6}$  & $0.52~^{+0.35}_{-0.05}$ & -- & This work \\
   2020lfn & $58995.79 \pm 0.23$  & $74.4~^{+0.2}_{-0.7}$  & $0.51~^{+0.03}_{-0.04}$ & -- & This work \\
   2020nqo & $59026.90 \pm 0.76$  & $100.7~^{+0.4}_{-0.6}$  & $0.81~^{+0.03}_{-0.07}$ & -- & This work \\
   2020ovk & $59041.42 \pm 0.46$  & $104.0~^{+1.7}_{-1.1}$  & $0.66~^{+0.18}_{-0.07}$ & -- & This work \\
   2020rhg & $59068.49 \pm 0.98$  & $89.4~^{+5.1}_{-3.6}$  & $0.80~^{+0.12}_{-0.13}$ & -- & This work \\
   2020rka & $59070.45 \pm 0.47$  & $75.0~^{+3.3}_{-0.5}$  & $0.74~^{+0.20}_{-0.35}$ & -- & This work \\
   2020rth & $59075.95 \pm 0.23$  & $122.5~^{+0.4}_{-1.6}$  & $0.87~^{+0.08}_{-0.02}$ & -- & This work \\
   2021cwe & $59256.96 \pm 0.74$  & $97.1~^{+0.7}_{-1.1}$  & $0.26~^{+0.08}_{-0.08}$ & -- & This work \\
   2021ech & $59274.32 \pm 0.42$  & $114.7~^{+0.1}_{-0.8}$  & $1.02~^{+0.09}_{-0.08}$ & -- & This work \\
   2021enz & $59275.85 \pm 0.20$  & $108.4~^{+1.6}_{-0.4}$  & $0.58~^{+0.06}_{-0.04}$ & -- & This work \\
   2021gvv & $59293.97 \pm 0.73$  & $98.4~^{+2.7}_{-5.9}$  & $0.51~^{+0.05}_{-0.06}$ & -- & This work \\
   2021hse & $59302.92 \pm 0.23$  & $100.5~^{+1.1}_{-0.8}$  & $0.32~^{+0.26}_{-0.07}$ & -- & This work \\
   2021jsf & $59313.95 \pm 1.70$  & $96.9~^{+1.2}_{-1.7}$  & $0.53~^{+0.33}_{-0.09}$ & -- & This work \\
   2021kfo & $59320.42 \pm 0.44$  & $99.1~^{+0.3}_{-1.7}$  & $0.53~^{+0.22}_{-0.04}$ & -- & This work \\
   2021qiu & $59380.92 \pm 0.23$  & $114.4~^{+0.4}_{-3.4}$  & $0.78~^{+0.04}_{-0.05}$ & -- & This work \\
   2022jnh & $59704.86 \pm 1.19$  & $115.0~^{+6.0}_{-3.6}$  & $0.91~^{+0.22}_{-0.05}$ & -- & This work \\
   2022jps & $59704.86 \pm 1.21$  & $82.5~^{+0.4}_{-1.5}$  & $0.54~^{+0.36}_{-0.03}$ & -- & This work \\
   2022jpw & $59705.42 \pm 1.43$  & $136.4~^{+1.8}_{-0.1}$  & $0.86~^{+0.24}_{-0.03}$ & -- & This work \\
   2022kbm & $59713.33 \pm 0.43$  & $112.8~^{+1.4}_{-12.3}$  & $0.71~^{+0.07}_{-0.16}$ & -- & This work \\
   2022knz & $59718.81 \pm 0.22$  & $85.7~^{+0.5}_{-4.3}$  & $0.83~^{+0.13}_{-0.08}$ & -- & This work \\
   2022mxv & $59745.92 \pm 1.22$  & $142.0~^{+1.6}_{-4.2}$  & $0.48~^{+0.11}_{-0.01}$ & -- & This work \\
   2022ojo & $59761.51 \pm 1.44$  & $113.9~^{+1.2}_{-0.7}$  & $1.43~^{+0.04}_{-0.04}$ & -- & This work \\
   2022okz & $59762.00 \pm 0.27$  & $81.6~^{+1.9}_{-1.0}$  & $0.39~^{+0.06}_{-0.07}$ & -- & This work \\
   2022omr & $59766.92 \pm 0.72$  & $124.6~^{+0.2}_{-1.9}$  & $0.71~^{+0.08}_{-0.06}$ & -- & This work \\
   2022ovb & $59773.40 \pm 0.47$  & $101.6~^{+2.6}_{-0.5}$  & $0.71~^{+0.02}_{-0.03}$ & -- & This work \\
   2022rgd & $59803.93 \pm 0.23$  & $119.6~^{+4.2}_{-0.7}$  & $0.92~^{+0.27}_{-0.38}$ & -- & This work \\
   2023eqx & $60037.43 \pm 2.44$  & $99.5~^{+0.6}_{-1.3}$  & $0.51~^{+0.07}_{-0.06}$ & -- & This work \\
   2023eyj & $60037.43 \pm 2.42$  & $99.8~^{+0.4}_{-3.6}$  & $1.01~^{+0.27}_{-0.04}$ & -- & This work \\
   2023iwz & $60079.88 \pm 0.73$  & $107.4~^{+1.1}_{-7.0}$  & $0.94~^{+0.08}_{-0.05}$ & -- & This work \\
   2023ksz & $60101.80 \pm 0.67$  & $100.9~^{+1.2}_{-0.8}$  & $0.83~^{+0.15}_{-0.18}$ & -- & This work \\
   2023kui & $60106.86 \pm 1.13$  & $114.4~^{+0.9}_{-2.9}$  & $0.81~^{+0.35}_{-0.27}$ & -- & This work \\
   2024fcp & $60393.91 \pm 1.18$  & $64.0~^{+1.1}_{-0.8}$  & $0.35~^{+0.12}_{-0.08}$ & -- & This work \\
   2024gvd & $60414.48 \pm 0.93$  & $103.4~^{+0.6}_{-1.3}$  & $0.65~^{+0.14}_{-0.11}$ & -- & This work \\
   2024grw & $60415.81 \pm 0.65$  & $83.3~^{+1.5}_{-0.5}$  & $0.65~^{+0.04}_{-0.05}$ & -- & This work \\
   2024pkt & $60503.99 \pm 0.26$  & $120.6~^{+0.4}_{-1.5}$  & $0.67~^{+0.03}_{-0.03}$ & -- & This work \\
   2024tzk & $60550.41 \pm 0.46$  & $111.1~^{+1.5}_{-2.3}$  & $1.17~^{+0.16}_{-0.09}$ & -- & This work \\
\enddata
% \tablecomments{Table 2 is published in its entirety in the electronic 
% edition of the {\it Astrophysical Journal}.  A portion is shown here 
% for guidance regarding its form and content. The {\tt\string \digitalasset}\ command highlights the Table title to visually indicate to the reader that there is data associated with this table.}
\tablecomments{
% Unless otherwise noted, $\texp$ and $\tPT$ are derived from the photometry data reference listed in the last column. $\grmid$ and $\BVmid$ for all the listed SNe are derived in this work.
Unless otherwise noted, $\texp$ and $\tPT$ are taken from the photometry reference in the last column.
$\grmid$ and $(B-V)_{\rm mid}$ are derived in this work for all SNe.
Superscripts indicate that $\texp$ and/or $\tPT$ are adopted from an alternative source or derived in this work as follows:
$^{a}$ \citet{Valenti2016-ao}; $^{b}$ \citet{Gutierrez2017-ia}; $^{c}$ \citet{Anderson2014-fe}; $^{d}$ This work.
}
% % \tablenotetext{e}{\citet{Pastorello2004-pz}}
% % \tablenotetext{f}{\citet{Faran2014-fc}}
% % \tablenotetext{j}{\citet{Galbany2016-zp}}
% \tablenotetext{a}{\citet{Valenti2016-ao}}
% \tablenotetext{b}{\citet{Gutierrez2017-ia}}
% \tablenotetext{c}{\citet{Anderson2014-fe}}
% % \tablenotetext{g}{\citet{Hiramatsu2021-er}}
% % \tablenotetext{h}{\citet{Anderson2024-md}}
% % \tablenotetext{i}{\citet{Shrestha2024-ty}}
% \tablenotetext{d}{This work}
\end{deluxetable}

\section{The inferred properties from light-curve models with a theoretically expected $\Mni$
\label{app:param_lowMNi}}
The properties of the explosions and progenitors for the ECSN candidates, inferred from the multicolor light curve models with a theoretical $\Mni$ estimate \citep{Wanajo2009-yo} are shown in Table~\ref{table:properties_ECSNe_lowMNi}.

\suppressfloats[t]
\begin{deluxetable*}{cccccccccc}
\tablewidth{0pt}
\tablecaption{Properties of the ECSN candidates, inferred from the multicolor light-curve modelings with a theoretically expected $\Mni$}
\label{table:properties_ECSNe_lowMNi}
\tablehead{
% \colhead{Number} & \colhead{Units} & \colhead{Label} & \colhead{Explanation}
\colhead{SN Name} & \colhead{\(\Menv\)} & \colhead{\(\XH\)} & \colhead{\(\Eexp\)} & \colhead{\(\Mdot\)} & \colhead{$\Rcsm$} & \colhead{\(\vi\)} & \colhead{\(\vf\)} & \colhead{\(\beta\)} & \colhead{CSM mass} \\
\colhead{} & \colhead{[$\Msun$]} & \colhead{} & \colhead{[\(10^{50}\) ergs]} & \colhead{[$\Myr$]} & \colhead{[$10^{14}$cm]} & \colhead{[km/s]} & \colhead{[km/s]} & \colhead{} & \colhead{[$\Msun$]}
}
\startdata
   ASASSN-14ha & 4.7 & 0.7 & 0.4 & $3 \times 10^{-3}$ & $3$ & 5 & 10 & 5 & 0.04 \\
   2018zd & 4.7 & 0.7 & 0.6 & $10^{-2}$ & $6$ & 5 & 10 & 5 & 0.27 \\
   2023axu & 4.7 & 0.5 & 1.4 & $10^{-2}$ & $10$ & 5 & 10 & 4 & 0.39 \\
   2019amt & 4.7 & 0.7 & 0.9 & $10^{-2}$ & $30$ & 1 & 10 & 5 & 1.64 \\
   2019lkx & 4.7 & 0.7 & 1.0 & $10^{-2}$ & $30$ & 5 & 10 & 1 & 0.99 \\
   2019nzy & 3.0 & 0.7 & 1.0 & $10^{-2}$ & $30$ & 5 & 10 & 1 & 1.00 \\
   2019pkh & 4.7 & 0.5 & 0.5 & $10^{-2}$ & $10$ & 5 & 10 & 3 & 0.37 \\
   2021cwe & 4.7 & 0.5 & 1.4 & $10^{-2}$ & $10$ & 5 & 10 & 4 & 0.39 \\
   2021hse & 4.7 & 0.5 & 1.3 & $10^{-2}$ & $10$ & 5 & 10 & 3 & 0.37 \\
   2022mxv & 4.7 & 0.7 & 2.7 & $10^{-2}$ & $30$ & 5 & 10 & 5 & 1.14 \\
\enddata
% \tablecomments{Table 2 is published in its entirety in the electronic 
% edition of the {\it Astrophysical Journal}.  A portion is shown here 
% for guidance regarding its form and content. The {\tt\string \digitalasset}\ command highlights the Table title to visually indicate to the reader that there is data associated with this table.}
\end{deluxetable*}

\facilities{IRSA}

% %% Similar to \facility{}, there is the optional \software command to allow 
% %% authors a place to specify which programs were used during the creation of 
% %% the manuscript. Authors should list each code and include either a
% %% citation or url to the code inside ()s when available.
% \software{astropy \citep{2013A&A...558A..33A,2018AJ....156..123A,2022ApJ...935..167A},  
%           Cloudy \citep{2013RMxAA..49..137F}, 
%           Source Extractor \citep{1996A&AS..117..393B}
%           }
\software{
\stella{} \citep{Blinnikov1993-vi,Blinnikov1998-ye,Blinnikov2000-xq},
AstroPy \citep{The-Astropy-Collaboration2013-xq,The-Astropy-Collaboration2018-uv,The-Astropy-Collaboration2022-yk},
NumPy \citep{harris2020array},
pandas \citep{mckinney-proc-scipy-2010, pandas_2_3_3},
SciPy \citep{2020SciPy-NMeth},
Matplotlib \citep{Hunter:2007, matplotlib_3_10_6}
}

\bibliography{main}{}
\bibliographystyle{aasjournalv7}
% \bibliographystyle{aasjournal}



%TC:endignore
\end{document}